\documentclass[12pt]{article}
\usepackage{amsfonts}
\usepackage{setspace}
\usepackage{graphicx}
\usepackage{amsthm}
\usepackage{amsmath}
\usepackage{amssymb}
\usepackage{appendix}
\usepackage{color}
\usepackage{sgame}
\usepackage{verbatim}
\usepackage{caption}
\usepackage{subcaption}
\usepackage{mathtools}
\usepackage{dsfont}
\usepackage{float}
\usepackage[margin=1in]{geometry}
\usepackage{natbib}
\usepackage[utf8]{inputenc}
\usepackage[english]{babel}
\usepackage{lscape}
\usepackage[colorlinks=true,allcolors=blue]{hyperref}
\usepackage{booktabs}
\usepackage{longtable}
\usepackage{multirow}
\usepackage{threeparttable}
\usepackage{dcolumn}
\usepackage{tikz}
\usepackage{xcolor}
\usepackage{istgame}

\usetikzlibrary{calc}
\newtheorem{claim}{Claim}

\newtheorem{corollary}{Corollary}

\newtheorem{definition}{Definition}

\newtheorem{lemma}{Lemma}

\newtheorem{proposition}{Proposition}

\DeclareMathOperator*{\argmax}{argmax}
\DeclareDocumentCommand\vpay{ m } {\begin{matrix} #1 \end{matrix}}
\begin{document}

\begin{singlespace}
\title{Cursed Sequential Equilibrium\thanks{%
Grants from the National Science Foundation (SES-0617820) and the Gordon and
Betty Moore Foundation (1158) supported this research. We are grateful
to Shengwu Li and Shani Cohen for recent correspondence that helped to
clarify the differences between the CSE and SCE approaches to the
generalization of cursed equilibrium for dynamic games. We thank
participants of the Caltech Theory Seminar and Colin Camerer for comments and
also thank Matthew Rabin for earlier discussions on the subject during his
visit at Caltech as a Moore Distinguished Scholar.}}
\author{Meng-Jhang Fong\thanks{%
Division of the Humanities and Social Sciences, California Institute of
Technology, Pasadena, CA 91125 USA. mjfong@caltech.edu} \and Po-Hsuan Lin%
\thanks{
Division of the Humanities and Social Sciences, California Institute of
Technology, Pasadena, CA 91125 USA. plin@caltech.edu} \and Thomas R. Palfrey%
\thanks{%
Corresponding Author: Division of the Humanities and Social Sciences,
California Institute of Technology, Pasadena, California 91125 USA.
trp@hss.caltech.edu. Fax: +16263958967 Phone: +16263954088} }
\date{April 11, 2023}
\maketitle

\begin{abstract}
This paper develops a framework to extend the strategic form analysis of
cursed equilibrium (CE) developed by \cite{eyster2005cursed} to multi-stage
games. The approach uses behavioral strategies rather than normal form mixed
strategies, and imposes sequential rationality. We define cursed sequential
equilibrium (CSE) and compare it to sequential equilibrium and standard
normal-form CE. We provide a general characterization of CSE and establish
its properties. We apply CSE to five applications in economics and political
science. These applications illustrate a wide range of differences between
CSE and Bayesian Nash equilibrium or CE: in signaling games; games with
preplay communication; reputation building; sequential voting;
and the dirty faces game where higher order beliefs play a key role. A common theme in several of these applications is
showing how and why CSE implies systematically different behavior than
Bayesian Nash equilibrium in dynamic games of incomplete information with
private values, while CE coincides with Bayesian Nash equilibrium for such
games.
\end{abstract}


\end{singlespace}

\thispagestyle{empty}

\newpage \setcounter{page}{1}

\section{Introduction}

\label{sec:intro}

Cursed equilibrium (CE) proposed by \cite{eyster2005cursed} is a leading
behavioral equilibrium concept that was developed to explain the
\textquotedblleft winner's curse\textquotedblright\ and related anomalies in
applied game theory. The basic idea behind CE is that individuals do not
fully take account of the dependence of other players' strategic actions on
private information. Cursed behavior of this sort has been detected in a
variety of contexts. \cite{capen1971competitive} first noted that in
oil-lease auctions, \textquotedblleft the winner tends to be the bidder who
most overestimates the reserves potential\textquotedblright\ (\cite%
{capen1971competitive}, p. 641). Since then, this observation of overbidding
relative to the Bayesian equilibrium benchmark, which can result in large
losses for the winning bidder, has been widely documented in laboratory
auction experiments
\citep{bazerman1983won, kagel1986winner, kagel1989first,
ForsytheEtAl1989BlindBidding,
dyer1989comparison, lind1991winner, kagel2009common,  ivanovEtAl2010,camererEtAl2016MaxGame}%
. In addition, the neglect of the connection between the opponents' actions
and private information is also found in non-auction environments, such as
bilateral bargaining games
\citep{samuelson1985negotiation,
holt1994loser,carrilloPalfrey2009compromise,carrillo2011no}, zero-sum
betting games with asymmetric information
\citep{rogers2009heterogeneous,
sovik2009strength}, and voting and jury decisions %
\citep{guarnaschelli2000experimental}.

While CE provides a tractable alternative to Bayesian Nash equilibrium and
can explain some anomalous behavior in games with a winner's-curse
structure, a significant limitation is that it is only developed as a
strategic form concept for simultaneous-move Bayesian games. Thus, when
applying the standard CE to dynamic games, the CE analysis is carried out on
the strategic form representation of the game, implying that CE cannot
distinguish behavior across dynamic games that differ in their timing of
moves but have the same strategic form. That is, players are assumed to
choose type-dependent contingent strategies simultaneously and \emph{not}
update their beliefs as the history of play unfolds. A further limitation
implied by the strategic form approach is that CE and standard Bayesian Nash
equilibrium make identical predictions in games with a private-values
information structure (\cite{eyster2005cursed}, Proposition 2). In this
paper we extend the CE in a simple and natural way to multi-stage games of
incomplete information. We call the new equilibrium concept \emph{Cursed
Sequential Equilibrium}\ (CSE).

In Section \ref{sec:model}, we present the framework and our extension of
cursed equilibrium to dynamic games. We consider the framework of
multi-stage games with observed actions, introduced by \cite%
{fudenberg1991perfect}, where players' private information is represented by
types, with the assumption that the set of available actions is independent
of their types at each public history. Our new solution concept is in the
same spirit of the cursed equilibrium---in our model, at each stage, players
will (partially) neglect the dependence of the other players' behavioral
strategies on their types, by placing some weight on the incorrect belief
that all types adopt the average \emph{behavioral} strategy. Specifically,
at each public history, this corresponds to the average distribution of
actions \emph{given the current belief about others' types at that stage}.
Therefore, as players update their beliefs about others' private information
via Bayes' rule, but with incorrect beliefs about the other players'
behavioral strategies, in later stages this can lead them to have \emph{%
incorrect} beliefs about the other players' average distribution of actions.

Following \cite{eyster2005cursed}'s notion of cursedness, we parameterize
the model by a single parameter $\chi\in [0,1]$ which captures the degree of
cursedness and define fully cursed ($\chi=1$) CSE analogously to fully
cursed ($\chi=1$) CE. Recall that in a fully cursed ($\chi=1$) CE, each type
of each player chooses a best reply to expected (cursed) equilibrium
distribution of other players' actions, averaged over the type-conditional
strategies of the other players, with this average distribution calculated
using the prior belief on types. Loosely speaking, a player best responds to
the average CE strategy of the others. In a $\chi$-CE, players are only
partially cursed, in the sense that each player best responds to a $\chi$%
-weighted linear combination of the \emph{average} $\chi$-CE strategy of the
others and the \emph{true} (type-dependent) $\chi$-CE strategy of the others.

The extension of this definition to multi-stage games with observed actions
is different from $\chi $-CE in two essential ways: (1) the game is analyzed
with behavioral strategies; and (2) we impose sequential rationality and
Bayesian updating. In a fully cursed ($\chi =1$) CSE, (1) implies at every
stage $t$ and each public history at $t$, each type of each player $i$
chooses a best reply to the expected (cursed) equilibrium distribution of
other players' stage-$t$ actions, averaged over the type-conditional stage-$%
t $ behavioral strategies of other players, with this average distribution
calculated using $i$'s current belief about types at stage $t$. That is,
player $i$ best responds to the average stage-$t$ CSE strategy of others.
Moreover, (2) requires that each player's belief at each public history is
derived by Bayes' rule wherever possible, and best replies are with respect
to the continuation values computed by using the fully cursed beliefs about
the behavioral strategies of the other players in current and future stages.
% ---assuming that all types of other players do not
% take account of the fact that the distribution might depend on the \emph{%
% average} strategy.

A $\chi$-CSE, for $\chi<1$, is then defined in analogously to $\chi$-CE,
except for using a $\chi$-weighted linear combination of the \emph{average} $%
\chi$-CSE \emph{behavioral strategies} of others and the \emph{true}
(type-dependent) $\chi$-CSE \emph{behavioral strategies} of others. Thus,
similar to the fully cursed CE, in a fully cursed ($\chi=1$) CSE, each
player believes other players' actions at each history are \emph{independent}
of their private information. On the other hand, $\chi=0$ corresponds to the
standard sequential equilibrium where players have correct perceptions about
other players' behavioral strategies and are able to make correct Bayesian
inferences.\footnote{%
For the off-path histories, similar to the idea of \cite{krepswilson1982},
we impose the \emph{$\chi$-consistency} requirement (see Definition \ref%
{def:chi_consistency}) so the assessment is approachable by a sequence of
totally mixed behavioral strategies. The only difference is that players'
beliefs are incorrectly updated by assuming others play the $\chi$-cursed
behavioral strategies. Hence, in our approach if $\chi=0$, a CSE is a
sequential equilibrium.}

%Finally, it is important to emphasize that we assume players are forward-looking and in equilibrium, they will make best response at every history. In other words,
%players are sequentially rational (to the
%incorrectly perceived behavioral strategies).
%For the off-path histories, similar to the
%idea of \cite{krepswilson1982}, we impose the
%\emph{$\chi$-consistency} requirement so the assessment
%is approachable by a sequence of totally mixed
%behavioral strategies. The only difference is that
%players' beliefs are incorrectly updated by
%assuming others play the $\chi$-cursed behavioral
%strategies.

After defining the equilibrium concept, in Section \ref{sec:general_property}
we explore some general properties of the model. We first prove the
existence of a cursed sequential equilibrium in Proposition \ref%
{prop:existence}. Intuitively speaking, CSE mirrors the standard sequential
equilibrium. The only difference is that players have incorrect beliefs
about the other players' behavioral strategies at each stage since they fail
to fully account for the correlation between others' actions and types at
every history. We prove in Proposition \ref{prop:upper_hemi} that the set of
CSE is upper hemi-continuous with respect to $\chi $. Consequently, every
limit point of a sequence of $\chi $-CSE points as $\chi $ converges to 0 is
a sequential equilibrium. This result bridges our behavioral solution
concept with the standard equilibrium theory. Finally, we also show in
Proposition \ref{prop:one_period} that $\chi $-CSE is equivalent to $\chi $%
-CE for one-stage games, demonstrating the connection between the two
behavioral solutions.

In multi-stage games, cursed beliefs about behavioral strategies will
distort the evolution of a player's beliefs about the other players' types.
As shown in Proposition \ref{prop:equiv_def}, a direct consequence of the
distortion is that in $\chi $-CSE players tend to update their beliefs about
others' types too passively. That is, there is some persistence in beliefs
in the sense that at each stage $t$, each $\chi $-cursed player's belief
about any type profile is at least $\chi $ times the belief about that type
profile at stage $t-1$. Among other things, this implies that if the \emph{%
prior} belief about the types is full support and $\chi >0$, the full
support property will persist at all histories, and players will (possibly
incorrectly) believe every profile of others' types is possible at every
history.

This dampened updating property plays an important role in our framework.
Not only does it contribute to the difference between CSE and the standard
CE through the updating process, but it also implies additional restrictions
on off-path beliefs. The effect of dampened updating is starkly illustrated
in the pooling equilibria of signaling games where every type of sender
behaves the same everywhere. In this case, Proposition \ref{prop:pooling_cse}
shows if an assessment associated with a pooling equilibrium is a $\chi $%
-CSE, then it also a $\chi^{\prime}$-CSE for all $\chi ^{\prime }\leq \chi $%
, but it is not necessarily a pooling equilibrium for all $\chi ^{\prime
}>\chi $. This contrasts with one of the main results about CE, that if a
pooling equilibrium is a $\chi $-CE for some $\chi $, then it is a $\chi
^{\prime }$-CE for all $\chi ^{\prime }\in \lbrack 0,1]$ (\cite%
{eyster2005cursed}, Proposition 3).

This suggests that perhaps the dampened updating property is an equilibrium
selection device that eliminates some pooling equilibrium, but actually this
is \emph{not} a general property. As we demonstrate later, the $\chi $-CE
and $\chi $-CSE sets can be non-overlapping, which we illustrate with a variety
of applications. The intuition is that in CSE, players generally do not have
correct beliefs about the opponents' average behavioral strategies. The
pooling equilibrium is just a special case where players have correct
beliefs.

In Section \ref{sec:application} we explore the implications of cursed
sequential equilibrium with five applications in economics and political
science. Section \ref{subsec:signaling} analyzes the $\chi $-CSE of
signaling games. Besides studying the theoretical properties of pooling $%
\chi $-CSE, we also analyze two simple signaling games that were studied in
a laboratory experiment \citep{brandts1993adjustment}. We show how varying
the degree of cursedness can change the set of $\chi $-CSE in these two
signaling games in ways that are consistent with the reported experimental
findings. Next, we turn to the exploration of how sequentially cursed
reasoning can influence strategic communication. To this end, we analyze the
$\chi $-CSE for a public goods game with communication %
\citep{palfrey1991testing,NilanjanEtAl2017communication} in Section \ref%
{subsec:pgg}, finding that $\chi$-CSE predicts there will be less effective
communication when players are more cursed.

%and the classic information transmission game by \cite%
%{crawford1982strategic} in Section \ref{subsec:sender_receiver}. In the
%information transmission game, we find that the $\chi $-CSE predicts \emph{%
%more} information transmission than is possible in the most informative
%Bayesian Nash equilibrium. This is consistent with experimental evidence of
%overcommunication in information transmission games reported in \cite%
%{cai2006overcommunication}.

Next, in Section \ref{subsec:centipede} we apply $\chi $-CSE to the
centipede game studied experimentally by \cite{mckelvey1992experimental}
where one of the players believes the other player might be an
\textquotedblleft altruistic\textquotedblright\ player who always passes.
This is a simple reputation-building game, where selfish types can gain by
imitating altruistic types in early stages of the game. The public goods
application and the centipede game are both private-values environments, so
these two applications clearly demonstrate how CSE departs from CE and the
Bayesian Nash equilibrium, and shows the interplay between sequentially
cursed reasoning and the learning of types in private-value models.

In strategic voting applications, conditioning on \textquotedblleft
pivotality\textquotedblright ---the event where your vote determines the
final outcome---plays a crucial role in understanding equilibrium voting
behavior. To illustrate how cursedness distorts the pivotal reasoning, in
Section \ref{subsec:voting} we study the three-voter two-stage agenda voting
game introduced by \cite{ordeshook1988agendas}. Since this is a private
value game, the predictions of the $\chi $-CE and the Bayesian Nash
equilibrium coincide for all $\chi $. That is, cursed equilibrium predicts
no matter how cursed the voters are, they are able to correctly perform
pivotal reasoning. On the contrary, our CSE predicts that cursedness will
make the voters less likely to vote strategically.
This is consistent with the empirical evidence about the prevalence of
sincere voting over sequential agendas when inexperienced voters have
incomplete information about other voters' preferences (%
\citealp{levine1977agenda}; \citealp{plottLevine1978agenda}; %
\citealp{EckelHolt1989agenda}).

Finally, in Section \ref{subsec:dirty} we study the relationship between
cursedness and epistemic reasoning by considering the two-person dirty faces
game previously studied by \cite{weber2001behavior} and \cite{bayer2007dirty}%
. In this game, $\chi $-CSE predicts cursed players are, to some extent,
playing a \textquotedblleft coordination\textquotedblright\ game where they
coordinate on a specific learning speed about their face types. Therefore,
from the perspective of CSE, the non-equilibrium behavior observed in
experiments can be interpreted as possibly due to a coordination failure
resulting from cognitive limitations.

The cursed sequential equilibrium extends the concept of cursed equilibrium
from static Bayesian games to multi-stage games with observed actions. This
generalization preserves the spirit of the original cursed equilibrium in a
simple and tractable way, and provides additional insights about the effect
of cursedness in dynamic games. A contemporaneous working paper by \cite%
{cohen2022sequential} is closely related to our paper. That paper adopts an
approach based on the coarsening of information sets to define sequential
cursed equilibrium (SCE) for extensive form games with perfect recall. The
SCE model captures a different kind of cursedness\footnote{%
We illustrate some implications of these differences in the application to
signaling games in Section \ref{subsec:signaling}. For a more detailed discussion of the differences between CSE and SCE, see \cite{fong2023cursednote}} that arises if a player
neglects the dependence of other players' unobserved (i.e., either future or
simultaneous) actions on the history of play in the game, which is different
from the dependence of other players' actions on their type (as in CE and
CSE). In the terminology of \cite{eyster2005cursed} (p. 1665), the
cursedness is with respect to endogenous information, i.e., what players
observe about the path of play. The idea is to treat the unobserved actions
of other players in response to different histories (endogenous information)
similarly to how cursed equilibrium treats players' types.
% A (fully) cursed player in
%SCE believes the distribution of unobserved actions of a player is constant
%across a suitably defined equivalence class (coarsening) of information
%sets.
A two-parameter model of partial cursedness is developed, and a series of
examples demonstrate that for plausible parameter values, the model is
consistent with some experimental findings related to the failure of
subjects to fully take account of unobserved hypothetical events, whereas
behavior is \textquotedblleft more rational\textquotedblright\ if subjects
make decisions after directly observing such events. At a more conceptual
level, our paper is related to several other behavioral solution concepts
developed for dynamic games, such as agent quantal response equilibrium
(AQRE) \citep{mckelvey1998quantal}, dynamic cognitive hierarchy theory (DCH)
\citep{lin2022cognitive,
lin2022cognitivemultii}, and the analogy-based expectation equilibrium
(ABEE) \citep{jehiel2005analogy, jehiel2008revisiting}, all of which modify
the requirements of sequential equilibrium in different ways than cursed
sequential equilibrium.

\section{The Model}

\label{sec:model}

Since CSE is a solution concept for dynamic games of incomplete information,
in this paper we will focus on the framework of multistage games with
observed actions \citep{fudenberg1991perfect}. Section \ref{subsec:game_def}
defines the formal structure of multi-stage games with observed actions,
followed by Section \ref{subsec:cse_def}, where the $\chi $-cursed
sequential equilibrium is formally developed.

\subsection{Multi-Stage Games with Observed Actions}

\label{subsec:game_def}

Let $N=\{1,\ldots ,n\}$ be a finite set of players. Each player $i\in N$ has
a \textit{type} $\theta _{i}$ drawn from a finite set $\Theta _{i}$. Let $%
\theta \in \Theta \equiv \times _{i=1}^{n}\Theta _{i}$ be the type profile
and $\theta _{-i}\in \Theta _{-i}\equiv \times _{j\neq i}\Theta _{j}$ be the
type profile without player $i$. All players share a common (full support)
prior distribution $\mathcal{F}(\cdot ):\Theta \rightarrow (0,1)$.
Therefore, for every player $i$, the belief of other players' types
conditional on his own type is
\begin{equation*}
\mathcal{F}(\theta _{-i}|\theta _{i})=\frac{\mathcal{F}(\theta _{-i},\theta
_{i})}{\sum_{\theta _{-i}^{\prime }\in \Theta _{-i}}\mathcal{F}(\theta
_{-i}^{\prime },\theta _{i})}.
\end{equation*}%
%
%
%
%
%
%If the types are independent across players, then for each player $i$, his
%belief of other players' types is $\mathcal{F}_{-i}(\theta _{-i})=\Pi
%_{j\neq i}\mathcal{F}_{j}(\theta _{j})$ where $\mathcal{F}_{j}(\theta _{j})$
%is the marginal distribution of player $j$'s type.
At the beginning of the game, players observe their own types, but not the
other players' types. That is, each player's type is his own private
information.

The game is played in stages $t=1,2,\ldots ,T$. %\footnote{%
%We focus on finite horizon games although $T$ can be arbitrarily large.}
In each stage, players simultaneously choose actions, which will be revealed
at the end of the stage. The feasible set of actions can vary with
histories, so games with alternating moves are also included. Let $\mathcal{H%
}^{t-1}$ be the set of all possible histories at stage $t$, where $\mathcal{H%
}^{0}=\{h_{\emptyset }\}$ and $\mathcal{H}^{T}$ is the set of terminal
histories. Let $\mathcal{H}=\cup _{t=0}^{T}\mathcal{H}^{t}$ be the set of
all possible histories of the game, and $\mathcal{H}\backslash \mathcal{H}%
^{T}$ be the set of non-terminal histories.

For every player $i$, the available information at stage $t$ is in $\Theta
_{i}\times \mathcal{H}^{t-1}$. Therefore, player $i$'s information sets can
be specified as $\mathcal{I}_{i}\in \mathcal{Q}_{i}=\{(h,\theta ):h\in
\mathcal{H}\backslash \mathcal{H}^{T},\theta _{i}\in \Theta _{i}\}$.
That is, a type $\theta_{i}$ player $i$'s information set at the public history $h^{t}$ can be defined as $\bigcup_{\theta_{-i} \in \Theta_{-i}} (\theta_i, \theta_{-i}, h^t)$.
With a slight abuse of notation, it will be denoted as $(\theta_i, h^t)$.
For the sake of simplicity, we assume that, at each history, the feasible set of
actions for every player is independent of their type %s.\footnote{%
%The consequence of this assumption is that players cannot signal their own
%type by choosing some action that is not available to some other types.}
and use $A_{i}(h^{t-1})$ to denote the feasible set of actions for player $i$
at history $h^{t-1}$. Let $A_{i}={\times}_{h\in \mathcal{H}\backslash
\mathcal{H}^{T}}A_{i}(h)$ denote player $i$'s feasible actions in all
histories of the game and $A=A_{1}\times \cdots \times A_{n}$.
%We assume that all players in the game have perfect recall (see %
%\citealp{krepswilson1982} for the definition).
In addition, we assume $A_{i}$ is finite for all $i\in N$ and $%
|A_{i}(h)|\geq 1$ for all $i\in N$ and any $h\in \mathcal{H}\backslash
\mathcal{H}^{T}$.

A behavioral strategy for player $i$ is a function $\sigma _{i}:\mathcal{Q}%
_{i}\rightarrow \Delta (A_{i})$ satisfying $\sigma _{i}(h^{t-1},\theta
_{i})\in \Delta (A_{i}(h^{t-1}))$. Furthermore, we use $\sigma
_{i}(a_{i}^{t}|h^{t-1},\theta _{i})$ to denote the probability player $i$
chooses $a_{i}^{t}\in A_{i}(h^{t-1})$. We use $a^{t}=(a_{1}^{t},\ldots
,a_{n}^{t})\in \times _{i=1}^{n}A_{i}(h^{t-1})\equiv A(h^{t-1})$ to denote
the action profile at stage $t$ and $a_{-i}^{t}$ to denote the action
profile at stage $t$ without player $i$. If $a^{t}$ is the action profile
realized at stage $t$, then $h^{t}=(h^{t-1},a^{t})$. Finally, each player $i$
has a payoff function $u_{i}:\mathcal{H}^{T}\times \Theta \rightarrow
\mathbb{R},$ and we let $u=(u_{1},\ldots ,u_{n})$ be the profile of payoff
functions. A multi-stage game with observed actions, $\Gamma$, is defined by
the tuple $\Gamma = \langle T, A, N, \mathcal{H}, \Theta, \mathcal{F},
u\rangle$.

\subsection{Cursed Sequential Equilibrium}

\label{subsec:cse_def}

In a multi-stage game with observed actions, a solution is defined by an
\textquotedblleft assessment,\textquotedblright\ which consists of a
(behavioral) strategy profile $\sigma $, and a belief system $\mu $. Since
action profiles will be revealed to all players at the end of each stage,
the belief system specifies, for each player, a conditional distribution
over the set of type profiles conditional on each history. Consider an
assessment $(\mu ,\sigma )$. Following the spirit of the cursed equilibrium,
for player $i$ at stage $t$, we define the \emph{average behavioral strategy
profile of the other players} as:
\begin{equation*}
\bar{\sigma}_{-i}(a_{-i}^{t}|h^{t-1},\theta _{i})=\sum_{\theta _{-i}\in
\Theta _{-i}}\mu _{i}(\theta _{-i}|h^{t-1},\theta _{i})\sigma
_{-i}(a_{-i}^{t}|h^{t-1},\theta _{-i})
\end{equation*}%
for any $i\in N$, $\theta_i \in \Theta_i $ and $h^{t-1}\in \mathcal{H}^{t-1}$%
.

In CSE, players have incorrect perceptions about other players' behavioral
strategies. Instead of thinking they are using $\sigma _{-i}$, a $\chi $%
-cursed\footnote{%
We assume throughout the paper that all players are equally cursed, so there
is no $i$ subscript on $\chi $. The framework is easily extended to allow
for heterogeneous degrees of cursedness.} type $\theta_i$ player $i$ would
believe the other players are using a $\chi $-weighted average of the
average behavioral strategy and the true behavioral strategy:\footnote{%
If $\chi =0$, players have correct beliefs about the other players'
behavioral strategies at every stage.}
\begin{equation*}
\sigma _{-i}^{\chi }(a_{-i}^{t}|h^{t-1},\theta _{-i},\theta _{i})=\chi \bar{%
\sigma}_{-i}(a_{-i}^{t}|h^{t-1},\theta _{i})+(1-\chi )\sigma
_{-i}(a_{-i}^{t}|h^{t-1},\theta _{-i}).
\end{equation*}

The beliefs of player $i$ about $\theta _{-i}$ are updated in the $\chi $%
-CSE via Bayes' rule, whenever possible, assuming other players are using
the $\chi $-cursed behavioral strategy rather than the true behavioral
strategy. We call this updating rule the $\chi $\emph{-cursed Bayes' rule}.\
Specifically, an assessment satisfies the $\chi $-cursed Bayes' rule if the
belief system is derived from the Bayes' rule while perceiving others are
using $\sigma _{-i}^{\chi }$ rather than $\sigma _{-i}$.

\begin{definition}
\label{def:cursed_bayes} $(\mu ,\sigma )$ satisfies $\chi $-cursed Bayes'
rule if the following rule is applied to update the posterior beliefs
whenever $\sum_{\theta _{-i}^{\prime }\in \Theta _{-i}}\mu _{i}(\theta
_{-i}^{\prime }|h^{t-1},\theta _{i})\sigma _{-i}^{\chi
}(a_{-i}^{t}|h^{t-1},\theta _{-i}^{\prime },\theta _{i})>0$:
\begin{equation*}
\mu _{i}(\theta _{-i}|h^{t},\theta _{i})=\frac{\mu _{i}(\theta
_{-i}|h^{t-1},\theta _{i})\sigma _{-i}^{\chi }(a_{-i}^{t}|h^{t-1},\theta
_{-i},\theta _{i})}{\sum_{\theta _{-i}^{\prime }\in \Theta _{-i}}\mu
_{i}(\theta _{-i}^{\prime }|h^{t-1},\theta _{i})\sigma _{-i}^{\chi
}(a_{-i}^{t}|h^{t-1},\theta _{-i}^{\prime },\theta _{i})}.
\end{equation*}
\end{definition}

% We place a consistency restriction, analogous to consistent assessments in
% sequential equilibrium, on how $\chi $-cursed beliefs are updated off the
% equilibrium path, i.e., when
% $$\sum_{\theta _{-i}^{\prime }\in \Theta
% _{-i}}\mu _{i}(\theta _{-i}^{\prime }|h^{t-1},\theta _{i})\sigma _{-i}^{\chi
% }(a_{-i}^{t}|h^{t-1},\theta _{-i}^{\prime },\theta _{i})=0.$$
Let $\Sigma ^{0} $ be the set of totally mixed behavioral strategy profiles,
and let $\Psi ^{\chi }$ be the set of assessments $(\mu ,\sigma )$ such that
$\sigma \in \Sigma ^{0}$ and $\mu $ is derived from $\sigma $ using $\chi $%
-cursed Bayes' rule.\footnote{%
In the following, we will use $\mu^\chi(\cdot)$ to denote the belief system
derived under $\chi$-cursed Bayes' Rule. Also, note that both $%
\sigma_{-i}^\chi$ and $\mu^\chi$ are induced by $\sigma$; that is, $%
\sigma_{-i}^{\chi}(\cdot) = \sigma_{-i}^{\chi}[\sigma](\cdot)$ and $%
\mu^{\chi}(\cdot) = \mu^{\chi}[\sigma](\cdot)$. For the ease of exposition,
we drop $[\sigma]$ when it does not cause confusion.} Lemma \ref%
{lemma:rearrange} below shows that another interpretation of the $\chi $%
-cursed Bayes' rule is that players have correct perceptions about $\sigma
_{-i}$ but are unable to make perfect Bayesian inference when updating
beliefs. From this perspective, player $i$'s cursed belief is simply a
linear combination of player $i$'s cursed belief at the beginning of that
stage (with $\chi $ weight) and the Bayesian posterior belief (with $1-\chi $
weight). Because $\sigma $ is totally mixed, there are no off-path histories.

\begin{lemma}
\label{lemma:rearrange} For any $(\mu, \sigma)\in \Psi^{\chi}$, $i\in N$, $%
h^{t} = (h^{t-1}, a^t)\in \mathcal{H}\backslash \mathcal{H}^{T}$ and $\theta
\in \Theta$,
\begin{align*}
\mu_i(\theta_{-i}| h^{t}, \theta_i) = \chi \mu_i(\theta_{-i}| h^{t-1},
\theta_i) + (1-\chi)\left[\frac{\mu_i(\theta_{-i}|h^{t-1},\theta_i)%
\sigma_{-i}(a_{-i}^t|h^{t-1},\theta_{-i})}{\sum_{\theta^{\prime}_{-i}}\mu_i(%
\theta^{\prime}_{-i}|h^{t-1},\theta_i)\sigma_{-i}(a_{-i}^t|h^{t-1},\theta^{%
\prime}_{-i})} \right]
\end{align*}
\end{lemma}

\begin{proof}
See Appendix \ref{appendix_proof_general}.
\end{proof}

This is analogous to Lemma 1 of \cite{eyster2005cursed}.
%The key difference is that in $\chi $-CSE, $\chi $%
%-cursed players will update their beliefs with this $\chi $-cursed Bayes'
%rule at every history, while players in $\chi $-CE only make this cursed
%inference at the initial history.
Another insight provided by Lemma \ref{lemma:rearrange} is that even if
player types are independently drawn, i.e., $\mathcal{F}(\theta )=\Pi
_{i=1}^{n}\mathcal{F}_{i}(\theta _{i})$, players' cursed beliefs about other
players' types are generally \emph{not} independent across players. That is,
in general, $\mu _{i}(\theta _{-i}|h^{t},\theta _{i})\neq \Pi _{j\neq i}\mu
_{ij}(\theta _{j}|h^{t},\theta _{i}).$ The belief system will preserve the
independence only when the players are either fully rational ($\chi=0$) or
fully cursed ($\chi=1$).

%\textbf{(TP) SHOULD WE DROP THIS PARAGRAPH?? This is because even if }$\mu
%_{i}(\theta _{-i}|h^{t-1},\theta _{i})$\textbf{\ and the Bayesian posterior
%are both independent across players, that independence is generally not
%preserved under the linear combination of the two objects. In other words,
%with }$\chi $\textbf{-cursed Bayes' rule, every player }$i$\textbf{\ will
%incorrectly believe any other player }$j$\textbf{'s action conveys some
%private information about player }$k\neq i,j$\textbf{. The intuition behind
%this is that a }$\chi $\textbf{-cursed player incorrectly thinks all other
%players will play the average behavioral strategy---which will be affected
%by all other players---with probability }$\chi $\textbf{. As a result, when
%assuming the prior belief }$F$\textbf{\ to be independent, the belief system
%of }$\chi $\textbf{-CSE will generically violate the \textquotedblleft
%no-signaling-what-you-don't-know\textquotedblright\ condition introduced by
%\cite{fudenberg1991perfect}.}

Finally, we place a consistency restriction, analogous to consistent
assessments in sequential equilibrium, on how $\chi $-cursed beliefs are
updated off the equilibrium path, i.e., when
\begin{equation*}
\sum_{\theta _{-i}^{\prime }\in \Theta _{-i}}\mu _{i}(\theta _{-i}^{\prime
}|h^{t-1},\theta _{i})\sigma _{-i}^{\chi }(a_{-i}^{t}|h^{t-1},\theta
_{-i}^{\prime },\theta _{i})=0.
\end{equation*}

An assessment satisfies $\chi$-consistency if it is in the closure of $%
\Psi^\chi$.

\begin{definition}
\label{def:chi_consistency} $(\mu ,\sigma )$ satisfies $\chi $-consistency
if there is a sequence of assessments $\{(\mu _{k},\sigma _{k})\}\subseteq
\Psi ^{\chi }$ such that $\lim_{k\rightarrow \infty }(\mu _{k},\sigma
_{k})=(\mu ,\sigma )$.
\end{definition}

For any $i \in N$, $\chi \in [0,1]$, $\sigma$, and $\theta \in \Theta$, let $%
\rho_i^\chi(h^T|h^t, \theta,\sigma_{-i}^\chi,\sigma_i)$ be player $i$'s
perceived conditional realization probability of terminal history $h^T \in
\mathcal{H}^T$ at history $h^t \in \mathcal{H}\backslash\mathcal{H}^T$ if
the type profile is $\theta$ and player $i$ uses the behavioral strategy $%
\sigma_i$ whereas perceives other players' using the cursed behavioral
strategy $\sigma_{-i}^\chi$. At every non-terminal history $h^t$, a $\chi$%
-cursed player in $\chi$-CSE will use $\chi$-cursed Bayes' rule (Definition %
\ref{def:cursed_bayes}) to derive the posterior belief about the other
players' types. Accordingly, a type $\theta_i$ player $i$'s conditional
expected payoff at history $h^t$ is given by:
\begin{equation*}
\mathbb{E}u_i(\sigma|h^t, \theta_i) = \sum_{\theta_{-i} \in \Theta_{-i}}
\sum_{h^T \in \mathcal{H}^T} \mu_i(\theta_{-i}|h^t, \theta_i)
\rho_i^\chi(h^T|h^t, \theta,\sigma_{-i}^\chi,\sigma_i) u_i(h^T, \theta_i,
\theta_{-i}).
\end{equation*}

\begin{definition}
An assessment $(\mu^*, \sigma^*)$ is a $\chi$-cursed sequential equilibrium
if it satisfies $\chi$-consistency and $\sigma_i^*(h^t, \theta_i)$ maximizes
$\mathbb{E}u_i(\sigma^*|h^t, \theta_i)$ for all $i$, $\theta_i$, $h^t \in
\mathcal{H}\backslash\mathcal{H}^T$. % $$
\end{definition}

%From the definition of $\chi$-CSE, we can observe that although players have
%incorrect perceptions about other players' behavioral strategies, they can
%perfectly foresee their future actions. Therefore, similar to the standard
%equilibrium model, $\chi$-CSE can also be solved by backward induction.

\section{General Properties of \texorpdfstring{$\chi $}--CSE}

\label{sec:general_property}

In this section, we characterize some general theoretical properties of $%
\chi $-CSE. The first result is the existence of the $\chi $-CSE. The
definition of $\chi $-CSE mirrors the definition of the sequential
equilibrium by \cite{krepswilson1982}---the only difference is that players
in $\chi $-CSE update their beliefs by $\chi $-cursed Bayes' rule and best
respond to $\chi $-cursed (behavioral) strategies. Therefore, one can prove
the existence of $\chi $-CSE in a similar way as in the standard argument of
the existence of sequential equilibrium.

\begin{proposition}
\label{prop:existence} For any $\chi\in[0,1]$ and any finite multi-stage
game with observed actions, there is at least one $\chi$-CSE.
\end{proposition}

\begin{proof}
We briefly sketch the proof here, and the details can be found in
Appendix \ref{appendix_proof_general}.

Fix any $\chi\in[0,1]$. For any $i\in N$ and any information set
$\mathcal{I}_i = (h^{t-1}, \theta_i)$, player $i$ has to choose every
action $a_i^t\in A_i(h^{t-1})$ with probability at
least $\epsilon$.
Since there are no off-path histories, the belief system is uniquely
pinned down by $\chi$-cursed Bayes' rule and a $\chi$-CSE
exists in this $\epsilon$-constrained game. We denote this $\chi$-CSE as $(\mu^\epsilon, \sigma^\epsilon)$.
By compactness, there is a converging sub-sequence of assessments such that
$(\mu^\epsilon, \sigma^\epsilon) \rightarrow (\mu^*, \sigma^*)$ as $\epsilon\rightarrow 0$,
which is a $\chi$-CSE, as desired.
\end{proof}

Let $\Phi(\chi)$ be the correspondence that maps $\chi\in [0,1]$ to the set
of $\chi$-CSE. Proposition \ref{prop:existence} guarantees $\Phi(\chi)$ is
non-empty for any $\chi\in[0,1]$. Because $\chi$-cursed Bayes' rule changes
continuously in $\chi$, we can further prove in Proposition \ref%
{prop:upper_hemi} that $\Phi(\chi)$ is an upper hemi-continuous
correspondence.

\begin{proposition}
\label{prop:upper_hemi} $\Phi(\chi)$ is upper hemi-continuous with respect
to $\chi$.
\end{proposition}

\begin{proof}
The proof follows a standard argument.
See Appendix \ref{appendix_proof_general} for details.
\end{proof}

As shown in Corollary \ref{coro:limit_0}, a direct consequence of upper
hemi-continuity is that every limit point of a sequence of $\chi $-CSE when $%
\chi \rightarrow 0$ is a sequential equilibrium. This result bridges our
behavioral equilibrium concept with standard equilibrium theory.

\begin{corollary}
\label{coro:limit_0} Every limit point of a sequence of $\chi $-CSE with $%
\chi $ converging to 0 is a sequential equilibrium.
\end{corollary}

\begin{proof}
By Proposition \ref{prop:upper_hemi}, we know $\Phi(\chi)$ is upper
hemi-continuous at $0$. Consider of a sequence of $\chi$-CSE. As
$\chi\rightarrow0$, the limit point remains a CSE, which is a
sequential equilibrium at $\chi=0$.
This completes the proof.
\end{proof}

Finally, by a similar argument to \cite{krepswilson1982}, for any $\chi\in[%
0,1]$, $\chi$-CSE is also upper hemi-continuous with respect to payoffs. In
other words, our $\chi $-CSE preserves the continuity property of sequential
equilibrium.

The next result is the characterization of a necessary condition for $\chi $%
-CSE. As seen from Lemma \ref{lemma:rearrange}, players update their beliefs
more passively in $\chi $-CSE than in the standard equilibrium---they put $%
\chi $-weight on their beliefs formed in previous stage. %prior beliefs.
To formalize this, we define the $\chi $\textit{-dampened updating property}
in Definition \ref{def:passive}. An assessment satisfies this property if at
\emph{any} non-terminal history, the belief puts at least $\chi $ weight on
the belief in previous stage---both on and off the equilibrium path. In
Proposition \ref{prop:equiv_def}, we show that $\chi $-consistency implies
the $\chi $-dampened updating property.

\begin{definition}
\label{def:passive} An assessment $(\mu ,\sigma )$ satisfies the $\chi $%
-dampened updating property if for any $i\in N$, $\theta \in \Theta $ and $%
h^{t}=(h^{t-1},a^{t})\in \mathcal{H}\backslash \mathcal{H}^{T}$,
\begin{equation*}
\mu _{i}(\theta _{-i}|h^{t},\theta _{i})\geq \chi \mu _{i}(\theta
_{-i}|h^{t-1},\theta _{i}).
\end{equation*}
\end{definition}

%The following lemma about updating when $(\mu ,\sigma )\in \Psi ^{\chi }$ is
%used in the next proposition:
%
%\begin{lemma}
%\label{lemma:rearrange copy(1)} For any $(\mu ,\sigma )\in \Psi ^{\chi }$, $%
%i\in N$, $h^{t}=(h^{t-1},a^{t})\in \mathcal{H}\backslash \mathcal{H}^{T},$ $%
%\chi \in \lbrack 0,1]$ and $\theta \in \Theta $,
%\begin{equation*}
%\mu _{i}(\theta _{-i}|h^{t},\theta _{i})=\chi \mu _{i}(\theta
%_{-i}|h^{t-1},\theta _{i})+(1-\chi )\left[ \frac{\mu _{i}(\theta
%_{-i}|h^{t-1},\theta _{i})\sigma _{-i}(a_{-i}^{t}|h^{t-1},\theta _{-i})}{%
%\sum_{\theta _{-i}^{\prime }}\mu _{i}(\theta _{-i}^{\prime }|h^{t-1},\theta
%_{i})\sigma _{-i}(a_{-i}^{t}|h^{t-1},\theta _{-i}^{\prime })}\right]
%\end{equation*}
%\end{lemma}
%
%\begin{proof}
%See Appendix \ref{appendix_proof_general}.
%\end{proof}
%
%Lemma \ref{lemma:rearrange} is indeed parallel to Lemma 1 of \cite%
%{eyster2005cursed}. The key difference is that in $\chi $-CSE, $\chi $%
%-cursed players will update their beliefs with this $\chi $-cursed Bayes'
%rule at every subgame, while players in $\chi $-CE only make this cursed
%inference at the initial history.

\begin{proposition}
\label{prop:equiv_def} $\chi $-consistency implies $\chi $-dampened updating
for any $\chi \in \lbrack 0,1]$.
\end{proposition}

\begin{proof}
See Appendix \ref{appendix_proof_general}.
\end{proof}

It follows that if assessment $(\mu ,\sigma )$ satisfies the $\chi $%
-dampened updating property, then for any player $i$, any history $h^{t}$
and any type profile $\theta $, player $i$'s belief about $\theta _{-i}$ is
bounded by
\begin{equation*}
\chi \mu _{i}(\theta _{-i}|h^{t-1},\theta _{i})\leq \mu _{i}(\theta
_{-i}|h^{t},\theta _{i})\leq 1-\chi \sum_{\theta _{-i}^{\prime }\neq \theta
_{-i}}\mu _{i}(\theta _{-i}^{\prime }|h^{t-1},\theta _{i}).
\end{equation*}%
One can see from this condition that when $\chi $ increases, the feasible
range of $\mu _{i}(\theta _{-i}|h^{t},\theta _{i})$ shrinks, and the
restriction on the belief system becomes more stringent. Moreover, if the
history $h^{t}$ is an off-path history of $(\mu ,\sigma )$, then this
condition characterizes the feasible set of off-path beliefs, which shrinks
as $\chi $ increases.

An important implication of this observation is that $\Phi (\chi )$ is not
lower hemi-continuous with respect to $\chi $. The intuition is that for
some $\chi $-CSE that contains off-path histories, the off-path beliefs to
support the equilibrium might not be $\chi $-consistent for sufficiently
large $\chi $. In this case, the $\chi $-CSE is not attainable by a sequence
of $\chi _{k} $-CSE where $\chi _{k}$ converges to $\chi $ from above,
causing the lack of lower hemi-continuity.\footnote{%
An example is provided in Section \ref{subsec:signaling} (see Footnote 7).}

Lastly, another implication of $\chi$-dampened updating property is that for
each player $i$, history $h^t$ and type profile $\theta$, the belief $%
\mu_i(\theta_{-i}| h^t, \theta_i)$ has a lower bound that is \emph{%
independent} of the strategy profile. The lower bound is characterized in
Corollary \ref{coro:lower_bound}. This result implies that when $\chi >0$, $%
\mathcal{F}(\theta _{-i}|\theta _{i})>0$ implies $\mu _{i}(\theta
_{-i}|h^{t},\theta _{i})>0$ for all $h^{t}$, so that if prior beliefs are
bounded away from zero, beliefs are always bounded away from 0 as well. In
other words, when $\chi >0$, because of the $\chi $-dampened updating,
beliefs will always have full support even if at off-path histories.

\begin{corollary}
\label{coro:lower_bound} For any $\chi$-consistent assessment $(\mu, \sigma)$%
, $i\in N$, $\theta \in \Theta$ and $h^t\in\mathcal{H}\backslash\mathcal{H}%
^T $,
\begin{align*}
\mu_i(\theta_{-i}| h^t, \theta_i)\geq \chi^t\mathcal{F}(\theta_{-i}|\theta_i)
\end{align*}
\end{corollary}

\begin{proof}
See Appendix \ref{appendix_proof_general}.
\end{proof}

%Note that the restriction of the consistency requirement would bite only
%when the belief system is not full support, i.e., only when some player
%thinks some type profile is \emph{infinitely more likely} than some other
%type profile. Yet, Corollary \ref{coro:lower_bound} suggests when $\chi>0$,
%the likelihood ratio of any two type profiles at any information set is
%always finite. In other words, $\chi$-consistency will not eliminate any
%belief system that satisfies $\chi$-dampened updating property. This
%observation provides a guide for our analysis as the $\chi$-dampened
%updating property is a much easier condition to work with than the $\chi$%
%-consistency condition.

If the game has only one stage, then the dampened updating property has no
effect, in which case $\chi $-CSE and $\chi $-CE are equivalent solution
concepts. This is formally stated and proved in Proposition \ref%
{prop:one_period}.

\begin{proposition}
\label{prop:one_period} For any one-stage game and for any $\chi $, $\chi $%
-CSE and $\chi $-CE are equivalent.
\end{proposition}

\begin{proof}

For any one-stage game, the only public history is the initial history $h_\emptyset$.
Thus, in any $\chi$-CSE, for each player $i\in N$ and type profile $\theta\in \Theta$,
player $i$'s belief about other players' types at this history is
$$\mu_i(\theta_{-i}| h_\emptyset, \theta_i) = \mathcal{F}(\theta_{-i}|\theta_i).$$
Since the game has only one stage, the outcome is simply $a^1 = (a_1^1, \ldots,
a_n^1)$, the action profile at stage 1. Moreover, given any behavioral strategy
profile $\sigma$, player $i$ believes $a^1$ will be the outcome with probability
$$\sigma_i(a_i^1 | h_\emptyset, \theta_i)\times
\left[\chi\bar{\sigma}_{-i}(a_{-i}^1 | h_\emptyset, \theta_i)
+ (1-\chi)\sigma_{-i}(a_{-i}^1 | h_\emptyset, \theta_{-i})\right].$$
Therefore, if $\sigma$ is the behavioral strategy profile of a $\chi$-CSE in an
one-stage game, then for each player $i$, type $\theta_i\in \Theta_i$ and each $a_i^1\in
A_i(h_\emptyset)$ such that $\sigma_i(a_i^1 | h_\emptyset, \theta_i)>0$,
\begin{align*}
    a_i^1 \in \argmax_{a_i^{1'}\in A_i(h_\emptyset)}&\sum_{\theta_{-i}\in
    \Theta_{-i}}\mathcal{F}(\theta_{-i}|\theta_i) \; \times \\
    &\left\{\sum_{a_{-i}^1\in A_{-i}(h_\emptyset)}
    \left[\chi\bar{\sigma}_{-i}(a_{-i}^1 | h_\emptyset, \theta_i)
    + (1-\chi)\sigma_{-i}(a_{-i}^1 | h_\emptyset, \theta_{-i})\right]\right\}
    u_i(a_i^{1'}, a_{-i}^1, \theta_i, \theta_{-i}),
\end{align*}
which coincides with the maximization problem of $\chi$-CE. This completes
the proof.
\end{proof}

From the proof of Proposition \ref{prop:one_period}, one can see that in
one-stage games players have \emph{correct} perceptions about the average
strategy of others. Therefore, the maximization problem of $\chi $-CSE
coincides with the problem of $\chi $-CE. For general multi-stage games,
because of the $\chi$-dampened updating property, players will update
beliefs incorrectly and thus their perceptions about other players' future
moves can also be distorted.

\section{Applications}

\label{sec:application}

In this section, we will explore $\chi $-CSE in five applications of
multi-stage games with observed actions, in order to illustrate the range of
effects it can have and to show how it is different from the $\chi $-CE and
sequential equilibrium.
%The key difference between $\chi $-CE and $\chi $-CSE is that $\chi $-CE is
%developed for simultaneous move Bayesian games and player are cursed with
%respect to their beliefs about the normal form strategies of the other
%players. In contrast, $\chi $-CSE analyzes dynamic games where players are
%cursed with respect to their beliefs about other players' behavioral
%strategies at current and future stages of the game, sequential rationality
%is imposed, and players beliefs about other players' types evolve over time
%in a cursed fashion because of their misunderstanding of the relationship
%between types and actions in the behavioral strategies of the other players.
%Among other things, this difference implies that, in contrast to $\chi $-CE
%where cursedness only affects games where the other player types affect
%one's own payoffs, $\chi $-CSE has bite in games with pure private values,
%resulting in systematically different equilibrium behavior.

Our first application is the sender-receiver signaling game, which is
practically the simplest possible multi-stage game. From our analysis, we
will see both the theoretical and empirical implications of our $\chi$-CSE.

\subsection{Pooling Equilibria in Signaling Games}

\label{subsec:signaling}

We first make a general observation about pooling equilibria in multi-stage
games. Player $j$ follows a \emph{pooling strategy} if for every
non-terminal history, $h^{t}$, all types of player $j$ take the same action $%
a_{j}^{t+1}\in A_{j}(h^{t})$. Conceptually, since every type of player $j$
takes the same action, players other than $j$ cannot make any inference
about $j$'s type from $j$'s actions. A \emph{pooling }$\chi $\emph{-CSE} is
a $\chi $-CSE where every player follows a pooling strategy. Hence, every
player has correct beliefs about any other player's future move because
every type of every player chooses the same action.

Since in any pooling $\chi $-CSE, players can correctly anticipate other
players' future moves no matter how cursed they are, one may naturally
conjecture that a pooling $\chi$-CSE is also a $\chi ^{\prime } $-CSE for
any $\chi ^{\prime }\in [0,1]$. As shown by \cite{eyster2005cursed}, this is
true for one-stage Bayesian games: if a pooling strategy profile is a $\chi $%
-cursed equilibrium, then it is also a $\chi ^{\prime }$-cursed equilibrium
for any $\chi ^{\prime }\in[0,1]$. Surprisingly, this result does not extend
to multi-stage games. Proposition \ref{prop:pooling_cse} shows if a pooling
behavioral strategy profile is a $\chi $-CSE, then it remains a $\chi
^{\prime }$-CSE only for $\chi ^{\prime }\leq \chi $, which is a weaker
result than \cite{eyster2005cursed}.

This result is driven by the $\chi$-dampened updating property which
restricts the set of off-path beliefs. As discussed above, when $\chi $ gets
larger, the set of feasible off-path beliefs shrinks, eliminating some
pooling $\chi $-CSE.

\begin{proposition}
\label{prop:pooling_cse} A pooling $\chi $-CSE is a $\chi ^{\prime }$-CSE
for $\chi ^{\prime }\leq \chi $.
\end{proposition}

\begin{proof}
See Appendix \ref{appendix_proof}.
\end{proof}

The proof strategy is similar to the one in \cite{eyster2005cursed}
Proposition 3. Given a $\chi $-CSE behavioral strategy profile, we can
separate the histories into on-path and off-path histories. For on-path
histories in a pooling equilibrium, since all types of players make the same
decisions, players cannot make any inference about other players' types.
Therefore, for on-path histories, their beliefs are the prior beliefs, which
are independent of $\chi $. On the other hand, for off-path histories, as
shown in Proposition \ref{prop:equiv_def}, a necessary condition for $\chi $%
-CSE is that the belief system has to satisfy the $\chi$-dampened updating
property. When $\chi $ gets larger, this requirement becomes more stringent,
and hence some pooling $\chi $-CSE may break down.

Example 1 is a signaling game where the sender has only two types and two
messages, and the receiver has only two actions. This example demonstrates
the implication of Proposition \ref{prop:pooling_cse} and shows the lack of
lower hemi-continuity; i.e., it is possible for a pooling behavioral
strategy profile to be a $\chi $-CSE, but not a $\chi ^{\prime }$-CSE for $%
\chi ^{\prime }>\chi $. We will also use this example to illustrate how the
notion of cursedness in sequential cursed equilibrium proposed by \cite%
{cohen2022sequential} departs from our CSE. \bigskip

\begin{figure}[htbp]
\centering
\par
\begin{tikzpicture}[font=\footnotesize,edge from parent/.style={draw,thick}]
% Two node styles: solid and hollow
\tikzstyle{solid node}=[circle,draw,inner sep=1.2,fill=black];
\tikzstyle{hollow node}=[circle,draw,inner sep=1.2];
% Specify spacing for each level of the tree
\tikzstyle{level 1}=[level distance=15mm,sibling distance=70mm]
\tikzstyle{level 2}=[level distance=15mm,sibling distance=25mm]
\tikzstyle{level 3}=[level distance=20mm,sibling distance=12mm]
% The Tree
\node(0)[hollow node]{}
child{node[solid node]{}
child{node[solid node]{}
child{node[below]{$2,2$} edge from parent node[left]{$L$}}
child{node[below]{$-1,4$} edge from parent node[right]{$R$}}
edge from parent node[above left]{$A$}
}
child{node[solid node]{}
child{node[below]{$4,-1$} edge from parent node(s)[left]{$L$}}
child{node[below]{$1,0$} edge from parent node(t)[right]{$R$}}
edge from parent node[above right]{$B$}
}
edge from parent node[above left]{$\theta_1$}
edge from parent node[below right]{$[\frac{1}{4}]$}
}
child{node[solid node]{}
child{node[solid node]{}
child{node[below]{$2,1$} edge from parent node(m)[left]{$L$}}
child{node[below]{$-1,0$} edge from parent node(n)[right]{$R$}}
edge from parent node[above left]{$A$}
}
child{node[solid node]{}
child{node[below]{$4,-2$} edge from parent node[left]{$L$}}
child{node[below]{$1,0$} edge from parent node[right]{$R$}}
edge from parent node[above right]{$B$}
}
edge from parent node[above right]{$\theta_2$}
edge from parent node[below left]{$[\frac{3}{4}]$}
};
% information sets
\draw[loosely dotted,very thick](0-1-1)to[out=-15,in=195](0-2-1);
\draw[loosely dotted,very thick](0-1-2)to[out=25,in=165](0-2-2);
% movers
\node[above,yshift=2]at(0){Nature};
\foreach \x in {1,2} \node[above,yshift=2]at(0-\x){1};
\node at($(0-1-2)!.4!(0-2-1)$)[below, yshift=-14]{2};
\node at($(0-1-2)!.7!(0-2-1)$)[above, yshift=6]{2};
\end{tikzpicture}
\caption{Game Tree for Example 1}
\label{fig:counter_ex_tree}
\end{figure}
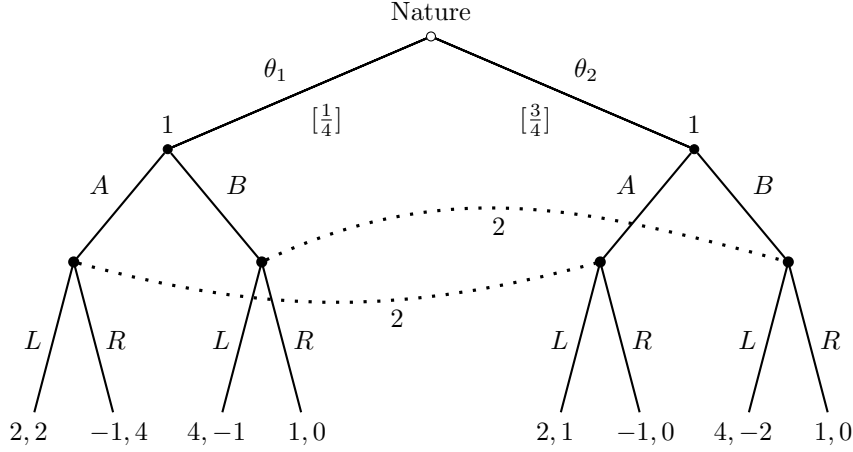

\noindent \textbf{Example 1.} The sender has two possible types drawn from
the set $\Theta =\{\theta _{1},\theta _{2}\}$ with $\Pr (\theta _{1})=1/4$.
The receiver does not have any private information. After the sender's type
is drawn, the sender observes his type and decides to send a message $m\in
\{A,B\}$, or any mixture between the two. After that, the receiver decides
between action $a\in \{L,R\}$ or any mixture between the two, and the game
ends. The game tree is illustrated in Figure \ref{fig:counter_ex_tree}.

If we solve for the $\chi $-CE of the game (or the sequential equilibria),
we find that there are two pooling equilibria for every value of $\chi $. In
the first pooling $\chi $-CE, both sender types choose $A$; the receiver
chooses $L$ in response to $A$ and $R$ at the off-path history $B$. In the
second pooling $\chi $-CE, both sender types pool at $B$ and the receiver
chooses $R$ at both histories. By Proposition 3 of \cite{eyster2005cursed},
these two equilibria are in fact pooling $\chi $-CE for all $\chi \in
\lbrack 0,1]$. The intuition is that in a pooling $\chi $-CE, players are
not able to make any inference about other players' types from their actions
because the average normal form strategy is the same as the type-conditional
normal form strategy. Therefore, their beliefs are independent of $\chi $,
and hence a pooling $\chi $-CE will still be an equilibrium for any $\chi
\in \lbrack 0,1]$.

However, as summarized in Claim \ref{claim:counter_example} below, the $\chi
$-CSE imposes \emph{stronger} restrictions than $\chi $-CE in this example,
in the sense that when $\chi $ is sufficiently large, the second pooling
equilibrium cannot be supported as a $\chi $-CSE. The key reason is that
when the game is analyzed in its normal form, the $\chi $-dampened updating
property shown in Proposition \ref{prop:equiv_def} does not have any bite,
allowing both pooling equilibria to be supported as a $\chi $-CE for any
value of $\chi $. Yet, in the $\chi $-CSE analysis, the additional
restriction of $\chi $-dampened updating property eliminates some extreme
off-path beliefs, and hence, eliminates the second pooling $\chi $-CSE
equilibrium for sufficiently large $\chi $. For simplicity, we use a
four-tuple $[(m(\theta_1), m(\theta_2)); (a(A), a(B))]$ to denote a
behavioral strategy profile.

\begin{claim}
\label{claim:counter_example} In this example, there are two pure pooling $%
\chi$-CSE, which are:

\begin{itemize}
\item[1.] $[(A,A); (L,R)]$ is a pooling $\chi$-CSE for any $\chi\in[0,1]$.

\item[2.] $[(B,B); (R,R)]$ with $\mu _{2}(\theta _{1}|A)\in \left[ \frac{1}{3%
},1-\frac{3}{4}\chi \right] $ is a pooling $\chi $-CSE if and only if $\chi
\leq 8/9$.
\end{itemize}
\end{claim}

\begin{proof}
See Appendix \ref{appendix_proof}.
\end{proof}

From previous discussion, we know in general, the sets of $\chi$-CSE and $%
\chi$-CE are non-overlapping because of the nature of sequential distortion
of beliefs in $\chi$-CSE. Yet, a pooling $\chi$-CSE is an exception. In a
pooling $\chi$-CSE, players can correctly anticipate others' future moves,
so a pooling $\chi$-CSE will mechanically be a pooling $\chi$-CE. In cases
such as this, we can find that $\chi$-CSE is a \emph{refinement} of $\chi$%
-CE.\footnote{%
Note that the $\chi $-CSE correspondence $\Phi (\chi )$ is not lower
hemi-continuous with respect to $\chi $. To see this, we consider a sequence
of $\{\chi_k\}$ where $\chi_k = \frac{8}{9} + \frac{1}{9k} $ for $k\geq 1$.
From the analysis of Claim \ref{claim:counter_example}, we know $[(B,B);
(R,R)] \not\in\Phi(\chi_k)$ for any $k\geq 1$. However, in the limit where $%
\chi_k \rightarrow 8/9$, $[(B,B); (R,R)]$ with $\mu _{2}(\theta _{1}|A) =
1/3 $ is indeed a CSE. That is, $[(B,B); (R,R)]$ is not approachable by this
sequence of $\chi _{k}$-CSE.}

\bigskip

\noindent\textbf{Remark.} This game is useful for illustrating some of the
differences between the notions of \textquotedblleft
cursedness\textquotedblright\ in $\chi $-CSE and the sequential cursed
equilibrium ($(\chi _{S},\psi _{S})$-SCE) proposed by \cite%
{cohen2022sequential}. The first distinction is that the $\chi $
and $\chi _{S}$ parameters capture substantively different sources of
distortion in a player's beliefs about the other players' strategies. In $%
\chi $-CSE, the degree of cursedness, $\chi $, captures how much a player
neglects the dependence of the other players' behavioral strategies on those
players' (exogenous) \textit{private information}, i.e, types, drawn by
nature, and as a result, mistakenly treats different types as behaving the
same with probability $\chi $. In contrast, in $(\chi _{S},\psi_S )$-SCE,
the cursedness parameter, $\chi _{S}$, captures how much a player neglects
the dependence of the other players' strategies on future moves of the
others, or current moves that are unobserved because of simultaneous play.
Thus, it is a neglect related to endogenous information. If player $i$
\textit{observes} a previous move by some other player $j$, then player $i$
correctly accounts for the dependence of player $j$'s chosen action on
player $j$'s private type, as would be the case in $\chi $-CSE only at the
boundary where $\chi =0$.

In the context of pooling equilibria in sender-receiver signaling games, if $%
\chi _{S}=1$, then in SCE the sender believes the receiver will respond the
same way both on and off the equilibrium path. This distorts how the sender
perceives the receiver's future action in response to an off-equilibrium
path message. In $\chi $-CSE, cursedness does not hinder the sender from
correctly perceiving the receiver's strategy since the receiver only has one
type. Take the strategy profile $[(A,A);(L,R)]$ for example, which is a
pooling $\chi $-CSE equilibrium for all $\chi \in \lbrack 0,1]$. However,
with $(\chi _{S},\psi_S )$-SCE, a sender misperceives that the receiver,
upon receiving the off-path message $B$, will, with probability $\chi _{S}$,
take the same action ($L$) as when receiving the on-path message $A$. If $%
\chi _{S}$ is sufficiently high, the sender will deviate to send $B$, which
implies that $[(A,A);(L,R)]$ cannot be supported as an equilibrium when $%
\chi _{S}$ is sufficiently large ($\chi _{S}$ $>1/3$). The distortion
induced by $\chi _{S}$ also creates an additional SCE if $\chi _{S}$ is
sufficiently large: $[(B,B);(L,R)]$. To see this, if $\chi _{S}=1$, then a
sender incorrectly believes that the receiver will continue to choose $R$ if
the sender deviates to $A$, rather than switching to $L$, and hence $B$ is
optimal for both sender types. However, $[(B,B);(L,R)]$ is \emph{not} a $%
\chi $-CSE equilibrium for any $\chi \in \lbrack 0,1]$, or a $\chi $-CE in
the sense of \cite{eyster2005cursed}, or a sequential equilibrium.

In the two possible pooling equilibria analyzed in the last paragraph, the
second SCE parameter, $\psi _{S}$, does not have any effect, but the role of
$\psi _{S}$ can be illustrated in the context of the $[(B,B);(R,R)]$
sequential equilibrium. This second SCE parameter, $\psi _{S}$, is
introduced to accommodate a player's possible failure to fully account for
the informational content from \textit{observed} events. The larger $(1-\psi
_{S})$ is, the greater extent a player neglects the informational content of
observed actions. Although the parameter $\psi_S $ has a similar flavor to $%
1-\chi $ in $\chi $-CSE, it is different in a number of ways. In particular
this parameter only has an effect via its interaction with $\chi _{S}$ and
thus does not independently arise. In the two parameter model, the overall
degree of cursedness is captured by the product, $\chi _{S}(1-\psi _{S})$,
and thus any cursedness effect of $\psi _{S}$ is shut down when $\chi _{S}=0$%
. For instance, under our $\chi$-CSE, the strategy profile $[(B,B);(R,R)]$
can only be supported as an equilibrium when $\chi $ is sufficiently small.
However, $[(B,B);(R,R)]$ can be supported as a $(\chi _{S},\psi_S )$-SCE
even when $(1-\psi _{S})=1$ as long as $\chi _{S}$ is sufficiently small. In
fact, when $\chi _{S}=0$, a $(\chi _{S},\psi_S )$-SCE is equivalent to
sequential equilibrium regardless of the value of $\psi_S $.
%This observation illustrates another difference between $\chi $-CSE and $(\chi _{S},\psi_S )$-SCE.

%Here we emphasize that it is purely a theoretical coincidence of this
%example that $\chi $-CSE is a refinement of $\chi $-CE. As we will see in
%other applications, the nature of $\chi $-CSE is in general completely
%distinct from $\chi $-CE, especially in games of \emph{pure private values}.

% Lastly, we can observe from this example that the $\chi $-CSE correspondence
% $\Phi (\chi )$ is not lower hemi-continuous with respect to $\chi $. To see
% this, we consider a sequence of $\{\chi_k\}$ where $\chi_k = \frac{8}{9} +
% \frac{1}{9k} $ for $k\geq 1$. From the analysis of Claim \ref%
% {claim:counter_example}, we know $[(B,B); (R,R)] \not\in\Phi(\chi_k)$ for
% any $k\geq 1$. However, in the limit where $\chi_k \rightarrow 8/9$, $%
% [(B,B); (R,R)]$ with $\mu _{2}(\theta _{1}|A) = 1/3$ is indeed a CSE. That
% is, $[(B,B); (R,R)]$ is not approachable by this sequence of $\chi _{k}$-CSE.

% \begin{remark}
% \label{remark:not_lower} $\Phi(\chi)$ is not lower hemi-continuous with
% respect to $\chi$.
% \end{remark}

\begin{figure}[htbp]
\centering
\par
\noindent%
\makebox[\textwidth]{
\begin{tikzpicture}[font=\footnotesize,edge from parent/.style={draw,thick}]
\tikzstyle{solid node}=[circle,draw,inner sep=1.2,fill=black];
\tikzstyle{hollow node}=[circle,draw,inner sep=1.2];
\tikzstyle{level 1}=[level distance=15mm,sibling distance=82mm]
\tikzstyle{level 2}=[level distance=15mm,sibling distance=41mm]
\tikzstyle{level 3}=[level distance=22mm,sibling distance=13mm]
\node(0)[hollow node]{}
child{node[solid node]{}
child{node[solid node]{}
child{node(payoff)[below]{$\vpay{45, 30\\ 30, 30}$} edge from parent node[left, yshift=-5]{$C$}}
child{node[below]{$\vpay{15,0\\ \phantom{1}0,0}$} edge from parent node[left, yshift=-5]{$D$}}
child{node[below]{$\vpay{30, 15\\ 50, 35}$} edge from parent node[right, yshift=-5]{$E$}}
edge from parent node[above left]{$I$}
}
child{node[solid node]{}
child{node[below]{$\vpay{30, 90\\ 45, 90}$} edge from parent node(s)[left, yshift=-5]{$C$}}
child{node[below]{$\vpay{\phantom{1}0, 15\\ 15, 15}$} edge from parent node[left, yshift=-5]{$D$}}
child{node[below]{$\vpay{\phantom{1}45, 15\\ 100, 30}$} edge from parent node(t)[right, yshift=-5]{$E$}}
edge from parent node[above right]{$S$}
}
edge from parent node[above left]{$\theta_1$}
edge from parent node[below right]{$[\frac{1}{2}]$}
}
child{node[solid node]{}
child{node[solid node]{}
child{node[below]{$\vpay{30, 30\\ 30, 30}$} edge from parent node(m)[left, yshift=-5]{$C$}}
child{node[below]{$\vpay{\phantom{3}0, 45\\ 30, 45}$} edge from parent node[left, yshift=-5]{$D$}}
child{node[below]{$\vpay{30, 15\\ 30, 0\phantom{0}}$} edge from parent node(n)[right, yshift=-5]{$E$}}
edge from parent node[above left]{$I$}
}
child{node[solid node]{}
child{node[below]{$\vpay{45, 0\\ 45, 0}$} edge from parent node[left, yshift=-5]{$C$}}
child{node[below]{$\vpay{15, 30\\ \phantom{0}0, 30}$} edge from parent node[left, yshift=-5]{$D$}}
child{node[below]{$\vpay{30, 15\\ \phantom{0}0, 15}$} edge from parent node[right, yshift=-5]{$E$}}
edge from parent node[above right]{$S$}
}
edge from parent node[above right]{$\theta_2$}
edge from parent node[below left]{$[\frac{1}{2}]$}
};
\draw[loosely dotted,very thick](0-1-1)to[out=-15,in=195](0-2-1);
\draw[loosely dotted,very thick](0-1-2)to[out=25,in=165](0-2-2);
\draw[<-](payoff)--+(-1,0)node[left]{$\displaystyle\binom{\text{BH 3}}{\text{BH 4}}$};
\node[above,yshift=2]at(0){Nature};
\foreach \x in {1,2} \node[above,yshift=2]at(0-\x){1};
\node at($(0-1-2)!.4!(0-2-1)$)[below, yshift=-14]{2};
\node at($(0-1-2)!.7!(0-2-1)$)[above, yshift=6]{2};
\end{tikzpicture}
}
\caption{Game Tree for \emph{BH 3} and \emph{BH 4} in \protect\cite%
{brandts1993adjustment}}
\label{fig:bh_signal_tree}
\end{figure}
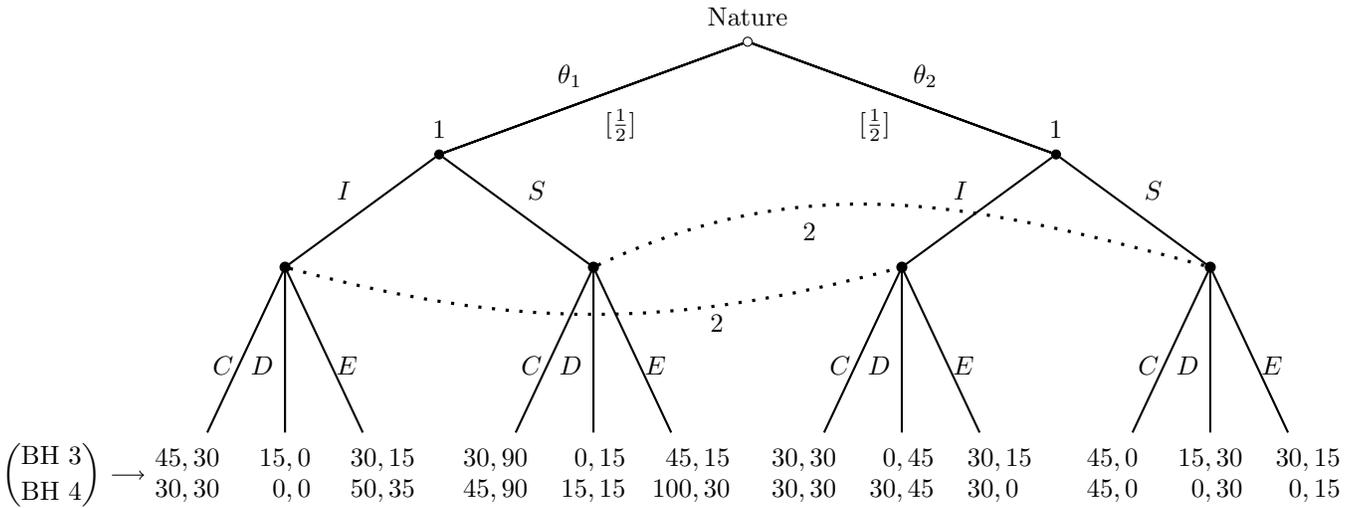

\noindent \textbf{Example 2.} Here we analyze two signaling games that were
studied experimentally by \cite{brandts1993adjustment} (\emph{BH 3} and
\emph{BH 4}) and show that $\chi $-CSE can help explain some of their
findings. In both Game \emph{BH 3} and Game \emph{BH 4}, the sender has two
possible types $\{\theta _{1},\theta _{2}\}$ which are equally likely. There
are two messages $m\in \{I,S\}$ available to the sender.\footnote{$I$ stands
for \textquotedblleft \textbf{I}ntuitive\textquotedblright\ and $S$ stands
for \textquotedblleft \textbf{S}equential but not
intuitive\textquotedblright, corresponding to the two pooling sequential
equilibria of the two games.} After seeing the message, the receiver chooses
an action from $a\in \{C,D,E\}$. The game tree and payoffs for both games
are summarized in Figure \ref{fig:bh_signal_tree}.

In both games, there are two pooling sequential equilibria. In the first
equilibrium, both sender types send message $I$, and the receiver will
choose $C$ in response to $I$ and choose $D$ in response to $S$. In the
second equilibrium, both\ sender types send message $S$, and the receiver
will choose $D$ in response to $I$ while choose $C$ in response to $S$. Both
are sequential equilibria, in both games, but only the first equilibrium
where the sender sends $I$ satisfies the intuitive criterion proposed by
\cite{cho1987signaling}.

Since the equilibrium structure is similar in both games, the sequential
equilibrium and the intuitive criterion predict the behavior should be the
same in both games. However, this prediction is strikingly rejected by the
data. \cite{brandts1993adjustment} report that in the later rounds of the
experiment, almost all type $\theta_1$ senders send $I$ in Game \emph{BH 3}
(97 \%), and yet all type $\theta_1$ senders send $S$ in Game \emph{BH 4} ($%
100\%$). In contrast, type $\theta_2$ senders behave similarly in both
games---$46.2\%$ and $44.1\%$ of type $\theta_2$ senders send $I$ in Games
\emph{BH 3} and \emph{BH 4}, respectively. Qualitatively speaking, the
empirical pattern reported by \cite{brandts1993adjustment} is that \emph{%
sender type $\theta_1$ is more likely to send $I$ in Game BH 3 than Game BH
4 while sender type $\theta_2$'s behavior is insensitive to the change of
games.}

To explain this finding, \cite{brandts1993adjustment} propose a descriptive
story based on \emph{naive receivers}. A naive receiver will think both
sender types are equally likely, regardless of which message is observed.
This naive reasoning will lead the receiver to choose $C$ in both games.
Given this naive response, a type $\theta_1$ sender has an incentive to send
$I$ in Game \emph{BH 3} and choose $S$ in Game \emph{BH 4}. (\cite%
{brandts1993adjustment}, p. 284 -- 285)

In fact, their story of naive reasoning echoes the logic of $\chi $-CSE.
When the receiver is fully cursed (or naive), he will ignore the correlation
between the sender's action and type, causing him to not update the belief
about the sender's type. Proposition \ref{prop:bh-signaling} characterizes
the set of $\chi $-CSE of both games. Following the notation in Example 1,
we use a four-tuple $[(m(\theta_1), m(\theta_2));(a(I),a(S))]$ to denote a
behavioral strategy profile.

\begin{proposition}
\label{prop:bh-signaling} The set of $\chi$-CSE of Game BH 3 and BH 4 are
characterized as below.

\begin{itemize}
\item In Game BH 3, there are three pure $\chi$-CSE:

\begin{itemize}
\item[1.] $[(I,I); (C,D)]$ is a pooling $\chi$-CSE if and only if $\chi\leq
4/7$.

\item[2.] $[(S,S); (D,C)]$ is a pooling $\chi$-CSE if and only if $\chi\leq
2/3$.

\item[3.] $[(I,S); (C,C)]$ is a separating $\chi$-CSE if and only if $%
\chi\geq 4/7$.
\end{itemize}

\item In Game BH 4, there are three pure $\chi$-CSE:

\begin{itemize}
\item[1.] $[(I,I); (C,D)]$ is a pooling $\chi$-CSE if and only if $\chi\leq
4/7$.

\item[2.] $[(S,S); (D,C)]$ is a pooling $\chi$-CSE if and only if $\chi\leq
2/3$.

\item[3.] $[(S,S); (C,C)]$ is a pooling $\chi$-CSE for any $\chi\in[0,1]$.
\end{itemize}
\end{itemize}
\end{proposition}

\begin{proof}
See Appendix \ref{appendix_proof}.
\end{proof}

As noted earlier for Example 1, by Proposition 3 of \cite{eyster2005cursed},
pooling equilibria (1) and (2) in games BH 3 and BH 4 survive as $\chi $-CE
for \textit{all} $\chi \in \lbrack 0,1]$. Hence, Proposition \ref%
{prop:bh-signaling} implies that $\chi $-CSE refines the $\chi $-CE pooling
equilibria for larger values of $\chi $. Moreover, $\chi $-CSE actually
eliminates \textit{all} pooling equilibria in BH 3 if $\chi >2/3$.
Proposition \ref{prop:bh-signaling} also suggests that for any $\chi \in
\lbrack 0,1]$, sender type $\theta _{2}$ will behave similarly in both
games, which is qualitatively consistent with the empirical pattern. In
addition, $\chi $-CSE predicts that a highly cursed ($\chi >2/3$) type $%
\theta _{1}$ sender will send different messages in different games---highly
cursed type $\theta _{1}$ senders will send $I$ and $S$ in Games \emph{BH 3}
and \emph{BH 4}, respectively. This is consistent with the empirical data.

\subsection{A Public Goods Game with Communication}

\label{subsec:pgg}

Our second application is a threshold public goods game with private
information and pre-play communication, variations of which have been
studied in laboratory experiments (\citealp{palfrey1991testing}; %
\citealp{NilanjanEtAl2017communication}). Here we consider the
\textquotedblleft unanimity\textquotedblright\ case where there are $N$
players and the threshold is also $N$.

Each player $i$ has a private cost parameter $c_{i}$, which is independently
drawn from a uniform distribution on $[0,K]$ where $K>1$. After each
player's $c_{i}$ is drawn, each player observes their own cost, but not the
others' costs. Therefore, $c_{i}$ is player $i$'s private information and
corresponds to $\theta _{i}$ in the general formulation.\footnote{%
This application has a continuum of types. The framework of analysis
developed for finite types is applied in the obvious way.} The game consists
of two stages. After the profile of cost parameters is drawn, the game will
proceed to stage 1 where each player simultaneously broadcasts a public
message $m_{i}\in \{0,1\}$ without any cost or commitment. After all players
observe the message profile from this first stage, the game proceeds to
stage 2 which is a unanimity threshold public goods game. Player $i$ has to
pay the cost $c_{i}$ if he contributes, but the public good will be provided
only if all players contribute.
The public good is worth a unit of payoff for every player.
Thus, if the public good is provided, each player's payoff will be $1-c_{i}$.

If there is no communication stage, the unique Bayesian Nash equilibrium is
that no player contributes, which is also the unique $\chi$-CE for any $\chi%
\in[0,1]$. In contrast, with the communication stage, there exists an
efficient sequential equilibrium where each player $i$ sends $m_{i}=1$ if
and only if $c_{i}\leq 1$ and contributes if and only if all players send $1$
in the first stage.\footnote{%
One can think of the first stage as a poll, where players are asked the
following question: \textquotedblleft Are you willing to contribute if
everyone else says they are willing to contribute?". The message $m_{i}=1$
corresponds to a \textquotedblleft yes" answer and the message $m_{i}=0$
corresponds to a \textquotedblleft no" answer.} Since this is a private
value game, the standard cursed equilibrium has no bite, and this efficient
sequential equilibrium is also a $\chi $-CE for all values of $\chi $, by
Proposition 2 of \cite{eyster2005cursed}. In the following, we demonstrate
that the prediction of $\chi $-CSE is different from CE (and sequential
equilibrium).

To analyze the $\chi $-CSE, consider a collection of \textquotedblleft
cutoff\textquotedblright\ costs, $\{C_{c}^{\chi },C_{0}^{\chi },C_{1}^{\chi
},\ldots ,C_{N}^{\chi }\}$. In the communication stage, each player
communicates the message $m_{i}=1$ if and only if $c_{i}\leq C_{c}^{\chi }$.
In the second stage, if there are exactly $0\leq k\leq N$ players sending $%
m_{i}=1$ in the first stage, then such a player would contribute in the
second stage if and only if $c_{i}\leq C_{k}^{\chi }$. A $\chi $-CSE is a
collection of these cost cutoffs such that the associated strategies are a $%
\chi $-CSE for the public goods game with communication. The most efficient
sequential equilibrium identified above for $\chi =0$ corresponds to cutoffs
with $C_{0}^{0}=C_{1}^{0}=\cdots =C_{N-1}^{0}=0$ and $C_{c}^{0}=C_{N}^{0}=1$.

There are in fact multiple equilibria in this game with communication. In
order to demonstrate how the cursed belief can distort players' behavior,
here we will focus on the $\chi $-CSE that is similar to the most efficient
sequential equilibrium identified above, where $C_{0}^{\chi }=C_{1}^{\chi
}=\cdots =C_{N-1}^{\chi }=0$ and $C_{c}^{\chi }=C_{N}^{\chi }$.
%Therefore, we only need to solve for the two
%cutoffs $C_{N}^{\chi }$ and $C_{c}^{\chi }$.
The resulting $\chi $-CSE is given in Proposition \ref{prop:pgg_com}.

\begin{proposition}
\label{prop:pgg_com} In the public goods game with communication, there is a
$\chi $-CSE where

\begin{itemize}
\item[1.] $C_0^\chi = C_1^\chi = \cdots = C_{N-1}^\chi=0$, and

\item[2.] there is a unique $C^{\ast }(N,K,\chi )\leq 1$ s.t. $C_{c}^{\chi
}=C_{N}^{\chi }=C^{\ast }(N,K,\chi )$ that solves:
\begin{equation*}
C^{\ast }(N,K,\chi )-\chi \left[ \frac{C^{\ast }(N,K,\chi )}{K}\right]
^{N-1}=1-\chi .
\end{equation*}
\end{itemize}
\end{proposition}

\begin{proof}
See Appendix \ref{appendix_proof}.
\end{proof}

To provide some intuition, we sketch the proof by analyzing the two-person
game, where the $\chi $-CSE is characterized by four cutoffs $\{C_{c}^{\chi
},C_{0}^{\chi },C_{1}^{\chi },C_{2}^{\chi }\}$, with $C_{0}^{\chi
}=C_{1}^{\chi }=0$ and $C_{c}^{\chi }=C_{2}^{\chi }$. If players use the
strategy that they would send message 1 if and only if the cost is less than
$C_{c}^{\chi }$, then by Lemma \ref{lemma:rearrange}, at the history where
both players send 1, player $i$'s cursed posterior belief density would be
\begin{equation*}
\mu _{i}^{\chi }(c_{-i}|\{1,1\})=%
\begin{cases}
\chi \cdot \left( \frac{1}{K}\right) +(1-\chi )\cdot \left( \frac{1}{%
C_{c}^{\chi }}\right) \qquad \mbox{ if }c_{-i}\leq C_{c}^{\chi } \\
\chi \cdot \left( \frac{1}{K}\right) \qquad \qquad \qquad \qquad \quad \;\;%
\mbox{ if }c_{-i}>C_{c}^{\chi }.%
\end{cases}%
\end{equation*}

Notice that cursedness leads a player to put some probability weight on a
type that is not compatible with the history. Namely, for $\chi $-cursed
players, when seeing another player sending 1, they still believe the other
player might have $c_{-i}>C_{c}^{\chi } $. When $\chi $ converges to 1, the
belief simply collapses to the prior belief as fully cursed players never
update their beliefs. On the other hand, when $\chi $ converges to 0, the
belief converges to $1/C_{c}^{\chi }$, which is the correct Bayesian
inference.

Given this cursed belief density, the optimal cost cutoff to contribute, $%
C_{2}^{\chi }$, solves
\begin{equation*}
C_{2}^{\chi }=\int_{0}^{C_{2}^{\chi }}\mu _{i}^{\chi
}(c_{-i}|\{1,1\})dc_{-i}.
\end{equation*}%
Finally, at the first stage cutoff equilibrium, the $C_{c}^{\chi }$ type of
player would be indifferent between sending 1 and 0 at the first stage.
Therefore, $C_{c}^{\chi }$ satisfies
\begin{equation*}
0=\left( \frac{C_{c}^{\chi }}{K}\right) \left\{ -C_{c}^{\chi
}+\int_{0}^{C_{2}^{\chi }}\mu _{i}^{\chi }(c_{-i}|\{1,1\})dc_{-i}\right\} .
\end{equation*}%
After substituting $C_{c}^{\chi }=C_{2}^{\chi }$, we obtain the $\chi $-CSE:
\begin{equation*}
C_{c}^{\chi }=C_{2}^{\chi }=\frac{K-K\chi }{K-\chi }.
\end{equation*}

From this expression, one can see that the cutoff $C_{c}^{\chi }$ (as well
as $C_{2}^{\chi }$) is decreasing in $\chi $ and $K$. When $\chi \rightarrow
0$, $C_{c}^{\chi }$ converges to $1$, which is the cutoff of the sequential
equilibrium. On the other hand, when $\chi \rightarrow 1$, $C_{c}^{\chi }$
converges to $0$, so there is no possibility for communication when players
are fully cursed. Similarly, when $K\rightarrow 1$, $C_{c}^{\chi }$
converges to $1$, which is the cutoff of the sequential equilibrium, while $%
\lim_{K\rightarrow \infty }C_{c}^{\chi }=1-\chi $.

These comparative statics results with respect to $\chi $ and $K$ are not
just a special property of the $N=2$ case, but hold for all $N>1$.
Furthermore, there is a similar effect of increasing $N$ that results in a
lower cutoff (less effective communication). These properties of $C^{\ast
}(N,K,\chi)$ \ are summarized in Corollary \ref{coro:pgg_limit}.

\begin{corollary}
\label{coro:pgg_limit} The efficient $\chi $-CSE predicts the following
comparative statics for all $N\geq 2$ and $K>1$:

\begin{itemize}
\item[1.] $C^{\ast }(N,K,0)=1$ and $C^{\ast }(N,K,1)=0$.

\item[2.] $C^{\ast }(N,K,\chi )$ is strictly decreasing in $N$, $K$, and $%
\chi $ for any $\chi\in(0,1)$.

\item[3.] For all $\chi \in \lbrack 0,1]$, $\lim_{N\rightarrow \infty
}C^{\ast }(N,K,\chi )=\lim_{K\rightarrow \infty }C^{\ast }(N,K,\chi )=1-\chi
.$
\end{itemize}
\end{corollary}

\begin{proof}
See Appendix \ref{appendix_proof}.
\end{proof}

\begin{figure}[htbp]
\centering
\noindent%
\makebox[\textwidth]{
\includegraphics[width=1.25\textwidth]{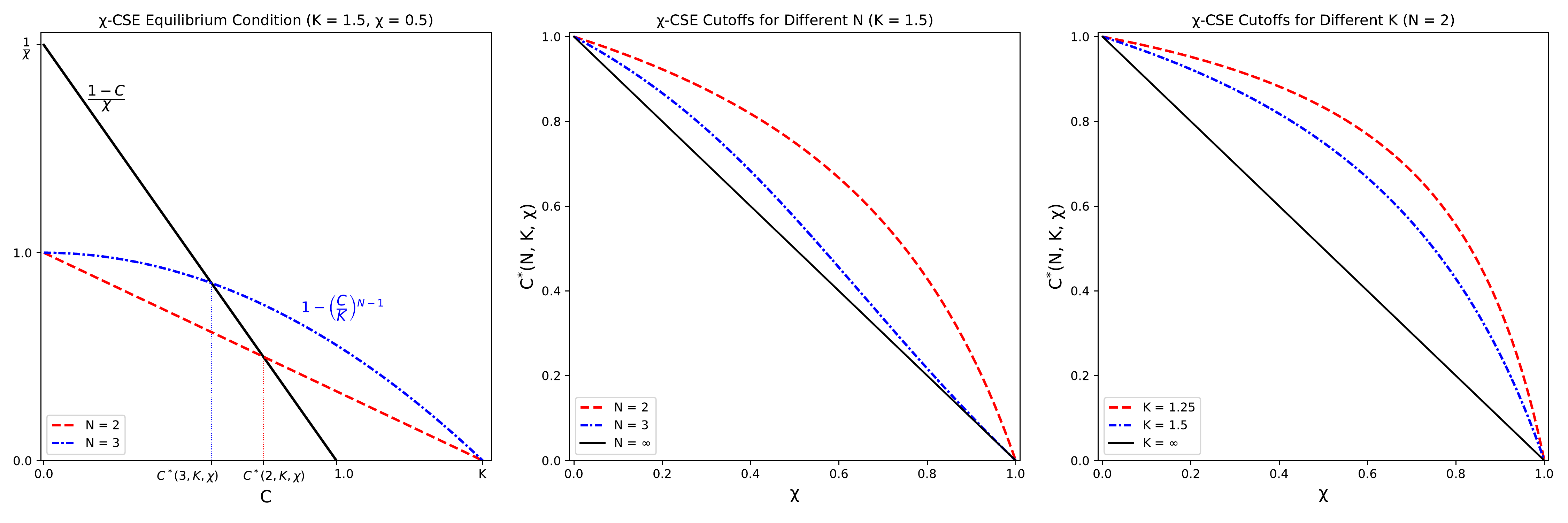}
}
\caption{(Left) Illustration of the $\protect\chi$-CSE equilibrium condition
when $K=1.5$ and $\protect\chi=0.5$. (Middle) The $\protect\chi$-CSE cutoff $%
C^*(N,K,\protect\chi)$ for $N=2, 3$ and for $N\rightarrow \infty$ when $%
K=1.5 $. (Right) The $\protect\chi$-CSE cutoff $C^*(N,K,\protect\chi)$ for $%
K=1.25, 1.5$ and for $K\rightarrow \infty$ when $N=2$. }
\label{fig:cursed_density}
\end{figure}

These properties are illustrated in Figure \ref{fig:cursed_density}. The
left panel illustrates the equilibrium condition for $C^{\ast}$ in a graph
where the horizontal axis is $C\in \lbrack 0,K]$. We can rewrite the
characterization of $C^{\ast }(N,K,\chi )$ in Proposition \ref{prop:pgg_com}
as a solution for $C$ to the following equation:
\begin{equation*}
\frac{1-C}{\chi }=1-\left[ \frac{C}{K}\right] ^{N-1}.
\end{equation*}%
The left panel displays the LHS of this equation, $\frac{1-C}{\chi }$, as
the downward sloping line that connects the points $(0,\frac{1}{\chi })$ and
$(1,0)$. The RHS is displayed for $N=2$ and $N=3$ by the two curves that
connect the points $(0,1)$ and $(K,0)$. The equilibrium, $C^{\ast }(N,K,\chi
)$, is given by the (unique) intersection of the LHS and RHS curves. It is
easy to see from this graph that $C^{\ast }(N,K,\chi )$ is strictly
decreasing in $N$, $K$, and $\chi $. When $N$ increases, the RHS increases
for all $C\in (0,K)$, resulting in an intersection at a lower value of $C$.
When $K$ increases, again the RHS increases for all $C\in (0,K)$, and also
the intercept of the RHS on the horizontal axis increases, leading to a
similar effect; and when $\chi $ increases, the intercept of the LHS on the
horizontal axis decreases, resulting in an intersection at a lower value of $%
C$. In addition, when $N$ grows without bound, the RHS
%becomes increasingly concave,
approaches a constant function equal to $1$ for $C<K$, resulting in a
limiting intersection at $C^{\ast }(\infty ,K,\chi )=1-\chi $. This is
illustrated in the middle panel of Figure \ref{fig:cursed_density}, which
graphs $C^{\ast }(2,1.5,\cdot )$, $C^{\ast }(3,1.5,\cdot )$, and $C^{\ast
}(\infty ,1.5,\cdot )$. A similar effect occurs for $K\rightarrow \infty $,
illustrated in the right panel of Figure \ref{fig:cursed_density}, which
displays $C^{\ast }(2,1.25,\cdot )$, $C^{\ast }(2,1.5,\cdot )$, and $C^{\ast
}(2,\infty ,\cdot )$.

An interesting takeaway of this analysis is that in the public goods game
with communication, \emph{cursedness limits information transmission}: $\chi
$-CSE predicts when players are more cursed (higher $\chi $), it will be
harder for them to effectively communicate in the first stage for efficient
coordination in the second stage. Moreover, Corollary \ref{coro:pgg_limit}
shows that this $\chi $-CSE varies systematically with \emph{all three
parameters of the model}: $N,K$, \emph{and} $\chi $. In contrast, in the
standard $\chi $-CE, players best respond to the \emph{average
type-contingent strategy} rather than the average behavioral strategy. Since
it is a private value game, %types are private values,
players do not care about the distribution of types, only the distribution
of actions. Thus, the prediction of standard CE coincides with the
equilibrium prediction \emph{for all values of $N,K$, and $\chi $}. This
seems behaviorally implausible and is also suggestive of an experimental
design that varies the two parameters $N$ and $K$, since\emph{\ }the
qualitative effects of changing these parameters are identified.

\subsection{Reputation Building: The Centipede Game with Altruists}

\label{subsec:centipede}

\begin{figure}[htbp]
\centering
\par
\begin{tikzpicture}[font=\footnotesize,scale=1]
% Two node styles: solid and hollow
\tikzstyle{solid node}=[circle,draw,inner sep=1.2,fill=black];
\tikzstyle{hollow node}=[circle,draw,inner sep=1.2];
\tikzstyle{level 1}=[level distance=25mm,sibling distance=20mm]
% The Tree
\node(0)[hollow node]{}
child[grow=down]{node[solid node]{}edge from parent node[left]{$T1$}}
child[grow=right]{node(1)[solid node]{}
child[grow=down]{node[solid node]{}edge from parent node[left]{$T2$}}
child[grow=right]{node(2)[solid node]{}
child[grow=down]{node[solid node]{}edge from parent node[left]{$T3$}}
child[grow=right]{node(3)[solid node]{}
child[grow=down]{node[solid node]{}edge from parent node[left]{$T4$}}
child[grow=right]{node(4)[solid node]{}
edge from parent node[above]{$P4$}
}
edge from parent node[above]{$P3$}
}
edge from parent node[above]{$P2$}
}
edge from parent node[above]{$P1$}
};
% Movers
\foreach \x in {0,2,4}
\node[above]at(\x){1};
\foreach \x in {1,3}
\node[above]at(\x){2};
% payoffs
\node[below]at(0-1){$4,1$};
\node[below]at(1-1){$2,8$};
\node[below]at(2-1){$16,4$};
\node[below]at(3-1){$8,32$};
\node[right]at(4){$64,16$};
\end{tikzpicture}
\caption{Four-stage Centipede Game}
\label{fig:centipede}
\end{figure}
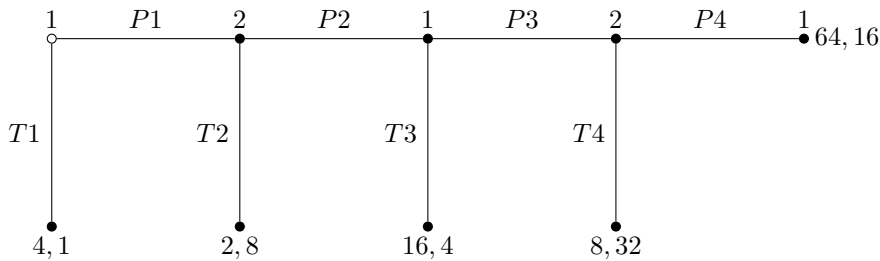

In order to further demonstrate the difference between $\chi $-CE and $\chi $%
-CSE, in this section we consider a variation of the centipede game with
private information, as analyzed in \cite{mckelvey1992experimental} and \cite%
{kreps1990}. This game is an illustration of reputation-building, where a
selfish player imitates an altruistic type in order to develop a reputation
for passing, which in turn entices the opponent to pass and leads to higher
payoffs.

There are two players and four stages, and the game tree is shown in Figure %
\ref{fig:centipede}. In stage one, player one can choose either Take $(T1)$
or Pass $(P1).$ If she chooses action $T1,$ the game ends and the payoffs to
players one and two are $4$ and $1$, respectively. If she chooses the action
$P1,$ the game continues and player two has a choice between take $(T2)$ and
pass $(P2).$ If he chooses $T2,$ the game ends and the payoffs to players
one and two are $2$ and $8$, respectively. If he chooses $P2,$ the game
continues to the third stage where player one chooses between $T3$ and $P3.$
Similar to the previous stages, if she chooses $T3,$ the payoffs to players
one and two are $16$ and $4$, respectively. If she chooses $P3,$ the game
proceeds to the last stage where player two chooses between $T4$ and $P4$.
If player two chooses $T4$ the payoffs are $8$ and $32$, respectively. If
player two alternatively chooses $P4$, the payoffs are $64$ and $16$,
respectively.

There are two types of player one, selfish and altruistic. Selfish players
are assumed to have a utility function that is linear in their own payoff.
Altruistic players are assumed to have a utility function that is linear in
the sum of the two payoffs. For the sake of simplicity, we assume that
player two has only one type, selfish. The common knowledge probability that
player one is altruistic is $\alpha $. Player one knows her own type, but
player two does not. Therefore, player one's type is her private
information. In the following, we will focus on the interesting case where $%
\alpha \leq 1/7$.\footnote{%
If $\alpha >\frac{1}{7}$, player two always chooses $P2$ in the second stage
since the probability of encountering altruistic player one is sufficiently
high. Selfish player one would thus chooses $P1$ in the first stage and
choose $T3$ in the third stage.}

Because this is a game of incomplete information with \textit{private values,%
} the standard $\chi $-CE is equivalent to the Bayesian Nash equilibrium of
the game for all $\chi \in \lbrack 0,1]$, and yields the same take
probabilities as the Bayesian equilibrium. Since altruistic player one wants
to maximize the sum of the payoffs, it is optimal for her to always pass.
The equilibrium behavior is summarized in Claim \ref{claim:centipede}.

\begin{claim}
\label{claim:centipede} In the Bayesian Nash equilibrium, selfish player one
will choose $P1$ with probability $\frac{6\alpha}{1-\alpha}$ and choose $T3$
with probability 1; player two will choose $P2$ with probability $\frac{1}{7}
$ and choose $T4$ with probability 1.
\end{claim}

\begin{proof}
See Appendix \ref{appendix_proof}.
\end{proof}

It is useful to see exactly why, in this example (and more generally) the
standard $\chi $-CE is the same as the perfect Bayesian equilibrium. In
particular, why it is not the case that cursed beliefs will change player
two's updating process after observing $P1$ at stage one. Belief updating is
\emph{not} a property of the standard $\chi $-CE as the analysis is in the
strategic form, and thus is solved as a BNE of the game in the reduced
normal form.\footnote{%
The analysis is similar for the unreduced normal form.} Table \ref%
{tab:cent_reduced_table} summarizes the payoff matrices in the reduced
normal form of centipede game for selfish and altruistic type.

\begin{table}[htbp]
\caption{Reduced Normal Form Centipede Game Payoff Matrix}
\label{tab:cent_reduced_table}\centering
\begin{tabular}{|c|ccc|c|c|ccc|}
\cline{1-4}\cline{6-9}
selfish $(1-\alpha)$ & $T_2$ & $P_2 T_4$ & $P_2 P_4$ &  & altruistic $%
(\alpha)$ & $T_2$ & $P_2 T_4$ & $P_2 P_4$ \\ \cline{1-4}\cline{6-9}
$T_1$ & $4,1$ & $4,1$ & $4,1$ &  & $T_1$ & $5,1$ & $5,1$ & $5,1$ \\
$P_1 T_3$ & $2,8$ & $16,4$ & $16,4$ &  & $P_1 T_3$ & $10,8$ & $20,4$ & $20,4$
\\
$P_1 P_3$ & $2,8$ & $8,32$ & $64,16$ &  & $P_1 P_3$ & $10,8$ & $40,32$ & $%
80,16$ \\ \cline{1-4}\cline{6-9}
\end{tabular}%
\end{table}

It is easily verified that at the Bayesian Nash equilibrium, selfish player
one would choose $T_1$ with probability $(1-7\alpha)/(1-\alpha)$ and choose $%
P_1 T_3$ with probability $6\alpha / (1-\alpha)$, while player two would
choose $T_2$ with probability $6/7$.

To solve the standard $\chi$-CE, let selfish player one choose $T_1$ with
probability $p$ and $P_1 T_3$ with probability $1-p$. Let player two choose $%
T_2$ with probability $q$ and $P_2 T_4$ with probability $1-q$. Notice that
for player two, $P_2 P_4$ is a dominated strategy and given this, it is also
sub-optimal for selfish player one to choose $P_1 P_3$. In this case,
selfish player one would choose $T_1$ if and only if
\begin{align*}
4 \geq 2q + 16(1-q) \iff q \geq 6/7,
\end{align*}
implying that selfish player one's best response correspondence in the
standard cursed analysis coincides with the Bayesian Nash equilibrium
analysis. On the other hand, to solve for player two's best responses we
need to first solve for the \emph{perceived} strategy. When player two is $%
\chi$-cursed, he would think that player one is using $\sigma_1^\chi(a|%
\theta)$ where $a\in \{T_1, P_1 T_3, P_1 P_3 \}$ and $\theta\in \{%
\mbox{selfish}, \mbox{ altruistic} \}$. Player one's true strategy is given
in Table \ref{tab:TrueStrat}.
\begin{table}[htbp]
\caption{Player 1's True Strategy}
\label{tab:TrueStrat}\centering
\begin{tabular}{ccc}
\hline
& \multicolumn{2}{c}{player one's type} \\ \cline{2-3}
$\sigma_1(a| \theta)$ & selfish & altruistic \\ \hline
$T_1$ & $p$ & 0 \\
$P_1 T_3$ & $1-p$ & 0 \\
$P_1 P_3$ & 0 & 1 \\ \hline
\end{tabular}%
\end{table}

In this case, player one's \emph{average strategy} is simply:
\begin{align*}
\bar{\sigma}_1(T_1) = (1-\alpha)p, \;\; \bar{\sigma}_1(P_1 T_3) =
(1-\alpha)(1-p), \;\; \bar{\sigma}_1(P_1 P_3) = \alpha.
\end{align*}
By definition, $\sigma_1^\chi(a|\theta) = \chi\bar{\sigma}_1(a) +
(1-\chi)\sigma_1(a|s)$ and hence we can find that $\sigma_1^\chi(a|\theta)$
is given in Table \ref{tab:CursedStrat}.
\begin{table}[htbp]
\caption{Cursed Perception of Player 1's Strategy}
\label{tab:CursedStrat}\centering
\begin{tabular}{ccc}
\hline
& \multicolumn{2}{c}{player one's type} \\ \cline{2-3}
$\sigma^\chi_1(a| \theta)$ & selfish & altruistic \\ \hline
$T_1$ & $p(1-\chi\alpha)$ & $p\chi(1-\alpha)$ \\
$P_1 T_3$ & $(1-p)(1-\chi\alpha)$ & $(1-p)\chi(1-\alpha)$ \\
$P_1 P_3$ & $\chi\alpha$ & $1-\chi+\chi\alpha$ \\ \hline
\end{tabular}%
\end{table}

From player two's perspective, given any action profile, player two's
expected payoff is not affected by whether player one is selfish or
altruistic. Hence, player two only cares about the \textit{marginal}
distribution of player one's actions. In this case, $\chi $-cursed player
two believes player one will choose $a\in \{T_{1},P_{1}T_{3},P_{1}P_{3}\}$
with probability $\bar{\sigma}_{1}(a)$. Therefore, it is optimal for player
two to choose $T_{2}$ if and only if
\begin{equation*}
\bar{\sigma}_{1}(T_{1})+8\left[ 1-\bar{\sigma}_{1}(T_{1})\right] \geq \bar{%
\sigma}_{1}(T_{1})+4\bar{\sigma}_{1}(P_{1}T_{3})+32\bar{\sigma}%
_{1}(P_{1}P_{3})\iff p\leq \frac{1-7\alpha }{1-\alpha },
\end{equation*}
implying player two's best responses in the standard cursed analysis also
coincides with the Nash best responses. As a result, one concludes that
standard $\chi $-CE would make exactly the same prediction as the Bayesian
Nash equilibrium regardless how cursed the players are.

In contrast, the $\chi $-CSE will exhibit distortions to the \emph{%
conditional beliefs} of player two, given that player one has passed,
because player two incorrectly takes into account how player one's choice to
pass depended on player one's private information. In particular, it is
harder to build a reputation, since a selfish type will have to imitate
altruists in such a way that the true posterior on altruistic type
conditional on a pass is higher than in the perfect Bayesian equilibrium,
because the updating by player two about player one's type is dampened
relative to this true posterior due to cursedness. This distorted belief
updating will result in \emph{less} passing by player one compared to the
Bayesian equilibrium. Formally, the $\chi $-CSE is described in Proposition %
\ref{prop:centipede}.

\begin{proposition}
\label{prop:centipede}

In the $\chi$-CSE, selfish player one will choose $P1$ with probability $%
q_1^\chi$ and choose $T3$ with probability 1; player two will choose $P2$
with probability $q_2^\chi$ and choose $T4$ with probability 1 where
\begin{align*}
q_1^\chi =
\begin{cases}
\left[\frac{7\alpha - 7\alpha\chi}{1-7\alpha\chi}-\alpha \right]\bigg/%
(1-\alpha) \qquad \mbox{ if } \chi \leq \frac{6}{7(1-\alpha)} \\
\qquad\qquad\quad 0 \qquad\qquad\qquad\;\; \mbox{ if } \chi > \frac{6}{%
7(1-\alpha)}%
\end{cases}
\;\;\mbox{ and, }
\end{align*}
\begin{align*}
q_2^\chi =
\begin{cases}
1/7 \qquad \mbox{ if } \chi \leq \frac{6}{7(1-\alpha)} \\
\; 0 \; \qquad\;\; \mbox{ if } \chi > \frac{6}{7(1-\alpha)}.%
\end{cases}%
\end{align*}
\end{proposition}

\begin{proof}
See Appendix \ref{appendix_proof}.
\end{proof}

In order to see how the cursedness affects the equilibrium behavior, here we
focus on the case of $\chi \leq \frac{6}{7(1-\alpha )}$ where selfish player
one and player two will both mix at stage one and two. Given selfish player
one chooses $P1$ with probability $q_{1}^{\chi }$, by Lemma \ref%
{lemma:rearrange}, we know when the game reaches stage two, player two's
belief about player one being altruistic becomes
\begin{equation*}
\mu ^{\chi }=\chi \alpha +(1-\chi )\left[ \frac{\alpha }{\alpha +(1-\alpha
)q_{1}^{\chi }}\right] .
\end{equation*}%
Here we see that when $\chi $ is larger, player two will update his belief
more slowly. Therefore, in order to maintain indifference at the mixed
equilibrium, selfish player one has to pass with \emph{lower} probability so
that $P1$ is a more informative signal to player two. As a result, to make
player two indifferent between $T2$ and $P2$, the following condition must
hold at the equilibrium:
\begin{equation*}
\mu ^{\chi }=\frac{1}{7}\iff q_{1}^{\chi }=\left[ \frac{7\alpha -7\alpha
\chi }{1-7\alpha \chi }-\alpha \right] \bigg/(1-\alpha ).
\end{equation*}

\begin{figure}[htbp]
\centering
\includegraphics[width=\textwidth]{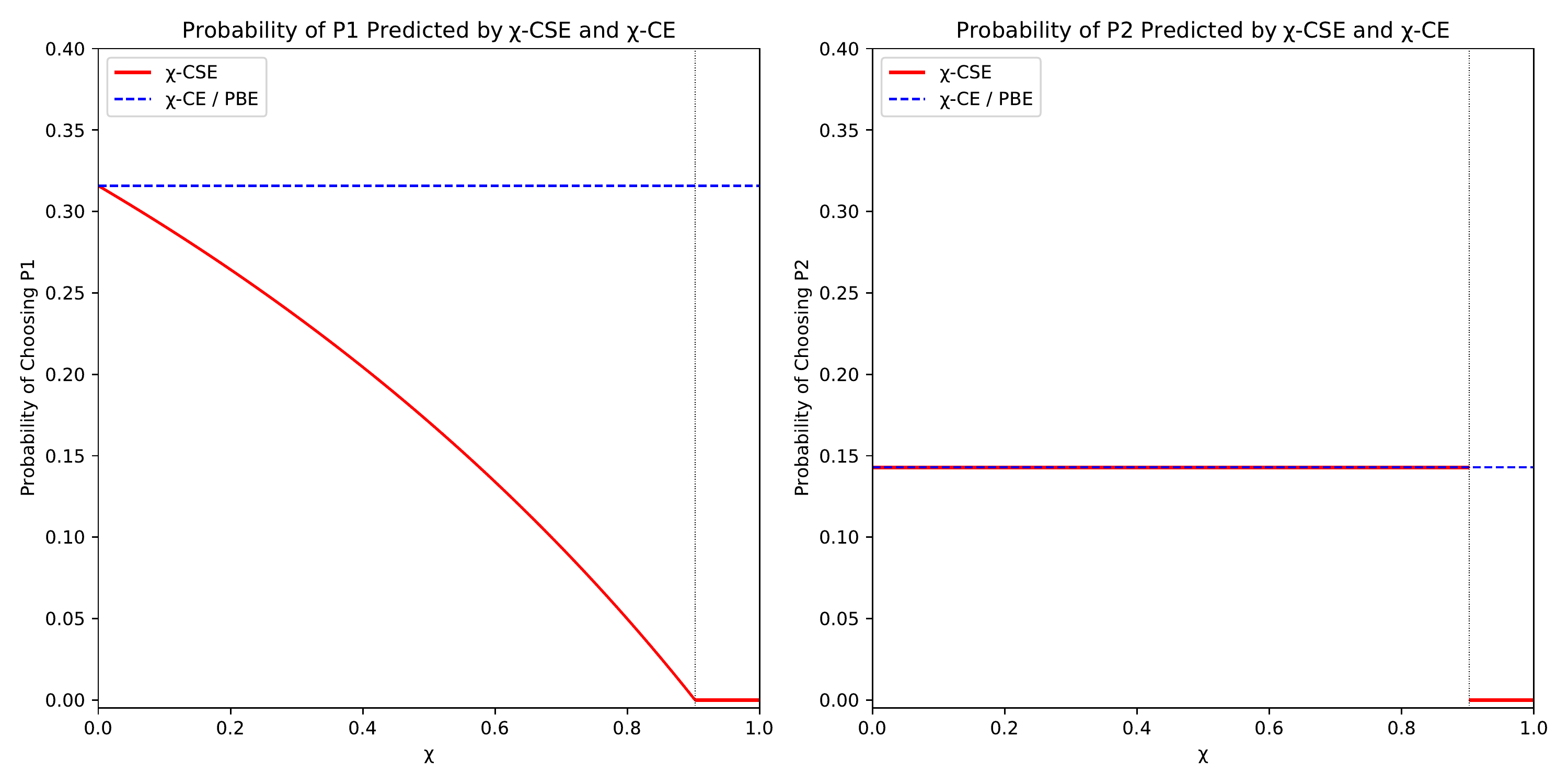}
\caption{$\protect\chi$-CSE of the centipede game with altruistic players $(%
\protect\alpha=0.05)$}
\label{fig:centipede_CSE}
\end{figure}

To conclude this section, in Figure \ref{fig:centipede_CSE}, we plot the
probabilities of choosing $P1$ and $P2$ at $\chi $-CSE when there is a five
percent chance that player one is an altruist (i.e., $\alpha =0.05$). From
our analysis above, we can find that both the standard equilibrium theory
and $\chi $-CE predict selfish player one chooses $P1$ with probability and
player two chooses $P2$ with probability $0.14$. Moreover, these
probabilities are independent of $\chi $. However, $\chi $-CSE predicts when
players are more cursed, selfish player one is less likely to choose $P1$.
When players are sufficiently cursed ($\chi \geq 0.91$), selfish player one
and player two will never pass---i.e., behave as if there were no altruistic
players.

\subsection{Sequential Voting over Binary Agendas}

\label{subsec:voting}

In this section, we apply the concept of $\chi $-CSE to the model of
strategic binary amendment voting with incomplete information studied by
\cite{ordeshook1988agendas}. Let $N=\{1,2,3\}$ denote the set of voters.
These three voters will vote\ over three possible alternatives in $%
X=\{a,b,c\}$. Voting takes place in a two-stage agenda. In the first stage,
voters vote between $a$ and $b$. In the second stage, voters vote between $c$
and the majority rule winner of the first stage. The majority rule winner of
the second stage is the outcome.

Each voter $i$ has three possible private-value types where $\Theta \in
\{\theta _{1},\theta _{2},\theta _{3}\}$ is the set of possible types. Each
voter's type is independently drawn from a common prior distribution of
types, $p$. In other words, the probability of a voter being type $\theta
_{k}$ is $p_{k}$. Each voter's type is their own private information. Each
voter has the same type-dependent payoff function, which is denoted by $%
u(x|\theta )$ for any $x\in X$ and $\theta \in \Theta $. We summarize the
payoff function with the following table.

\begin{table}[htbp]
\centering
\begin{tabular}{ccccc}
&  & \multicolumn{3}{c}{$x$} \\ \cline{2-5}
\multicolumn{1}{c|}{} & \multicolumn{1}{c|}{$u(x|\theta)$} & $a$ & $b$ &
\multicolumn{1}{c|}{$c$} \\ \cline{2-5}
\multicolumn{1}{c|}{} & \multicolumn{1}{c|}{$\theta_1$} & 1 & $v$ &
\multicolumn{1}{c|}{0} \\
\multicolumn{1}{c|}{$\theta$} & \multicolumn{1}{c|}{$\theta_2$} & 0 & 1 &
\multicolumn{1}{c|}{$v$} \\
\multicolumn{1}{c|}{} & \multicolumn{1}{c|}{$\theta_3$} & $v$ & 0 &
\multicolumn{1}{c|}{1} \\ \cline{2-5}
\end{tabular}%
\end{table}
Notice that $v\in (0,1)$ is a parameter that measures the \textit{intensity}
of the second ranked outcome relative to the top ranked outcome. This
intensity parameter, $v,$ is assumed to be the same for all types of all
voters. Because this is a game of private values, the standard $\chi $-CE
and the Bayesian Nash equilibrium coincide.

We use $a_{i}^{1}(\theta )$ to denote type $\theta $ voter $i$'s action at
stage 1. As is standard in majority voting games we will focus on the
analysis of symmetric pure-strategy equilibria where voters do not use
weakly dominated strategies. In other words, we will consider $%
a_{i}^{t}(\cdot )=a_{j}^{t}(\cdot )$ for all $i,j\in N$, and will drop the
subscript.

In this PBE (and $\chi $-CE) all voters will vote sincerely in equilibrium
\textit{except for} type $\theta _{1}$ voters at stage 1. To see this, first
note that voting insincerely in the last stage is dominated and thus
eliminated, so all types of voters vote for their preferred alternative on
the last ballot. Second, voting sincerely in both stages is a dominant
strategy for a type $\theta _{2}$ voter, who prefers any lottery between $b$
and $c$ to either $a$ or $c$. Third, voting sincerely in both stages is also
dominant for a type $\theta _{3}$ voter in the sense that, in the event that
neither of the other two voters are type $\theta _{3}$, then any lottery
between $a$ and $c$ is better than a vote between $b$ and $c$ since $b$
(i.e., type $\theta _{3}$'s least preferred alternative) will win.\footnote{%
When there is another type $\theta _{3}$ voter, the first ballot does not
matter since their most preferred alternative $c$ will always win in the
second stage.}

The PBE (and $\chi$-CE) prediction about a type $\theta_1$ voter's strategy
at stage 1 is summarized in the following claim.

\begin{claim}
\label{claim:voting} The symmetric (undominated pure) PBE strategy for type $%
\theta_1$ voters in the first stage can be characterized as follows.

\begin{itemize}
\item[1.] $a^1(\theta_1)=b$ is a PBE strategy if and only if $v\geq\frac{p_1%
}{p_1+p_2}$.

\item[2.] $a^1(\theta_1)=a$ is a PBE strategy if and only if $v\leq\frac{p_1%
}{p_1+p_3}$.
\end{itemize}
\end{claim}

\begin{proof}
See \cite{ordeshook1988agendas}.
\end{proof}

Claim \ref{claim:voting} shows that, if $v$ is relatively large, only type $%
\theta_1$ voting sophisticatedly for $b$ instead of sincerely for $a$ can be
supported by a PBE. Conditional on being pivotal, voting for $b$ in the
first stage guarantees an outcome of $b$ and thus guarantees getting $v$,
while voting for $a$ leads to a lottery between $a$ and $c$. As a result,
when $v$ is sufficiently high, a type $\theta _{1}$ voter will have an
incentive to strategically vote for $b$ to avoid the risk of having $c$
elected in the last stage.

The analysis of a cursed sequential equilibrium is different from the
standard cursed equilibrium in strategic form because the cursedness affects
belief updating over the stages of the game, and players anticipate future
play of the game. Because of the dynamics and the anticipation of future
cursed behavior, such cursed behavior at later stages of a game can feedback
and affect strategic behavior earlier in the game.

In the context of the two-stage binary amendment strategic voting model,
cursed behavior and belief updating mean that voters in the first stage use
the expected cursed beliefs in the second stage to compute the continuation
values in the two continuation games of the second stage, either a vote
between $a$ and $c$ or a vote between $b$ and $c$. Because they have a
cursed understanding about the relationship between types and voting in the
first stage, this affects their predictions about which alternative wins in
the second stage, conditional on which alternative wins in the first stage.

It is noteworthy that, given any $\chi \in \lbrack 0,1]$, all voters will
still vote sincerely in $\chi $-CSE \textit{except for} type $\theta _{1}$
voters at stage $1$. As implied by Proposition \ref{prop:one_period}, a
voter in the last stage would act as if solving a maximization problem of $%
\chi $-CE but under an (incorrectly) updated belief.
%Moreover, since this is a private
%value game, in $\chi $-CE a voter would act as if he perceives the other
%voters' behavioral strategies correctly in the last stage (Proposition $2$
%of \citealp{eyster2005cursed}).
Therefore, we can follow the same arguments as solving for the undominated
Bayesian equilibrium and conclude that type $\theta _{2}$ and $\theta _{3}$
voters as well as type $\theta _{1}$ voters at stage $2$ will vote sincerely
under a $\chi $-CSE.

Proposition \ref{prop:voting_sophisticatedly} establishes that the set of
parameters $v$ and $p$ that can support a $\chi$-CSE in which type $\theta_1$
voters vote sophisticatedly for $b$ shrinks as $\chi$ increases.

\begin{proposition}
\label{prop:voting_sophisticatedly} If $a^1(\theta_1) = b$ can be supported
by a symmetric $\chi$-CSE, then it can also be supported by a symmetric $%
\chi^{\prime}$-CSE for all $\chi^{\prime}\leq \chi$.
\end{proposition}

\begin{proof}
See Appendix \ref{appendix_proof}.
\end{proof}

The intuition behind strategic voting over agendas mainly comes from the
information content of \emph{hypothetical} pivotal events. However, a cursed
voter does not (fully) take such information into consideration, and thus
becomes overly optimistic about his favorite alternative $a$ being elected
in the second stage. Therefore, a type $\theta _{1}$ voter has a stronger
incentive to deviate from sophisticated voting to sincere voting in stage $1$
as $\chi $ increases.

Interestingly, the set of $v$ and $p$ that can support a $\chi$-CSE in which
type $\theta_1$ voters vote sincerely for $a$ does not necessarily expand as
the level of cursedness becomes higher, as characterized in Proposition \ref%
{prop:voting_sincerely}.

\begin{proposition}
\label{prop:voting_sincerely} Given $p$ and $v \in (0,1)$, there exists $%
\Tilde{\chi}(p,v)$ such that

\begin{itemize}
\item[1.] If $v > \frac{p_1}{p_1+p_3}$, then $a^1(\theta_1) = a$ is a $\chi$%
-CSE strategy if and only if $\chi \geq \Tilde{\chi}(p,v)$;

\item[2.] If $v < \frac{p_1}{p_1+p_3}$, then $a^1(\theta_1) = a$ is a $\chi$%
-CSE strategy if and only if $\chi \leq \Tilde{\chi}(p,v)$.
\end{itemize}
\end{proposition}

\begin{proof}
See Appendix \ref{appendix_proof}.
\end{proof}

Thus, Proposition \ref{prop:voting_sincerely} shows that, when $\chi $ is
sufficiently large, there are some values of $(v,p)$ that cannot support
sincere voting for type $\theta _{1}$ voters under PBE (and $\chi $-CE) but
can support it under $\chi $-CSE. Alternatively, there also exist some
values of $(v,p)$ that can support sincere voting under PBE but fail to
support it under $\chi $-CSE when $\chi $ is large.

\begin{figure}[htbp]
\centering
\includegraphics[width=\textwidth]{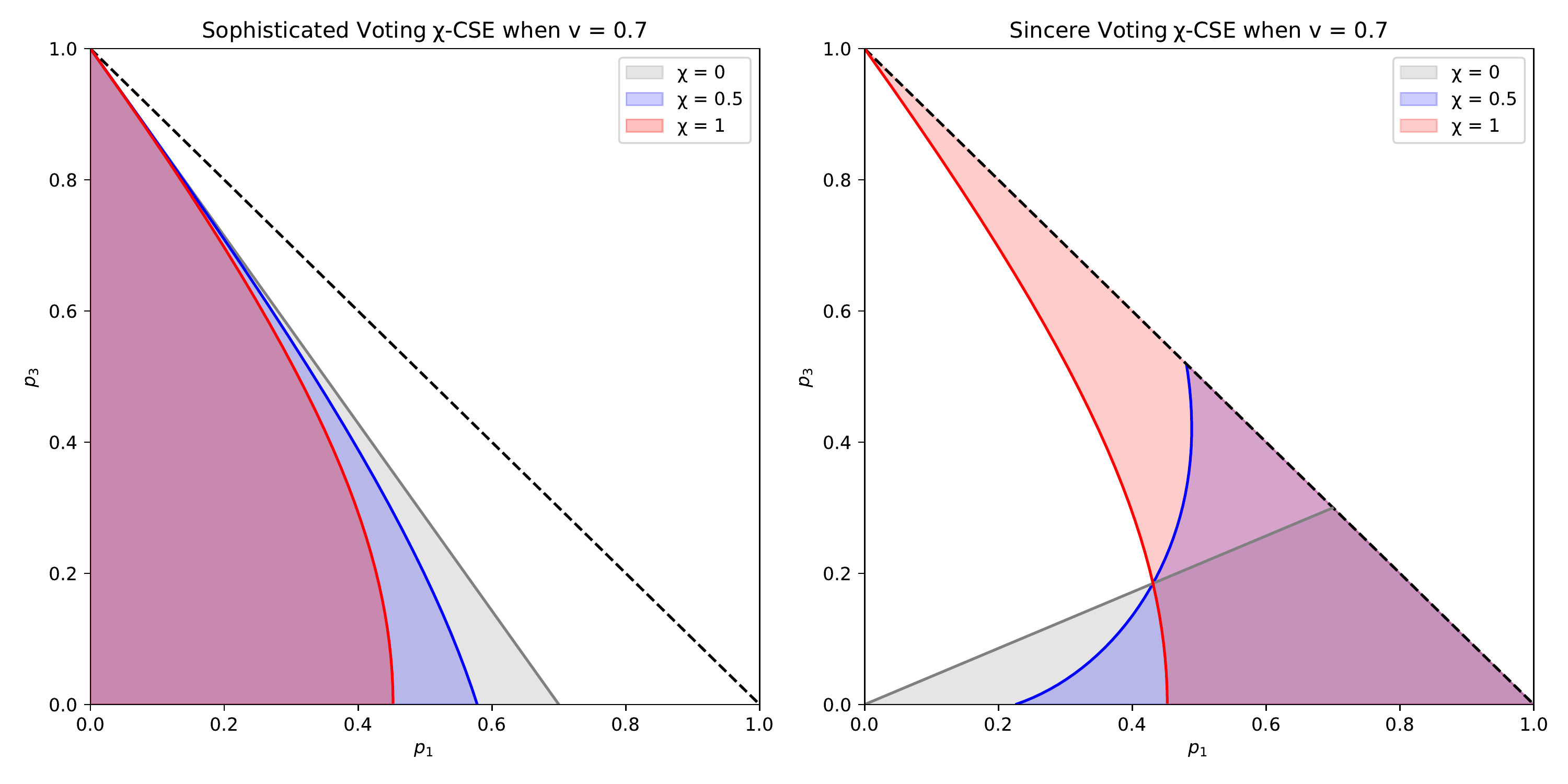}
\caption{$\protect\chi$-CSE for Sophisticated (left) and Sincere (right)
Voting When $v = 0.7$}
\label{fig:sequential_voting}
\end{figure}

To illustrate this, Figure \ref{fig:sequential_voting} plots the set of $p$
(fixing $v=0.7$) that can support a $\chi $-CSE for type $\theta _{1}$
voters at stage $1$ to vote sophisticatedly for $b$ and sincerely for $a$.
The left panel of Figure \ref{fig:sequential_voting} shows that a
sophisticated voting $\chi $-CSE becomes harder to be supported as $\chi $
increases, as indicated by Proposition \ref{prop:voting_sophisticatedly}.
For example, when $p\equiv (p_{1},p_{2},p_{3})=(0.6,0.3,0.1)$, type $\theta
_{1}$ voters will not vote for second preferred alternative $b$ if $\chi
>0.18$.

On the other hand, the right panel of Figure \ref{fig:sequential_voting}
shows that, while type $\theta _{1}$ voters who sincerely vote for $a$ at
stage $1 $ cannot be supported under PBE when $p_{3}$ is large, they may
emerge in a $\chi $-CSE with sufficiently high $\chi $. Also note that when $%
p_{2}$ is large, sincere voting by type $\theta _{1}$ voters is no longer a $%
\chi $-CSE with high $\chi $. In such a sincere voting equilibrium, a fully
rational type $\theta _{1}$ voter knows there will be only one type $\theta
_{2}$ voter among the other two voters when being pivotal. As a result,
whether to sincerely vote for $a$ is determined by the ratio of $p_{1}$ to $%
p_{3}$. When $p_{3}$ is large, sincere voting at stage $1$ will likely lead
to zero payoff for type $\theta _{1}$ voters and thus cannot be a PBE
strategy. However, cursed type $\theta _{1}$ voters will take the
possibility of having two type $\theta _{2}$ voters into account since they
are not correctly conditioning on pivotality. As a result, when $p_{2}$ is
large, sincere voting at stage $1$ will likely lead to zero payoff for type $%
\theta _{1}$ voters, and thus cannot be a $\chi $-CSE strategy with high $%
\chi $, while voting sophisticatedly for $b$ can likely secure a payoff of $%
v $.

\subsection{The Dirty Faces Game}

\label{subsec:dirty}

The dirty faces game was first described by \cite{littlewood1953littlewood}
to study the relationship between common knowledge and behavior.\footnote{%
The dirty faces game has also been reframed as the \textquotedblleft
cheating wives puzzle\textquotedblright\ \citep{gamow1958forty}, the
\textquotedblleft cheating husbands puzzle\textquotedblright\ %
\citep{moses1986cheating}, the \textquotedblleft muddy children
puzzle\textquotedblright\ \citep{barwise1981scenes} and %
\citep{halpern1990knowledge}, and the \textquotedblleft red hat
puzzle\textquotedblright\ \citep{hardin2008introduction}.} There are several
different variants of this game, but here we focus on a simplified version,
the two-person dirty faces game, which was theoretically analyzed by \cite%
{fudenberg1991game} and \cite{lin2022cognitivemultii} and was experimentally
studied by \cite{weber2001behavior} and \cite{bayer2007dirty}.

Let $N=\{1,2\}$ be the set of players. For each $i\in N$, let $x_{i}\in
\{O,X\}$ represent whether player $i$ has a clean face $(O)$ or a dirty face
$(X)$. Each player's face type is independently and identically determined
by a commonly known probability $p=\Pr (x_{i}=X)=1-\Pr (x_{i}=O)$. Once the
face types are drawn, each player $i$ can observe the other player's face $%
x_{-i}$ but not their own face.\footnote{%
To fit into the framework, each player's \textquotedblleft
type\textquotedblright\ (their own private information) can be specified as
\textquotedblleft other players' faces.\textquotedblright\ That is, $\theta
_{i}=x_{-i}$.} If there is at least one player with a dirty face, a public
announcement of this fact is broadcast to both players at the beginning of
the game. Let $\omega \in \{0,1\}$ denote whether there is an announcement
or not. If there is an announcement ($\omega =1$), all players are informed
there is at least one dirty face but not the identities. When $\omega =0$,
it is common knowledge to both players that their faces are clean and the
game becomes trivial. Hence, in the following, we will focus only on the
interesting case where $\omega =1$.

There are a finite number of $T\geq 2$ stages. In each stage, each player $i$
simultaneously chooses $s_{i}\in \{U,D\}$. The game ends as soon as either
player (or both) chooses $D,$ or at the end of stage $T$ in case neither
player has chosen $D$. Actions are revealed at the end of each stage.
Payoffs depend on own face types and action. If a player chooses $D$, he
will get $\alpha >0$ if he has a dirty face while receive $-1$ if he has a
clean face. We assume that
\begin{equation}
p\alpha -(1-p)<0\iff 0<\bar{\alpha}\equiv \frac{\alpha }{(1-p)(1+\alpha )}<1,
\label{payoff_assumption}
\end{equation}%
where $p\alpha -(1-p)$ is the expected payoff of $D$ when the belief of
having a dirty face is $p$. Thus, Assumption (\ref{payoff_assumption})
guarantees it is strictly dominated to choose $D$ at stage 1 when observing
a dirty face. In other words, players will be rewarded when correctly
inferring the dirty face but penalized when wrongly claiming the dirty face.

The payoffs are discounted with a common discount factor $\delta \in (0,1)$.
To summarize, conditional on reaching stage $t$, each player's payoff
function (which depends on their own face and action) can be written as:
\begin{equation*}
u_{i}(s_{i}|t,x_{i}=X)=%
\begin{cases}
\delta ^{t-1}\alpha \;\; & \mbox{if}\;\;s_{i}=D \\
0 & \mbox{if}\;\;s_{i}=U%
\end{cases}%
\;\;\;\;\mbox{ and }\;\;\;\;u_{i}(s_{i}|t,x_{i}=O)=%
\begin{cases}
-\delta ^{t-1}\;\; & \mbox{if}\;\;s_{i}=D \\
0 & \mbox{if}\;\;s_{i}=U.%
\end{cases}%
\end{equation*}%
Therefore, a two-person dirty faces game is defined by a tuple $\langle
p,T,\alpha ,\delta \rangle $.

Since the game ends as soon as some player chooses $D$, the information sets
of the game can be specified by the face type the player observes and the
stage number. Thus a behavioral strategy can be represented as:
\begin{equation*}
\sigma :\{O,X\}\times \{1,\ldots ,T\}\rightarrow \lbrack 0,1],
\end{equation*}%
which is a mapping from information sets to the probability of choosing $D$,
where $\{O,X\}$ corresponds to a player's observation of the other player's
face.

There is a unique Nash equilibrium. When observing a clean face, a player
would immediately know his face is dirty. Hence, it is strictly dominant to
choose $D$ at stage 1 in this case. On the other hand, when observing a
dirty face, because of Assumption (\ref{payoff_assumption}), it is optimal
for the player to choose $U$ at stage 1. However, if the game proceeds to
stage 2, the player would know his face is dirty because the other player
would have chosen $D$ at stage 1 if his face were clean and the game would
not have reached stage 2. This result is independent of the payoffs, the
timing, the discount factor, and the (prior) probability of having a dirty
face. The only assumption for this argument is common knowledge of
rationality.

Alternatively, when players are \textquotedblleft cursed,\textquotedblright\
they are not able to make perfect inferences from the other player's
actions. Specifically, since a cursed player has incorrect perceptions about
the relationship between the other player's actions and their private
information after seeing the other player choose $U$ in stage 1, a cursed
player does not believe they have a dirty face for sure. At the extreme when
$\chi =1$, fully cursed players never update their beliefs. In the
following, we will compare the predictions of the standard $\chi $-CE and
the $\chi $-CSE. A surprising result is that there is always a \emph{unique}
$\chi $-CE, but there can be multiple $\chi $-CSE.

For the sake of simplicity, we will focus on the characterization of pure
strategy equilibrium in the following analysis. Since the game ends when
some player chooses $D$, we can equivalently characterize a stopping
strategy as a mapping from the observed face type to a stage in $%
\{1,2,\ldots ,T,T+1\}$ where $T+1$ corresponds to the strategy of never
stopping. Furthermore, both $\chi $-CE and $\chi $-CSE will be symmetric
because if players were to stop at different stages, least one of the
players would have a profitable deviation. Finally, we use $\hat{\sigma}%
^{\chi }(x_{-i})$ and $\tilde{\sigma}^{\chi }(x_{-i})$ to denote the
equilibrium stopping strategies of $\chi $-CE and $\chi $-CSE, respectively.

We characterize the $\chi $-CE in Proposition \ref{prop:dirty_static}. Since
$\chi $-CE is defined for simultaneous move Bayesian games, to solve for the
$\chi $-CE, we need to look at the corresponding normal form where players
simultaneously choose $\{1,2,\ldots ,T,T+1\}$ given the observed face type.

\begin{proposition}
\label{prop:dirty_static} The $\chi$-cursed equilibrium can be characterized
as follows.

\begin{itemize}
\item[1.] If $\chi > \bar{\alpha}$, the only $\chi$-CE is that both players
choose:
\begin{align*}
\hat{\sigma}^\chi(O)= 1 \;\;\;\; \mbox{ and } \;\;\;\; \hat{\sigma}^\chi(X)=
T+1.
\end{align*}

\item[2.] If $\chi < \bar{\alpha}$, the only $\chi$-CE is that both players
choosing
\begin{align*}
\hat{\sigma}^\chi(O)= 1 \;\;\;\; \mbox{ and } \;\;\;\; \hat{\sigma}^\chi(X)=
2.
\end{align*}
\end{itemize}
\end{proposition}

\begin{proof}
See Appendix \ref{appendix_proof}.
\end{proof}

Proposition \ref{prop:dirty_static} shows that $\chi $-CE makes an extreme
prediction---when observing a dirty face, players would either choose $D$ at
stage 2 (the equilibrium prediction) or never choose $D$. Moreover, the
prediction of $\chi $-CE is unique for $\chi \neq \bar{\alpha }$. As
characterized in the next Proposition \ref{prop:dynamic_dirty}, for extreme
values of $\chi $, the prediction of $\chi $-CSE coincides with $\chi $-CE.
But for intermediate values of $\chi $, there can be multiple $\chi $-CSE.

\begin{proposition}
\label{prop:dynamic_dirty} The pure strategy $\chi$-CSE can be characterized
as follows.

\begin{itemize}
\item[1.] $\tilde{\sigma}^\chi(O) =1 $ for all $\chi \in [0,1]$.

\item[2.] Both players choosing $\tilde{\sigma}^\chi(X)=T+1$ is a $\chi$-CSE
if and only if $\chi\geq \bar{\alpha}^{\frac{1}{T+1}}$.

\item[3.] Both players choosing $\tilde{\sigma}^\chi(X)=2$ is a $\chi$-CSE
if and only if $\chi\leq \bar{\alpha}$.

\item[4.] For any $3\leq t \leq T$, both players choosing $\tilde{\sigma}%
^\chi(X)=t$ is a $\chi$-CSE if and only if
\begin{equation*}
\left(\frac{1-\kappa(\chi)}{1-p}\right)^{\frac{1}{t-2}}\leq\chi\leq \bar{%
\alpha}^{\frac{1}{t-1}} \;\;\;\; \mbox{where}
\end{equation*}
\begin{align*}
\kappa(\chi) \equiv \frac{[(1+\alpha)(1+\delta\chi)-\alpha\delta] - \sqrt{%
[(1+\alpha)(1+\delta\chi)-\alpha\delta]^2 - 4\delta\chi(1+\alpha)}}{%
2\delta\chi(1+\alpha)}.
\end{align*}
\end{itemize}
\end{proposition}

\begin{proof}
See Appendix \ref{appendix_proof}.
\end{proof}

\subsubsection*{Illustrative Example}

\begin{figure}[htbp]
\centering
\includegraphics[width=\textwidth]{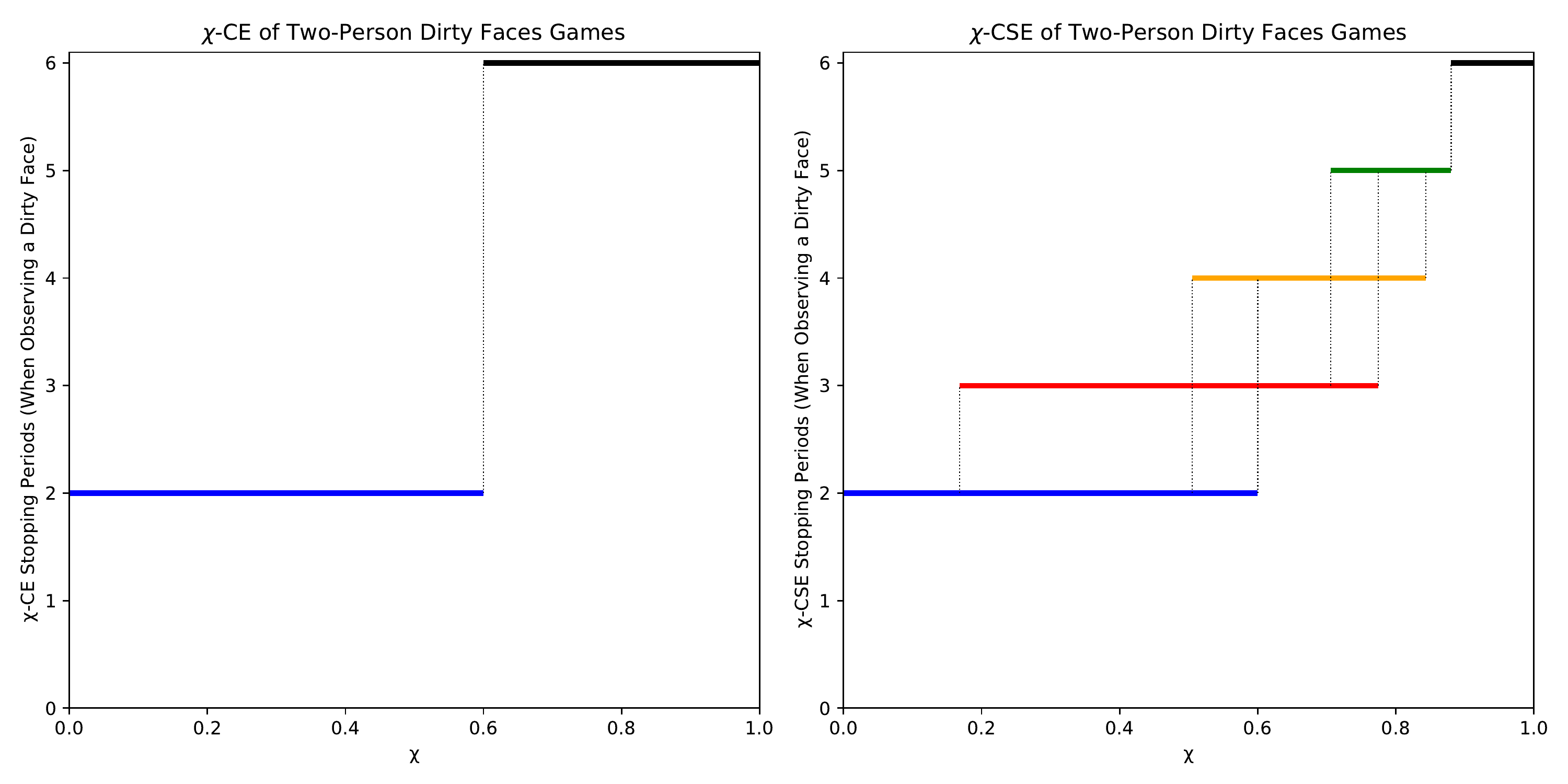}
\caption{$\protect\chi$-CE vs. $\protect\chi$-CSE When $(\protect\alpha,
\protect\delta, p, T) = \left(\frac{1}{4}, \frac{4}{5}, \frac{2}{3},
5\right) $}
\label{fig:dirty_face}
\end{figure}

In order to illustrate the sharp contrast between the predictions of $\chi$%
-CE and $\chi$-CSE, here we consider an illustrative example where $%
\alpha=1/4$, $\delta=4/5$, $p=2/3$ and the horizon of the game is $T=5$. As
characterized by Proposition \ref{prop:dirty_static}, $\chi$-CE predicts
players will choose $\hat{\sigma}^\chi(X)=2$ if $\chi\leq \bar{\alpha}=0.6$;
otherwise, they will choose $\hat{\sigma}^\chi(X)=6$, i.e., they never
choose $D$ when observing a dirty face. As demonstrated in the left panel of
Figure \ref{fig:dirty_face}, $\chi$-CE is (generically) unique and it
predicts players will either behave extremely sophisticated or unresponsive
to the other player's action at all.

In contrast, as characterized by Proposition \ref{prop:dynamic_dirty}, there
can be multiple $\chi $-CSE. As shown in the right panel of Figure \ref%
{fig:dirty_face}, when $\chi \leq \bar{\alpha}=0.6$, both players stopping
at stage 2 is still an equilibrium, but it is not unique except for very low
values of $\chi $. For $0.168\leq \chi \leq 0.505$, both players stopping at
stage 3 is also a $\chi $-CSE, and for $0.505\leq \chi \leq 0.6$, there are
three pure strategy $\chi $-CSE where both players stop at stage 2, 3, or 4,
respectively.

The existence of multiple $\chi $-CSE in which both players stop at $t>2$
highlights a player's learning process in a multi-stage game, which does not
happen in strategic form cursed equilibrium. In the strategic form, a player
has no opportunity to learn about the other player's type in middle stages.
Thus, when level of cursedness is not low enough to support a $\chi $-CE
with stopping at stage 2, both players would never stop. However, in a $\chi
$-CSE of the multi-stage game, a cursed player would still learn about his
own face being dirty as the game proceeds, even though he might not be
confident enough to choose $D$ at stage 2. If $\chi $ is not too large, the
expected payoff of choosing $D$ would eventually become positive at some
stage before the last stage $T$.\footnote{%
The upper bound of the inequality in Proposition \ref{prop:dynamic_dirty}
characterizes the stages at which stopping yields positive expected payoffs.}
For some intermediate values of $\chi $, there might be multiple stopping
stages which yield positive expected payoffs. In this case, the dirty faces
game becomes a special type of coordination games where both players
coordinate on stopping strategies, resulting in the existence of multiple $%
\chi $-CSE.\footnote{%
Note that players with low levels of cursedness would not coordinate on
stopping at late stages since the discount factor shrinks the informative
value of waiting (i.e., both choosing $U$). This result is characterized by
the lower bound of the inequality in Proposition \ref{prop:dynamic_dirty}.}

\section{Concluding Remarks}

\label{sec:conclusion}

In this paper, we formally developed Cursed Sequential Equilibrium, which
extends the strategic form cursed equilibrium \citep{eyster2005cursed} to
multi-stage games, and illustrated the new equilibrium concept with a series
of applications. While the standard CE has no bite in private value games,
we show that cursed beliefs can actually have significant consequences for
dynamic private value games. In the private value games we consider, our
cursed sequential equilibrium predicts (1) under-contribution caused by
under-communication in the public goods game with communication, (2) low
passing rate in the presence of altruistic players in the centipede game,
and (3) less sophisticated voting in the sequential two-stage binary agenda
game. We also illustrate the distinction between CE and CSE in some
non-private value games. In simple signaling games, $\chi $-CSE implies
refinements of pooling equilibria that are not captured by traditional
belief-based refinements (or $\chi $-CE), and are qualitatively consistent
with some experimental evidence. Lastly, we examine the dirty face game,
showing that the CSE further expands the set of equilibrium and predicts
stopping in middle stages of the game. We summarize our findings from these
applications in Table \ref{tab:summary_application}.

%Finally, we demonstrate the tractability of our CSE in
%the Crawford-Sobel information transmission game, which is a dynamic game with continuous
%types and thus difficult to analyze in the original CE framework.

\begin{table}[htbp]
\caption{Summary of Findings in Section \protect\ref{sec:application}}
\label{tab:summary_application}\centering
\begin{tabular}{@{}ccccc}
\toprule &  &
\begin{tabular}{@{}c}
Private-Value \\
Game%
\end{tabular}
& $\chi $-CE vs. BNE & $\chi $-CSE vs. $\chi $-CE \\
\midrule
\begin{tabular}{@{}c}
Signaling Games with \\
Pooling Equilibrium%
\end{tabular}
&  & No & $\neq$ & $\chi $-CSE $\subset \chi $-CE \\
&  &  &  &  \\
\begin{tabular}{@{}c}
Public Goods Game \\
with Communication%
\end{tabular}
&  & Yes & $=$ & $\neq$ \\
&  &  &  &  \\
\begin{tabular}{@{}c}
Centipede Game \\
with Altruists%
\end{tabular}
&  & Yes & $=$ & $\neq$ \\
&  &  &  &  \\
\begin{tabular}{@{}c}
Sequential Voting \\
Game%
\end{tabular}
&  & Yes & $=$ & $\neq$ \\
&  &  &  &  \\
\begin{tabular}{@{}c}
Dirty Faces Game \\
\;%
\end{tabular}
&  & No & $\neq$ & $\neq$ \\
\bottomrule &  &  &  &
\end{tabular}%
\end{table}

The applications we consider are only a small sample of the possible dynamic
games where CSE could be usefully applied. One prominent class of problems
where it would be interesting to study the dynamic effects of cursedness is
social learning. For example, in the standard information cascade model of
\cite{bikhchandani1992theory}, we conjecture that the effect would be to
delay the formation of an information cascade because players will partially
neglect the information content of prior decision makers. Laboratory
experiments report evidence that subjects underweight the information
contained in prior actions relative to their own signal %
\citep{goeree2007self}. A related class of problems involves information
aggregation through sequential voting and bandwagon effects (%
\citealp{callander2007sequentialVoting}; %
\citealp{aliEtAl2008sequentialVotingExperiment}; %
\citealp{aliKartik2012sequentialVoting}). A natural conjecture is that CSE
will impede information transmission in committees and juries as later
voters will under-appreciate the information content of the decisions by
early voters. This would dampen bandwagon effects. The centipede example we
studied suggests than CSE might have broader implications for behavior in
reputation-building games, such as the finitely repeated prisoner's dilemma
or entry deterrence games such as the chain store paradox.

The generalization of CE to dynamic games presented in this paper is limited
in several ways. First, the CSE framework is formally developed for finite
multi-stage games with observed actions. We do not extend CSE for games with
continuous types but we do provide one application that shows how such an
extension is possible. However, a complete generalization to continuous
types (or continuous actions) would require more technical development and
assumptions. We also assume that the number of stages is finite, and
extending this to infinite horizon multi-stage games would be a useful
exercise. Extending CSE to allow for imperfect monitoring in the form
of private histories is another interesting direction to pursue. The SCE approach in \cite%
{cohen2022sequential} allows for cursedness with respect to both public and
private endogenous information, which leads to some important differences
from our CSE approach. In CSE, we find that subjects are limited in their
ability to make correct inferences about hypothetical events, but the
mechanism is different from SCE, which introduces a second free parameter
that modulates cursedness with respect to hypothetical events. For a more
detailed discussion of these and other differences and overlaps between CSE
and SCE, see \cite{fong2023cursednote}.

As a final remark, our analysis of applications of $\chi$-CSE suggests some
interesting experiments. For instance, $\chi$-CSE predicts in the public
goods game with communication, when either the number of players ($N$) or
the largest possible contribution cost ($K$) increases, pre-play
communication will be less effective, while the prediction of sequential
equilibrium and $\chi$-CE is independent of $N$ and $K$. In other words, in
an experiment where $N$ and $K$ are manipulated, significant treatment
effects in this direction would provide evidence supporting $\chi$-CSE over $%
\chi$-CE. Also, $\chi$-CSE makes qualitatively testable predictions in the
sequential voting games and the dirty faces games, which have not been
extensively studied in laboratory experiments. In the sequential voting
game, it would be interesting to test how sensitive strategic (vs. sincere)
voting behavior is to preference intensity ($v$) and the type distribution.
In the dirty faces game, it would be interesting to design an experiment to
identify the extent to which deviations from sequential equilibrium are
related to the coordination problem that arises in $\chi$-CSE.

\newpage
\begin{singlespace}
\bibliographystyle{ecta}
\bibliography{reference}

\newpage \appendix

\section{Omitted Proofs of Section \protect\ref{sec:model} and \protect\ref%
{sec:general_property}}

\label{appendix_proof_general}

\subsection*{Proof of Lemma \protect\ref{lemma:rearrange}}

By definition \ref{def:cursed_bayes}, for any $(\mu ,\sigma )\in \Psi ^{\chi
}$, any history $h^{t-1}$, any player $i$ and any type profile $\theta
=(\theta _{i},\theta _{-i})$,
\begin{align*}
& \sum_{\theta _{-i}^{\prime }}\mu _{i}(\theta _{-i}^{\prime
}|h^{t-1},\theta _{i})[\chi \bar{\sigma}_{-i}(a_{-i}^{t}|h^{t-1},\theta
_{i})+(1-\chi )\sigma _{-i}(a_{-i}^{t}|h^{t-1},\theta _{-i}^{\prime })] \\
=\;& \chi \underbrace{\left[ \sum_{\theta _{-i}^{\prime }}\mu _{i}(\theta
_{-i}^{\prime }|h^{t-1},\theta _{i})\right] }_{=1}\bar{\sigma}%
_{-i}(a_{-i}^{t}|h^{t-1},\theta _{i})+(1-\chi )\underbrace{\left[
\sum_{\theta _{-i}^{\prime }}\mu _{i}(\theta _{-i}^{\prime }|h^{t-1},\theta
_{i})\sigma _{-i}(a_{-i}^{t}|h^{t-1},\theta _{-i}^{\prime })\right] }_{=\bar{%
\sigma}_{-i}(a_{-i}^{t}|h^{t-1},\theta _{i})} \\
=\;& \bar{\sigma}_{-i}(a_{-i}^{t}|h^{t-1},\theta _{i}).
\end{align*}%
Therefore, since $(\mu ,\sigma )\in \Psi ^{\chi }$, with some rearrangement,
it follows that
\begin{align*}
\mu _{i}(\theta _{-i}|h^{t},\theta _{i})& =\frac{\mu _{i}(\theta
_{-i}|h^{t-1},\theta _{i})\sigma _{-i}^{\chi }(a_{-i}^{t}|h^{t-1},\theta
_{-i},\theta _{i})}{\sum_{\theta _{-i}^{\prime }\in \Theta _{-i}}\mu
_{i}(\theta _{-i}^{\prime }|h^{t-1},\theta _{i})\sigma _{-i}^{\chi
}(a_{-i}^{t}|h^{t-1},\theta _{-i}^{\prime },\theta _{i})} \\
& =\frac{\mu _{i}(\theta _{-i}|h^{t-1},\theta _{i})[\chi \bar{\sigma}%
_{-i}(a_{-i}^{t}|h^{t-1},\theta _{i})+(1-\chi )\sigma
_{-i}(a_{-i}^{t}|h^{t-1},\theta _{-i})]}{\bar{\sigma}%
_{-i}(a_{-i}^{t}|h^{t-1},\theta _{i})} \\
& =\chi \mu _{i}(\theta _{-i}|h^{t-1},\theta _{i})+(1-\chi )\left[ \frac{\mu
_{i}(\theta _{-i}|h^{t-1},\theta _{i})\sigma _{-i}(a_{-i}^{t}|h^{t-1},\theta
_{-i})}{\sum_{\theta _{-i}^{\prime }}\mu _{i}(\theta _{-i}^{\prime
}|h^{t-1},\theta _{i})\sigma _{-i}(a_{-i}^{t}|h^{t-1},\theta _{-i}^{\prime })%
}\right] .
\end{align*}%
This completes the proof. $\blacksquare $

\subsection*{Proof of Proposition \protect\ref{prop:existence}}

The proof is similar to the proof for sequential equilibrium and proceeds in
three steps. First, for any finite multi-stage games with observed actions, $%
\Gamma $, we construct an $\epsilon $-perturbed game $\Gamma ^{\epsilon }$
that is identical to $\Gamma $ but every player in every information set has
to play any available action with probability at least $\epsilon $. Second,
we defined a cursed best-response correspondence for $\Gamma ^{\epsilon }$
and prove that the correspondence has a fixed point by Kakutani's fixed
point theorem. Finally, in step 3, we use a sequence of fixed points in
perturbed games, with $\epsilon $ converging to 0, where the limit of this
sequence is a $\chi $-CSE.

\bigskip

\noindent\textbf{Step 1:}

Let $\Gamma^\epsilon$ be a game identical to $\Gamma$ but for each player $%
i\in N$, player $i$ must play any available action in every information set $%
\mathcal{I}_i = (\theta_i, h^t)$ with probability at least $\epsilon$ where $%
\epsilon < \frac{1}{\sum_{j=1}^n |A_j|}$. Let $\Sigma^\epsilon =
\times_{j=1}^n \Sigma_j^\epsilon$ be set of feasible behavioral strategy
profiles for players in the perturbed game $\Gamma^\epsilon$. For any
behavioral strategy profile $\sigma \in \Sigma^\epsilon$, let $%
\mu^\chi(\cdot) \equiv (\mu_i^\chi(\cdot))_{i=1}^n$ be the belief system
induced by $\sigma$ via $\chi$-cursed Bayes' rule. That is, for each player $%
i\in N$, information set $\mathcal{I}_i = (\theta_i, h^t)$ where $h^t =
(h^{t-1}, a^t)$ and type profile $\theta_{-i} \in \Theta_{-i}$,
\begin{align*}
\mu_i^\chi(\theta_{-i}|h^t, \theta_i) = & \; \chi
\mu_i^\chi(\theta_{-i}|h^{t-1}, \theta_i) \; + \\
&\qquad\qquad (1-\chi)\left[\frac{\mu_i^\chi(\theta_{-i}|h^{t-1}, \theta_i)
\sigma_{-i}(a_{-i}^t | h^{t-1}, \theta_{-i})}{\sum_{ \theta_{-i}^{\prime}\in
\Theta_{-i}}\mu_i^\chi (\theta^{\prime}_{-i}|h^{t-1}, \theta_i)
\sigma_{-i}(a_{-i}^t | h^{t-1}, \theta^{\prime}_{-i})} \right].
\end{align*}
Notice that the $\chi$-cursed Bayes' rule is only defined on the framework
of multi-stage games with observed actions. As $\sigma$ is fully mixed, the
belief system is uniquely pinned down.

Finally, let $B^{\epsilon }:\Sigma^{\epsilon }\rightrightarrows \Sigma
^{\epsilon }$ be the cursed best response correspondence which maps any
behavioral strategy profile $\sigma \in \Sigma^{\epsilon }$ to the set of $%
\epsilon $-constrained behavioral strategy profiles $\tilde{\sigma}\in
\Sigma^{\epsilon }$ that are best replies given the belief system $\mu
^{\chi }(\cdot)$.

\bigskip \noindent\textbf{Step 2:}

Next, fix any $0<\epsilon <\frac{1}{\sum_{j=1}^{n}|A_{j}|}$ and show that $%
B^{\epsilon }$ has a fixed point by Kakutani's fixed point theorem. We check
the conditions of the theorem:

\begin{itemize}
\item[1.] It is straightforward that $\Sigma^\epsilon$ is compact and convex.

\item[2.] For any $\sigma \in \Sigma^\epsilon$, as $\mu^\chi(\cdot)$ is
uniquely pinned down by $\chi$-cursed Bayes' rule, it is straightforward
that $B^\epsilon(\sigma)$ is non-empty and convex.

\item[3.] To verify that $B^{\epsilon }$ has a closed graph, take any
sequence of $\epsilon $-constrained behavioral strategy profiles $%
\{\sigma^{k}\}_{k=1}^{\infty }\subseteq \Sigma^{\epsilon }$ such that $%
\sigma^{k}\rightarrow \sigma\in \Sigma^{\epsilon }$ as $k\rightarrow \infty $%
, and any sequence $\{\tilde{\sigma}^{k}\}_{k=1}^{\infty }$ such that $%
\tilde{\sigma}^{k}\in B^{\epsilon }(\sigma^{k}) $ for any $k$ and $\tilde{%
\sigma}^{k}\rightarrow \tilde{\sigma}$. We want to prove that $\tilde{\sigma}%
\in B^{\epsilon }(\sigma)$.

Fix any player $i\in N$ and information set $\mathcal{I}_{i}=(\theta
_{i},h^{t})$. For any $\sigma \in \Sigma ^{\epsilon }$, recall that $\sigma
_{-i}^{\chi }(\cdot )$ is player $i$'s $\chi $-cursed perceived behavioral
strategies of other players induced by $\sigma $. Specifically, for any type
profile $\theta \in \Theta $, non-terminal history $h^{t-1}$ and action
profile $a_{-i}^{t}\in A_{-i}(h^{t-1})$,
\begin{equation*}
\sigma _{-i}^{\chi }(a_{-i}^{t}|h^{t-1},\theta _{-i},\theta _{i})=\chi \bar{%
\sigma}_{-i}(a_{-i}^{t}|h^{t-1},\theta _{i})+(1-\chi )\sigma
_{-i}(a_{-i}^{t}|h^{t-1},\theta _{-i}).
\end{equation*}%
Additionally, recall that $\rho _{i}^{\chi }(\cdot )$ is player $i$'s belief
about the terminal nodes (conditional on the history and type profile),
which is also induced by $\sigma $. Since $\mu ^{\chi }(\cdot )$ is
continuous in $\sigma $ we have thaat $\sigma _{-i}^{\chi }(\cdot )$ and $%
\rho _{i}^{\chi }(\cdot )$ are also continuous in $\sigma $.

We further define
\begin{align*}
\mathcal{S}_{\mathcal{I}_{i}}^{k}& \equiv \left\{ \sigma _{i}^{\prime }\in
\Sigma _{i}^{\epsilon }:\sigma _{i}^{\prime }(\;\cdot \;|\mathcal{I}_{i})=%
\tilde{\sigma}_{i}^{k}(\;\cdot \;|\mathcal{I}_{i})\right\} , \\
\mathcal{S}_{\mathcal{I}_{i}}& \equiv \left\{ \sigma _{i}^{\prime }\in
\Sigma _{i}^{\epsilon }:\sigma _{i}^{\prime }(\;\cdot \;|\mathcal{I}_{i})=%
\tilde{\sigma}_{i}(\;\cdot \;|\mathcal{I}_{i})\right\} .
\end{align*}%
Since $\tilde{\sigma}^{k}\in B^{\epsilon }(\sigma^{k})$, for any $\sigma
_{i}^{\prime }\in \Sigma _{i}^{\epsilon }$, we can obtain that

\begin{align*}
& \max_{\sigma _{i}^{\prime \prime }\in \mathcal{S}_{\mathcal{I}%
_{i}}^{k}}\left\{ \sum_{\theta _{-i}\in \Theta _{-i}}\sum_{h^{T}\in \mathcal{%
H}^{T}}\mu _{i}^{\chi }[\sigma^{k}](\theta _{-i}|h^{t},\theta _{i})\rho
_{i}^{\chi }(h^{T}|h^{t},\theta ,\sigma _{-i}^{\chi }[\sigma^{k}],\sigma
_{i}^{\prime \prime })u_{i}(h^{T},\theta _{i},\theta _{-i})\right\} \\
& \qquad \qquad \qquad \geq \sum_{\theta _{-i}\in \Theta
_{-i}}\sum_{h^{T}\in \mathcal{H}^{T}}\mu _{i}^{\chi }[\sigma^{k}](\theta
_{-i}|h^{t},\theta _{i})\rho _{i}^{\chi }(h^{T}|h^{t},\theta ,\sigma
_{-i}^{\chi }[\sigma^{k}],\sigma _{i}^{\prime })u_{i}(h^{T},\theta
_{i},\theta _{-i}).
\end{align*}%
By continuity, as we take limits on both sides, we can obtain that
\begin{align*}
& \max_{\sigma _{i}^{\prime \prime }\in \mathcal{S}_{\mathcal{I}%
_{i}}}\left\{ \sum_{\theta _{-i}\in \Theta _{-i}}\sum_{h^{T}\in \mathcal{H}%
^{T}}\mu _{i}^{\chi }[\sigma ](\theta _{-i}|h^{t},\theta _{i})\rho
_{i}^{\chi }(h^{T}|h^{t},\theta ,\sigma _{-i}^{\chi }[\sigma],\sigma
_{i}^{\prime \prime })u_{i}(h^{T},\theta _{i},\theta _{-i})\right\} \\
& \qquad \qquad \qquad \geq \sum_{\theta _{-i}\in \Theta
_{-i}}\sum_{h^{T}\in \mathcal{H}^{T}}\mu _{i}^{\chi }[\sigma ](\theta
_{-i}|h^{t},\theta _{i})\rho _{i}^{\chi }(h^{T}|h^{t},\theta ,\sigma
_{-i}^{\chi }[\sigma],\sigma _{i}^{\prime })u_{i}(h^{T},\theta _{i},\theta
_{-i}).
\end{align*}%
Therefore, $\tilde{\sigma}\in B^{\epsilon }(\sigma)$.
\end{itemize}

By Kakutani's fixed point theorem, $B^\epsilon$ has a fixed point.

\bigskip

\noindent\textbf{Step 3:}

For any $\epsilon $, let $\sigma ^{\epsilon }$ be a fixed point of $%
B^{\epsilon }$ and $\mu ^{\epsilon }$ be the belief system induced by $%
\sigma ^{\epsilon }$ via $\chi $-cursed Bayes' rule. We combine these two
components and let $(\mu ^{\epsilon },\sigma ^{\epsilon })$ be the induced
assessment. We now consider a sequence of $\epsilon \rightarrow 0,$ where $%
\{(\mu ^{\epsilon },\sigma ^{\epsilon })\}$ is the corresponding sequence of
assessments.

By compactness and the finiteness of $\Gamma $, the Bolzano-Weierstrass
theorem guarantees the existence of a convergent subsequence of the
assessments. As $\epsilon \rightarrow 0$, let $(\mu ^{\epsilon },\sigma
^{\epsilon })\rightarrow (\mu ^{\ast },\sigma ^{\ast })$. By construction,
the limit assessment $(\mu ^{\ast },\sigma ^{\ast })$ satisfies $\chi $%
-consistency and sequential rationality. Hence, $(\mu ^{\ast },\sigma ^{\ast
})$ is a $\chi $-CSE. $\blacksquare $

\subsection*{Proof of Proposition \protect\ref{prop:upper_hemi}}

To prove $\Phi (\chi )$ is upper hemi-continuous in $\chi $, consider any
sequence of $\{\chi _{k}\}_{k=1}^{\infty }$ such that $\chi _{k}\rightarrow
\chi ^{\ast }\in \lbrack 0,1]$, and any sequence of CSE, $\{(\mu ^{k},\sigma
^{k})\}$, such that $(\mu ^{k},\sigma ^{k})\in \Phi (\chi _{k})$ for all $k$%
. Let $(\mu ^{\ast },\sigma ^{\ast })$ be the limit assessment, i.e., $(\mu
^{k},\sigma ^{k})\rightarrow (\mu ^{\ast },\sigma ^{\ast })$. We need to
show that $(\mu ^{\ast },\sigma ^{\ast })\in \Phi (\chi ^{\ast })$.

To simplify notation, for any player $i\in N$, any information set $\mathcal{%
I}_{i}=(h^{t},\theta _{i})$, any $\sigma _{i}^{\prime }\in \Sigma _{i}$, and
any $\sigma \in \Sigma $, the expected payoff under the belief system $\mu
^{\chi }(\cdot )$ induced by $\sigma $ is denoted as:
\begin{equation*}
\mathbb{E}_{\mu ^{\chi }{[\sigma ]}}\left[ u_{i}(\sigma _{i}^{\prime
},\sigma _{-i}|h^{t},\theta _{i})\right] \equiv \sum_{\theta _{-i}\in \Theta
_{-i}}\sum_{h^{T}\in \mathcal{H}^{T}}\mu _{i}^{\chi }(\theta
_{-i}|h^{t},\theta _{i})\rho _{i}^{\chi }(h^{T}|h^{t},\theta ,\sigma
_{-i}^{\chi },\sigma _{i}^{\prime })u_{i}(h^{T},\theta _{i},\theta _{-i}).
\end{equation*}

Suppose $(\mu ^{\ast },\sigma ^{\ast })\not\in \Phi (\chi ^{\ast })$. Then
there exists some player $i\in N$, some information set $\mathcal{I}%
_{i}=(h^{t},\theta _{i})$, some $\sigma _{i}^{\prime }\in \Sigma _{i}$, and
some $\epsilon >0$ such that
\begin{equation}
\mathbb{E}_{\mu ^{\chi ^{\ast }}{[\sigma ^{\ast }]}}\left[ u_{i}(\sigma
_{i}^{\prime },\sigma _{-i}^{\ast }|h^{t},\theta _{i})\right] -\mathbb{E}%
_{\mu ^{\chi ^{\ast }}{[\sigma ^{\ast }}]}\left[ u_{i}(\sigma _{i}^{\ast
},\sigma _{-i}^{\ast }|h^{t},\theta _{i})\right] >\epsilon .  \tag{A}
\label{umc_eq_1}
\end{equation}%
Since $\mu ^{\chi }(\cdot )$ is continuous in $\chi $, it follows that for
any strategy profile $\sigma$, $\sigma _{-i}^{\chi }(\cdot )$ and $\rho
_{i}^{\chi }(\cdot )$ are both continuous in $\chi $. As a result, there
exists a sufficiently large $M_{1}$ such that for every $k\geq M_{1}$,
\begin{equation}
\bigg |\mathbb{E}_{\mu ^{\chi ^{k}}{[\sigma ^{k}]}}\left[ u_{i}(\sigma
_{i}^{k},\sigma _{-i}^{k}|h^{t},\theta _{i})\right] -\mathbb{E}_{\mu ^{\chi
^{\ast }}{[\sigma ^{\ast }]}}\left[ u_{i}(\sigma _{i}^{\ast },\sigma
_{-i}^{\ast }|h^{t},\theta _{i})\right] \bigg |<\frac{\epsilon }{3}.  \tag{B}
\label{umc_eq_2}
\end{equation}%
Similarly, there exists a sufficiently large $M_{2}$ such that for every $%
k\geq M_{2}$,
\begin{equation}
\bigg |\mathbb{E}_{\mu ^{\chi ^{k}}{[\sigma ^{k}]}}\left[ u_{i}(\sigma
_{i}^{\prime },\sigma _{-i}^{k}|h^{t},\theta _{i})\right] -\mathbb{E}_{\mu
^{\chi ^{\ast }}{[\sigma ^{\ast }]}}\left[ u_{i}(\sigma _{i}^{\prime
},\sigma _{-i}^{\ast }|h^{t},\theta _{i})\right] \bigg |<\frac{\epsilon }{3}.
\tag{C}  \label{umc_eq_3}
\end{equation}%
Therefore, for any $k\geq \max \{M_{1},M_{2}\}$, inequalities (\ref{umc_eq_1}%
), (\ref{umc_eq_2}) and (\ref{umc_eq_3}) imply:
\begin{equation*}
\mathbb{E}_{\mu ^{\chi ^{k}}[\sigma ^{k}]}\left[ u_{i}(\sigma _{i}^{\prime
},\sigma _{-i}^{k}|h^{t},\theta _{i})\right] -\mathbb{E}_{\mu ^{\chi
^{k}}[\sigma ^{k}]}\left[ u_{i}(\sigma _{i}^{k},\sigma
_{-i}^{k}|h^{t},\theta _{i})\right] >\frac{\epsilon }{3},
\end{equation*}%
implying that $\sigma _{i}^{\prime }$ is a profitable deviation for player $%
i $ at information set $\mathcal{I}_{i}=(h^{t},\theta _{i})$, which
contradicts $(\mu ^{k},\sigma ^{k})\in \Phi (\chi _{k})$. Therefore, $(\mu
^{\ast },\sigma ^{\ast })\in \Phi (\chi ^{\ast })$, as desired. $%
\blacksquare $

\subsection*{Proof of Proposition \protect\ref{prop:equiv_def}}

Fix any $\chi \in \lbrack 0,1]$ and let $(\mu ,\sigma )$ be a $\chi $%
-consistent assessment. We prove the result by contradiction. Suppose $(\mu
,\sigma )$ does not satisfy $\chi $-dampened updating property. Then there
exists $i\in N$, $\tilde{\theta}\in \Theta $ and a non-terminal history $%
h^{t}$ such that
\begin{equation*}
\mu _{i}(\theta _{-i}|h^{t},\tilde{\theta}_{i})<\chi \mu _{i}(\theta
_{-i}|h^{t-1},\tilde{\theta}_{i}).
\end{equation*}%
Since $(\mu ,\sigma )$ is $\chi $-consistent, there exists a sequence $%
\{(\mu ^{k},\sigma ^{k})\}\subseteq \Psi ^{\chi }$ such that $(\mu
^{k},\sigma ^{k})\rightarrow (\mu ,\sigma )$ as $k\rightarrow \infty $. By
Lemma \ref{lemma:rearrange}, we know for this $i,\tilde{\theta}$ and $h^{t}$%
,
\begin{align*}
\mu _{i}^{k}(\tilde{\theta}_{-i}|h^{t},\tilde{\theta}_{i})=& \chi \mu
_{i}^{k}(\tilde{\theta}_{-i}|h^{t-1},\tilde{\theta}_{i})+(1-\chi )\left[
\frac{\mu _{i}^{k}(\tilde{\theta}_{-i}|h^{t-1},\tilde{\theta}_{i})\sigma
_{-i}^{k}(a_{-i}^{t}|h^{t-1},\tilde{\theta}_{-i})}{\sum_{\theta
_{-i}^{\prime }}\mu _{i}^{k}(\theta _{-i}^{\prime }|h^{t-1},\tilde{\theta}%
_{i})\sigma _{-i}^{k}(a_{-i}^{t}|h^{t-1},\theta _{-i}^{\prime })}\right] \\
\geq & \chi \mu _{i}^{k}(\tilde{\theta}_{-i}|h^{t-1},\tilde{\theta}_{i}).
\end{align*}%
As we take the limit $k\rightarrow \infty $ on both sides, we can obtain
that
\begin{equation*}
\mu _{i}(\tilde{\theta}_{-i}|h^{t},\tilde{\theta}_{i})=\lim_{k\rightarrow
\infty }\mu _{i}^{k}(\tilde{\theta}_{-i}|h^{t},\tilde{\theta}_{i})\geq
\lim_{k\rightarrow \infty }\chi \mu _{i}^{k}(\tilde{\theta}_{-i}|h^{t-1},%
\tilde{\theta}_{i})=\chi \mu _{i}(\tilde{\theta}_{-i}|h^{t-1},\tilde{\theta}%
_{i}),
\end{equation*}%
which yields a contradiction. $\blacksquare $

\subsection*{Proof of Corollary \protect\ref{coro:lower_bound}}

We prove the statement by induction on $t$. For $t=1$, by Proposition \ref%
{prop:equiv_def},
\begin{equation*}
\mu _{i}(\theta _{-i}|h^{1},\theta _{i})\geq \chi \mu _{i}(\theta
_{-i}|h_{\emptyset },\theta _{i})=\chi \mathcal{F}(\theta _{-i}|\theta _{i}).
\end{equation*}%
Next, suppose there is $t^{\prime }$ such that the statement holds for all $%
1\leq t\leq t^{\prime }-1$. At stage $t^{\prime }$, by Proposition \ref%
{prop:equiv_def} and the induction hypothesis, we can find that
\begin{equation*}
\mu _{i}(\theta _{-i}|h^{t^{\prime }},\theta _{i})\geq \chi \mu _{i}(\theta
_{-i}|h^{t^{\prime }-1},\theta _{i})\geq \chi \left[ \chi ^{t^{\prime }-1}%
\mathcal{F}(\theta _{-i}|\theta _{i})\right] =\chi ^{t^{\prime }}\mathcal{F}%
(\theta _{-i}|\theta _{i}).
\end{equation*}%
This completes the proof. $\blacksquare $

\newpage

\section{Omitted Proofs of Section \protect\ref{sec:application}}

\label{appendix_proof}

\subsection*{\protect\ref{subsec:signaling}\; Pooling Equilibria in
Signaling Games}

\subsubsection*{Proof of Proposition \protect\ref{prop:pooling_cse}}

Let the assessment $(\mu ,\sigma )$ be a pooling $\chi $-CSE. We want to
show that for any $\chi ^{\prime }\leq \chi $, the assessment $(\mu ,\sigma
) $ is also a $\chi ^{\prime }$-CSE. Consider any non-terminal history $%
h^{t-1} $, any player $i$, any $a_{i}^{t}\in A_{i}(h^{t-1})$ and any $\theta
\in \Theta $. We can first observe that
\begin{align*}
\bar{\sigma}_{-i}(a_{-i}^{t}|h^{t-1},\theta _{i})=& \;\sum_{\theta
_{-i}^{\prime }}\mu _{i}(\theta _{-i}^{\prime }|h^{t-1},\theta _{i})\sigma
_{-i}(a_{-i}^{t}|h^{t-1},\theta _{-i}^{\prime }) \\
=& \;\sigma _{-i}(a_{-i}^{t}|h^{t-1},\theta _{-i})\left[ \sum_{\theta
_{-i}^{\prime }}\mu _{i}(\theta _{-i}^{\prime }|h^{t-1},\theta _{i})\right]
\\
=& \;\sigma _{-i}(a_{-i}^{t}|h^{t-1},\theta _{-i})
\end{align*}%
where the second equality holds because $\sigma $ is a pooling behavioral
strategy profile, so $\sigma _{-i}$ is independent of other players' types.
For this pooling $\chi $-CSE, let $G^{\sigma }$ be the set of on-path
histories and $\tilde{G}^{\sigma }$ be the set of off-path histories. We can
first show that for every $h\in G^{\sigma }$, $i\in N$ and $\theta \in
\Theta $,
\begin{equation*}
\mu _{i}(\theta _{-i}|h,\theta _{i})=\mathcal{F}(\theta _{-i}|\theta _{i}).
\end{equation*}%
This can be shown by induction on $t$. For $t=1$, any $h^{1}=(h_{\emptyset
},a^{1})$ and any $\theta \in \Theta $, by Lemma \ref{lemma:rearrange}, we
can obtain that
\begin{align*}
\mu _{i}(\theta _{-i}|h^{1},\theta _{i})=& \chi \mu _{i}(\theta
_{-i}|h_{\emptyset },\theta _{i})+(1-\chi )\left[ \frac{\mu _{i}(\theta
_{-i}|h_{\emptyset },\theta _{i})\sigma _{-i}(a_{-i}^{1}|h_{\emptyset
},\theta _{-i})}{\bar{\sigma}_{-i}(a_{-i}^{1}|h_{\emptyset },\theta _{i})}%
\right] \\
=& \chi \mathcal{F}(\theta _{-i}|\theta _{i})+(1-\chi )\mathcal{F}(\theta
_{-i}|\theta _{i})\underbrace{\left[ \frac{\sigma
_{-i}(a_{-i}^{1}|h_{\emptyset },\theta _{-i})}{\bar{\sigma}%
_{-i}(a_{-i}^{1}|h_{\emptyset },\theta _{i})}\right] }_{=1} \\
=& \mathcal{F}(\theta _{-i}|\theta _{i}).
\end{align*}

Now, suppose there is $t^{\prime }$ such that the statement holds for $1\leq
t\leq t^{\prime }-1$. At stage $t^{\prime }$ and $h^{t^{\prime
}}=(h^{t^{\prime }-1},a^{t^{\prime }})\in G^{\sigma }$, by Lemma \ref%
{lemma:rearrange} and the induction hypothesis, we can again obtain that the
posterior belief is the prior belief
\begin{align*}
\mu _{i}(\theta _{-i}|h^{t^{\prime }},\theta _{i})=& \chi \mu _{i}(\theta
_{-i}|h^{t^{\prime }-1},\theta _{i})+(1-\chi )\left[ \frac{\mu _{i}(\theta
_{-i}|h^{t^{\prime }-1},\theta _{i})\sigma _{-i}(a_{-i}^{t^{\prime
}}|h^{t^{\prime }-1},\theta _{-i})}{\bar{\sigma}_{-i}(a_{-i}^{t^{\prime
}}|h^{t^{\prime }-1},\theta _{i})}\right] \\
=& \chi \mathcal{F}(\theta _{-i}|\theta _{i})+(1-\chi )\mathcal{F}(\theta
_{-i}|\theta _{i})\underbrace{\left[ \frac{\sigma _{-i}(a_{-i}^{t^{\prime
}}|h^{t^{\prime }-1},\theta _{-i})}{\bar{\sigma}_{-i}(a_{-i}^{t^{\prime
}}|h^{t^{\prime }-1},\theta _{i})}\right] }_{=1} \\
=& \mathcal{F}(\theta _{-i}|\theta _{i}).
\end{align*}%
Therefore, we have shown that players will not update their beliefs at every
on-path information set, so the belief system is independent of $\chi $.
Finally, for any off-path history $h^{t}\in \tilde{G}^{\sigma }$, by
Proposition \ref{prop:equiv_def}, we can find that the belief system
satisfies for any $\theta \in \Theta $,
\begin{equation*}
\mu _{i}(\theta _{-i}|h^{t},\theta _{i})\geq \chi \mu _{i}(\theta
_{-i}|h^{t-1},\theta _{i})\geq \chi ^{\prime }\mu _{i}(\theta
_{-i}|h^{t-1},\theta _{i}),
\end{equation*}%
implying that when $\chi ^{\prime }\leq \chi $, $\mu $ will still satisfy
the dampened updating property. Therefore, $(\mu ,\sigma )$ remains a $\chi
^{\prime }$-CSE. This completes the proof. $\blacksquare $

\subsubsection*{Proof of Claim \protect\ref{claim:counter_example}}

First observe that after player $1$ chooses $B$, it is strictly optimal for
player $2$ to choose $R$ for all beliefs $\mu _{2}(\theta _{1}|B)$, and
after player $1$ chooses $A$, it is optimal for player $2$ to choose $L$ if
and only if
\begin{equation*}
2\mu _{2}(\theta _{1}|A)+[1-\mu _{2}(\theta _{1}|A)]\geq 4\mu _{2}(\theta
_{1}|A)\iff \mu _{2}(\theta _{1}|A)\leq 1/3.
\end{equation*}

\noindent\textbf{Equilibrium 1.}

If both types of player 1 choose $A$, then $\mu _{2}(\theta _{1}|A)=1/4$, so
it is optimal for player 2 to choose $L$. Given $a(A)=L$ and $a(B)=R$, it is
optimal for both types of player 1 to choose $A$ as $2>1$. Hence $m(\theta
_{1})=m(\theta _{2})=A$, $a(A)=L$ and $a(B)=R$ is a pooling $\chi $-CSE for
any $\chi \in \lbrack 0,1]$.

\bigskip

\noindent\textbf{Equilibrium 2.}

In order to support $m(\theta _{1})=m(\theta _{2})=B$ to be an equilibrium,
player 2 has to choose $R$ at the off-path information set $A,$ which is
optimal if and only if $\mu _{2}(\theta _{1}|A)\geq 1/3$. In addition, by
Proposition \ref{prop:equiv_def}, we know in a $\chi $-CSE, the belief
system satisfies
\begin{equation*}
\mu _{2}(\theta _{2}|A)\geq \frac{3}{4}\chi \iff \mu _{2}(\theta _{1}|A)\leq
1-\frac{3}{4}\chi .
\end{equation*}%
Therefore, the belief system has to satisfy that $\mu _{2}(\theta _{1}|A)\in %
\left[ \frac{1}{3},1-\frac{3}{4}\chi \right] $, which requires $\chi \leq
8/9 $.

Finally, it is straightforward to verify that for any $\mu \in \left[ \frac{1%
}{3},1-\frac{3}{4}\chi \right] $, $\mu _{2}(\theta _{1}|A)=\mu $ satisfies $%
\chi $-consistency. Suppose type $\theta _{1}$ player 1 chooses $A$ with
probability $p$ and type $\theta _{2}$ player 1 chooses $A$ with probability
$q$ where $p,q\in (0,1)$. Given this behavioral strategy profile for player
1, by Lemma \ref{lemma:rearrange}, we have:
\begin{equation*}
\mu _{2}(\theta _{1}|A)=\frac{1}{4}\chi +(1-\chi )\left[ \frac{p}{p+3q}%
\right] .
\end{equation*}%
In other words, as long as $(p,q)$ satisfies
\begin{equation*}
q=\left[ \frac{4-4\mu -3\chi }{12-3\chi }\right] p,
\end{equation*}%
we can find that $\mu _{2}(\theta _{1}|A)=\mu $. Therefore, if $%
\{(p^{k},q^{k})\}\rightarrow (0,0)$ such that
\begin{equation*}
q^{k}=\left[ \frac{4-4\mu -3\chi }{12-3\chi }\right] p^{k},
\end{equation*}%
then $\mu _{2}^{k}(\theta _{1}|A)=\mu $ for all $k$. Hence, $%
\lim_{k\rightarrow \infty }\mu _{2}^{k}(\theta _{1}|A)=\mu $, suggesting
that $\mu _{2}(\theta _{1}|A)=\mu $ is indeed $\chi $-consistent. This
completes the proof. $\blacksquare $

\subsubsection*{Proof of Proposition \protect\ref{prop:bh-signaling}}

Here we provide a characterization of $\chi $-CSE of Game 1 and Game 2. For
the analysis of both games, we denote $\mu _{I}\equiv \mu _{2}(\theta
_{1}|m=I)$ and $\mu _{S}\equiv \mu _{2}(\theta _{1}|m=S)$.

\bigskip

\noindent\textbf{Analysis of Game BH 3.}

At information set $S$, given $\mu _{S}$, the expected payoffs of $C$, $D$, $%
E$ are $90\mu _{S}$, $30-15\mu _{S}$ and $15$, respectively. Therefore, for
any $\mu _{S}$, $E$ is never a best response. Moreover, $C$ is the best
response if and only if $90\mu _{S}\geq 30-15\mu _{S}$ or $\mu _{S}\geq 2/7$%
. Similarly, at information set $I$, given $\mu _{I}$, the expected payoffs
of $C$, $D$, $E$ are $30$, $45-45\mu _{I}$ and $15$, respectively.
Therefore, $E$ is strictly dominated, and $C$ is the best response if and
only if $30\geq 45-45\mu _{I}$ or $\mu _{I}\geq 1/3$. Now we consider four
cases.

\bigskip

\noindent \textbf{Case 1 [$m(\theta _{1})=I,m(\theta _{2})=S$]:}

By Lemma \ref{lemma:rearrange}, $\mu _{I}=1-\chi /2$ and $\mu _{S}=\chi /2$.
Moreover, since $\mu _{I}=1-\chi /2\geq 1/2$ for any $\chi $, player 2 will
choose $C$ at information set $I$. To support this equilibrium, player 2 has
to choose $C$ at information set $S$. In other words, $[(I,S);(C,C)]$ is
separating $\chi $-CSE if and only if $\mu _{S}\geq 2/7$ or $\chi \geq 4/7$.

\bigskip

\noindent \textbf{Case 2 [$m(\theta _{1})=S,m(\theta _{2})=I$]:}

By Lemma \ref{lemma:rearrange}, $\mu _{I}=\chi /2$ and $\mu _{S}=1-\chi /2$.
Because $\mu _{S}\geq 1-\chi /2\geq 1/2$, it is optimal for player 2 to
choose $C$ at information set $S$. To support this as an equilibrium, player
2 has to choose $D$ at information set $I$. Yet, in this case, type $\theta
_{2}$ player 1 will deviate to $S$. Therefore, this profile cannot be
supported as an equilibrium.

\bigskip

\noindent \textbf{Case 3 [$m(\theta _{1})=I,m(\theta _{2})=I$]:}

Since player 1 follows a pooling strategy, player 2 will not update his
belief at information set $I$, i.e., $\mu_I = 1/2$. $\chi$-dampened updating
property implies $\chi/2 \leq \mu_S \leq 1- \chi/2$. Since $\mu_I > 1/3$,
player 2 will choose $C$ at information set $I$. To support this profile to
be an equilibrium, player 2 has to choose $D$ at information set $S$, and
hence, it must be the case that $\mu_S \leq 2/7$. Coupled with the
requirement from $\chi$-dampened updating, the off-path belief has to
satisfy $\chi/2 \leq \mu_S \leq 2/7$. That is, $[(I,I); (C,D)]$ is pooling $%
\chi$-CSE if and only if $\chi/2 \leq 2/7$ or $\chi\leq 4/7$.

\bigskip

\noindent \textbf{Case 4 [$m(\theta _{1})=S,m(\theta _{2})=S$]:}

Similar to the previous case, since player 1 follows a pooling strategy,
player 2 will not update his belief at information set $S$, i.e., $\mu
_{S}=1/2$. Also, the $\chi $-dampened updating property suggests $\chi
/2\leq \mu _{I}\leq 1-\chi /2$. Because $\mu _{S}>2/7$, it is optimal for
player 2 to choose $C$ at information set $S$. To support this as an
equilibrium, player 2 has to choose $D$ at information set $I$. Therefore,
it must be that $\mu _{I}\leq 1/3$. Combined with the requirement of $\chi $%
-dampened updating, the off-path belief has to satisfy $\chi /2\leq \mu
_{I}\leq 1/3$. As a result, $[(S,S);(D,C)]$ is a pooling $\chi $-CSE if and
only if $\chi \leq 2/3$.

\bigskip

\noindent\textbf{Analysis of Game BH 4.}

At information set $I$, given $\mu _{I}$, the expected payoffs of $C$, $D$, $%
E$ are $30$, $45-45\mu _{I}$ and $35\mu _{I}$. Hence, $D$ is the best
response if and only if $\mu _{I}\leq 1/3$ while $E$ is the best response if
$\mu _{I}\geq 6/7$. For $1/3\leq \mu _{I}\leq 6/7$, $C$ is the best
response. On the other hand, since player 2's payoffs at information set $S$
are the same as in Game 1, player 2 will adopt the same decision
rule---player 2 will choose $C$ if and only if $\mu _{S}\geq 2/7$, and
choose $D$ if and only if $\mu _{S}\leq 2/7$. Now, we consider the following
four cases.

\bigskip

\noindent \textbf{Case 1 [$m(\theta _{1})=I,m(\theta _{2})=S$]:}

In this case, by Lemma \ref{lemma:rearrange}, $\mu _{I}=1-\chi /2$ and $\mu
_{S}=\chi /2$. To support this profile to be an equilibrium, player 2 has to
choose $E$ and $C$ at information set $I$ and $S$, respectively. To make it
profitable for player 2 to choose $E$ at information set $I$, it must be
that:
\begin{equation*}
\mu _{I}=1-\chi /2\geq 6/7\iff \chi \leq 2/7.
\end{equation*}%
On the other hand, player 2 will choose $C$ at information set $S$ if and
only if $\chi /2\geq 2/7$ or $\chi \geq 4/7$, which is not compatible with
the previous inequality. Therefore, this profile cannot be supported as an
equilibrium.

\bigskip

\noindent \textbf{Case 2 [$m(\theta _{1})=S,m(\theta _{2})=I$]:}

In this case, by Lemma \ref{lemma:rearrange}, $\mu _{I}=\chi /2$ and $\mu
_{S}=1-\chi /2$. To support this as an equilibrium, player 2 has to choose $%
D $ at both information sets. Yet, $\mu _{S}=1-\chi /2>2/7$, implying that
it is not a best reply for player 2 to choose $D$ at information set $S$.
Hence this profile also cannot be supported as an equilibrium.\bigskip

\noindent \textbf{Case 3 [$m(\theta _{1})=I,m(\theta _{2})=I$]:}

Since player 1 follows a pooling strategy, player 2 will not update his
belief at information set $I$, i.e., $\mu _{I}=1/2$. The $\chi $-dampened
updating property implies $\chi /2\leq \mu _{S}\leq 1-\chi /2$. Because $%
1/3<\mu _{I}=1/2<6/7$, player 2 will choose $C$ at information set $I$. To
support this profile as an equilibrium, player 2 has to choose $D$ at
information set $S$, and hence, it must be the case that $\mu _{S}\leq 2/7$.
Coupled with the requirement of $\chi $-dampened updating, the off-path
belief has to satisfy $\chi /2\leq \mu _{S}\leq 2/7$. That is, $%
[(I,I);(C,D)] $ is pooling $\chi $-CSE if and only if $\chi /2\leq 2/7$ or $%
\chi \leq 4/7$.

\bigskip

\noindent \textbf{Case 4 [$m(\theta _{1})=S,m(\theta _{2})=S$]:}

Similar to the previous case, since player 1 follows a pooling strategy,
player 2 will not update his belief at information set $S$, i.e., $\mu
_{S}=1/2$. Also, the $\chi $-dampened updating property implies $\chi /2\leq
\mu _{I}\leq 1-\chi /2$. Because $\mu _{S}>2/7$, it is optimal for player 2
to choose $C$ at information set $S$. To support this as an equilibrium,
player 2 can choose either $C$ or $D$ at information set $I$.

\noindent \textbf{Case 4.1:} To make it a best reply for player 2 to choose $%
D$ at information set $I$, it must be that $\mu _{I}\leq 1/3$. Combined with
the requirement from $\chi $-dampened updating, the off-path belief has to
satisfy $\chi /2\leq \mu _{I}\leq 1/3$. As a result, $[(S,S);(D,C)]$ is a
pooling $\chi $-CSE if and only if $\chi \leq 2/3$.

\noindent \textbf{Case 4.2:} To make it a best reply for player 2 to choose $%
C$ at information set $I$, it must be that $1/3\leq \mu _{I}\leq 6/7$.
Combined with the requirement from $\chi $-dampened updating, the off-path
belief has to satisfy
\begin{equation*}
\max \left\{ \frac{1}{2}\chi ,\;\frac{1}{3}\right\} \leq \mu _{I}\leq \min
\left\{ \frac{6}{7},1-\frac{1}{2}\chi \right\} .
\end{equation*}%
For any $\chi \in \lbrack 0,1]$, one can find $\mu _{I}$ that satisfies both
inequalities. Hence $[(S,S);(C,C)]$ is a pooling $\chi $-CSE for any $\chi $.

This completes the analysis of Game BH 3 and Game BH 4. $\blacksquare$

\subsection*{\protect\ref{subsec:pgg} \; A Public Goods Game with
Communication}

\subsubsection*{Proof of Proposition \protect\ref{prop:pgg_com}}

To prove this set of cost cutoffs form a $\chi $-CSE, we need to show that
there is no profitable deviation for any type at any subgame. First, at the
second stage where there are exactly $0\leq k\leq N-1$ players sending 1 in
the first stage, since no players will contribute, setting $C_{k}^{\chi }=0$
is indeed a best response. At the subgame where all $N$ players send 1 in
the first stage, we use $\mu _{i}^{\chi }(c_{-i}|N)$ to denote player $i$'s
cursed belief density. By Lemma \ref{lemma:rearrange}, the cursed belief
about all other players having a cost lower than $c$ is simply:
\begin{align*}
F^{\chi }(c)\equiv & \int_{\{c_{j}\leq c,\;\forall j\neq i\}}\mu _{i}^{\chi
}(c_{-i}^{\prime }|N)dc_{-i}^{\prime } \\
=&
\begin{cases}
\chi \left( c/K\right) ^{N-1}+(1-\chi )\left( c/C_{c}^{\chi }\right)
^{N-1}\quad \mbox{ if }c\leq C_{c}^{\chi } \\
1-\chi +\chi \left( c/C_{c}^{\chi }\right) ^{N-1}\qquad \qquad \qquad \;%
\mbox{
if }c>C_{c}^{\chi },%
\end{cases}%
\end{align*}%
and $C_{N}^{\chi }$ is the solution of the fixed point problem of $%
C_{N}^{\chi }=F^{\chi }(C_{N}^{\chi })$.

Moreover, in equilibrium, $C_{c}^{\chi }$ type of players would be
indifferent between sending 1 and 0 in the communication stage. Thus, given $%
C_{N}^{\chi }$, $C_{c}^{\chi }$ is the solution of the following equation
\begin{equation*}
0=\left( \frac{C_{c}^{\chi }}{K}\right) ^{N-1}\left[ -C_{c}^{\chi }+F^{\chi
}(C_{N}^{\chi })\right] .
\end{equation*}%
As a result, we obtain that in equilibrium, $C_{c}^{\chi }=C_{N}^{\chi
}=F^{\chi }(C_{N}^{\chi })\leq 1$ and denote this cost cutoff by $C^{\ast
}(N,K,\chi )$. Substituting it into $F^{\chi }(c)$, gives:
\begin{equation*}
C^{\ast }(N,K,\chi )-\chi \left[ \frac{C^{\ast }(N,K,\chi )}{K}\right]
^{N-1}=1-\chi .
\end{equation*}%
In the following, we show that for any $N\geq 2$ and $\chi $, the cutoff $%
C^{\ast }(N,K,\chi )$ is unique.

\bigskip

\noindent\textbf{Case 1:} When $N=2$, the cutoff $C^*(2, K,\chi)$ is the
unique solution of the linear equation
\begin{align*}
C^*(2, K,\chi) - \chi\left[\frac{C^*(2,K,\chi)}{K} \right] = 1 - \chi \iff
C^*(2,K,\chi) = \frac{K-K\chi}{K - \chi}.
\end{align*}

\bigskip

\noindent \textbf{Case 2:} For $N\geq 3$, we define the function $%
h(y):[0,1]\rightarrow \mathbb{R}$ where
\begin{equation*}
h(y)=y-\chi \left( \frac{y}{K}\right) ^{N-1}-(1-\chi ).
\end{equation*}%
It suffices to show that $h(y)$ has a unique root in $[0,1]$. When $\chi =0$%
, $h(y)=y-1$ which has a unique root at $y=1$. In the following, we will
focus on the case where $\chi >0$. Since $h(y)$ is continuous, $%
h(0)=-(1-\chi )<0$ and $h(1)=\chi \left[ 1-(1/K)^{N-1}\right] >0$, there
exists a root $y^{\ast }\in (0,1)$ by the intermediate value theorem.
Moreover, as we take the second derivative, we can find that for any $y\in
(0,1)$,
\begin{equation*}
h^{\prime \prime }(y)=-\left( \frac{\chi }{K^{N-1}}\right)
(N-1)(N-2)y^{N-3}<0,
\end{equation*}%
implying that $h(y)$ is strictly concave in $[0,1]$. Furthermore, $h(0)<0$
and $h(1)>0$, so the root is unique, as illustrated in the left panel of
Figure \ref{fig:cursed_density}. This completes the proof. $\blacksquare $

\subsubsection*{Proof of Corollary \protect\ref{coro:pgg_limit}}

By Proposition \ref{prop:pgg_com}, we know the cutoff $C^{\ast }(N,K,\chi
)\leq 1$ and it satisfies
\begin{equation*}
C^{\ast }(N,K,\chi )-\chi \left[ \frac{C^{\ast }(N,K,\chi )}{K}\right]
^{N-1}=1-\chi .
\end{equation*}%
Therefore, when $\chi =0$, the condition becomes $C^{\ast }(N,K,0)=1$. In
addition, when $\chi =1$, the condition becomes
\begin{equation*}
C^{\ast }(N,K,1)-\left[ \frac{C^{\ast }(N,K,1)}{K}\right] ^{N-1}=0,
\end{equation*}%
implying $C^{\ast }(N,K,1)=0$.

For $\chi \in (0,1)$, to prove $C^{\ast }(N,K,\chi )$ is strictly decreasing
in $N$, $K$ and $\chi$, we consider a function $g(y;N, K,
\chi):(0,1)\rightarrow \mathbb{R}$ where $g(y;N,K, \chi)=y-\chi \lbrack
y/K]^{N-1}.$ For any $y\in (0,1)$ and fix any $K$ and $\chi$, we can observe
that when $N \geq 2$,
\begin{equation*}
g(y;N+1,K)-g(y;N,K)=-\chi \left[ \frac{y}{K}\right] ^{N}+\chi \left[ \frac{y%
}{K}\right] ^{N-1}>0,
\end{equation*}%
so $g(\cdot;N,K,\chi)$ is strictly increasing in $N$. Therefore, the cutoff $%
C^{\ast }(N,K,\chi )$ is strictly decreasing in $N$. Similarly, for any $y
\in (0,1)$ and fix any $N$ and $\chi$, observe that when $K > 1$,
\begin{equation*}
\frac{\partial{g}}{\partial{K}} = \chi(N-1)\left(\frac{y^{N-1}}{K^N}\right)
> 0,
\end{equation*}
which implies that cutoff $C^{\ast }(N,K,\chi )$ is also strictly decreasing
in $K$. For the comparative statics of $\chi$, we can rearrange the
equilibrium condition where
\begin{align*}
\frac{1-C^*(N,K,\chi)}{\chi} = 1 - \left[\frac{C^*(N,K,\chi)}{K} \right]%
^{N-1}.
\end{align*}
Since LHS is strictly decreasing in $\chi$, the equilibrium cutoff is also
strictly decreasing in $\chi$. Finally, taking the limit on both sides of
the equilibrium condition, we obtain:
\begin{equation*}
\lim_{N\rightarrow \infty }C^{\ast }(N,K,\chi )= \lim_{K\rightarrow \infty
}C^{\ast }(N,K,\chi ) = 1-\chi.
\end{equation*}
This completes the proof. $\blacksquare $

\subsection*{\protect\ref{subsec:centipede} \; The Centipede Game with
Altruistic Types}

\subsubsection*{Proof of Claim \protect\ref{claim:centipede}}

By backward induction, we know selfish player two will choose $T4$ for sure.
Given that player two will choose $T4$ at stage four, it is optimal for
selfish player one to choose $T3$. Now, suppose selfish player one will
choose $P1$ with probability $q_{1}$ and player two will choose $P2$ with
probability $q_{2}$. Given this behavioral strategy profile, player two's
belief about the other player being altruistic at stage two is:
\begin{equation*}
\mu =\frac{\alpha }{\alpha +(1-\alpha )q_{1}}.
\end{equation*}%
In this case, it is optimal for selfish player two to pass if and only if
\begin{equation*}
32\mu +4(1-\mu )\geq 8\iff \mu \geq \frac{1}{7}.
\end{equation*}%
At the equilibrium, selfish player two is indifferent between $T2$ and $P2$.
If not, say $32\mu +4(1-\mu )>8$, player two will choose $P2$. Given that
player two will choose $P2$, it is optimal for selfish player one to choose $%
P1$, which makes $\mu =\alpha $ and $\alpha >1/7$. However, we know $\alpha
\leq 1/7$ which yields a contradiction. On the other hand, if $32\mu
+4(1-\mu )<8$, then it is optimal for player two to choose $T2$ at stage
two. As a result, selfish player one would choose $T1$ at stage one, causing
$\mu =1$. In this case, player two would deviate to choose $P2$, which again
yields a contradiction. To summarize, in equilibrium, player two has to be
indifferent between $T2$ and $P2$, i.e., $\mu =1/7$. As we rearrange the
equality, we can obtain that
\begin{equation*}
\frac{\alpha }{\alpha +(1-\alpha )q_{1}^{\ast }}=\frac{1}{7}\iff q_{1}^{\ast
}=\frac{6\alpha }{1-\alpha }.
\end{equation*}%
Finally, since the equilibrium requires selfish player one to mix at stage
one, selfish player one has to be indifferent between $P1$ and $T1$.
Therefore,
\begin{equation*}
4=16q_{2}^{\ast }+2(1-q_{2}^{\ast })\iff q_{2}^{\ast }=\frac{1}{7}.
\end{equation*}%
This completes the proof. $\blacksquare $

\subsubsection*{Proof of Proposition \protect\ref{prop:centipede}}

By backward induction, we know selfish player two will choose $T4$ for sure.
Given this, it is optimal for selfish player one to choose $T3$. Now,
suppose selfish player one will choose $P1$ with probability $q_{1}$ and
player two will choose $P2$ with probability $q_{2}$. Given this behavioral
strategy profile, by Lemma \ref{lemma:rearrange}, player two's \textit{cursed%
} belief about the other player being altruistic at stage 2 is:
\begin{equation*}
\mu ^{\chi }=\chi \alpha +(1-\chi )\left[ \frac{\alpha }{\alpha +(1-\alpha
)q_{1}}\right] .
\end{equation*}%
In this case, it is optimal for player two to pass if and only if
\begin{equation*}
32\mu ^{\chi }+4(1-\mu ^{\chi })\geq 8\iff \mu ^{\chi }\geq \frac{1}{7}.
\end{equation*}%
We can first show that in equilibrium, it must be that $\mu ^{\chi }\leq 1/7$%
. If not, then it is strictly optimal for player two to choose $P2$.
Therefore, it is optimal for selfish player one to choose $P1$ and hence $%
\mu ^{\chi }=\alpha \leq 1/7$, which yields a contradiction. In the
following, we separate the discussion into two cases.

\bigskip

\noindent\textbf{Case 1:} $\chi\leq \frac{6}{7(1-\alpha)}$

In this case, we argue that player two is indifferent between $P2$ and $T2$.
If not, then $32\mu^\chi + 4(1-\mu^\chi) < 8$ and it is strictly optimal for
player two to choose $T2$. This would cause selfish player one to choose $T1$
and hence $\mu^\chi = 1 - (1-\alpha)\chi$. This yields a contradiction
because
\begin{align*}
\mu^\chi = 1 - (1-\alpha)\chi < \frac{1}{7} \iff \chi> \frac{6}{7(1-\alpha)}.
\end{align*}
Therefore, in this case, player two is indifferent between $T2$ and $P2$ and
thus,
\begin{align*}
\mu^\chi = \frac{1}{7} &\iff \chi\alpha + (1-\chi)\left[\frac{\alpha}{\alpha
+ (1-\alpha)q^\chi_1} \right] =\frac{1}{7} \\
&\iff \chi + \frac{1-\chi}{\alpha + (1-\alpha)q_1^\chi} = \frac{1}{7\alpha}
\\
&\iff \alpha + (1-\alpha)q_1^\chi = (1-\chi) \bigg/ \left[\frac{1}{7\alpha}
- \chi \right] \\
&\iff q_1^\chi = \left[\frac{7\alpha - 7\alpha\chi}{1-7\alpha\chi}-\alpha %
\right]\bigg/(1-\alpha).
\end{align*}
Since the equilibrium requires selfish player one to mix at stage 1, selfish
player one has to be indifferent between $P1$ and $T1$. Therefore,
\begin{align*}
4 = 16q_2^\chi + 2(1-q_2^\chi) \iff q_2^\chi = \frac{1}{7}.
\end{align*}

\bigskip

\noindent\textbf{Case 2:} $\chi> \frac{6}{7(1-\alpha)}$

In this case, we know for any $q_{1}^{\chi }\in \lbrack 0,1]$,
\begin{equation*}
\mu ^{\chi }=\chi \alpha +(1-\chi )\left[ \frac{\alpha }{\alpha +(1-\alpha
)q_{1}^{\chi }}\right] \leq 1-(1-\alpha )\chi <\frac{1}{7},
\end{equation*}%
implying that it is strictly optimal for player two to choose $T2$, and
hence it is strictly optimal for selfish player one to choose $T1$ at stage
1. This completes the proof. $\blacksquare $

\subsection*{\protect\ref{subsec:voting} \; Sequential Voting over Binary
Agendas}

\subsubsection*{Proof of Proposition \protect\ref%
{prop:voting_sophisticatedly}}

Assuming that $a^1(\theta_1) = b$ and all other types of voters as well as
type $\theta_1$ at stage 2 vote sincerely, voter $i$'s $\chi $-cursed belief
in the second stage upon observing $a_{-i}^{1}=(a,b)$ is

\begin{equation*}
\mu _{i}^{\chi }(\theta _{-i}|a_{-i}^{1}=(a,b))=%
\begin{cases}
p_{1}p_{3}\chi +\frac{p_{1}}{p_{1}+p_{2}}(1-\chi )\qquad \mbox{ if }\theta
_{-i}=(\theta _{3},\theta _{1}) \\
p_{2}p_{3}\chi +\frac{p_{2}}{p_{1}+p_{2}}(1-\chi )\qquad \mbox{ if }\theta
_{-i}=(\theta _{3},\theta _{2}) \\
p_{k}p_{l}\chi \qquad \qquad \qquad \quad \;\;\;\;\;\;\;\mbox{otherwise.}%
\end{cases}%
\end{equation*}

As mentioned in Section \ref{subsec:voting}, a voter would act as if he
perceives the other voters' (behavioral) strategies correctly in the last
stage. However, misunderstanding the link between the other voters' types
and actions would distort a voter's belief updating process. In other words,
a voter would perceive the strategies correctly but form beliefs
incorrectly. As a result, the continuation value of the $a$ vs $c$ subgame
to a type $\theta_1$ voter is simply the voter's $\chi $-cursed belief,
conditional on being pivotal, about there being at least one type $\theta_1$
voter among his opponents. Similarly, the continuation value of the $b$ vs $c
$ subgame is equal to the voter's conditional $\chi $-cursed belief about
there being at least one type $\theta_1$ or $\theta_2$ voter among his
opponents multiplied by $v$. Therefore, the continuation values to a type $%
\theta _{1}$ voter in the two possible subgames of the second stage are (let
$\Tilde{p}_2 \equiv \textstyle\frac{p_{1}}{p_{1}+p_{2}}$):
\begin{align*}
& a\text{ vs }c:\quad \chi \left( 1-(1-p_{1})^{2}\right) +(1-\chi )\Tilde{p}%
_2 \\
& b\text{ vs }c:\quad \left( 1-p_{3}^{2}\chi \right) v
\end{align*}
It is thus optimal for a type $\theta_1$ voter to vote for $b$ in the first
stage if
\begin{align}
& \chi \left( 1-(1-p_{1})^{2}\right) +(1-\chi )\Tilde{p}_2\leq \left(
1-p_{3}^{2}\chi \right) v  \notag \\
\iff & [2p_{1}-p_{1}^{2}-\Tilde{p}_2+p_{3}^{2}v]\chi \leq v-\Tilde{p}_2
\label{inequ_sophisticated}
\end{align}

Notice that the statement would automatically hold when $\chi=0$. In the
following, we want to show that given $v$ and $p$, if condition %
\eqref{inequ_sophisticated} holds for some $\chi \in (0,1]$, then it will
hold for all $\chi^{\prime}\leq \chi$. As $\chi>0$, we can rewrite condition %
\eqref{inequ_sophisticated} as
\begin{align}  \label{inequ_sophisticated_2}
2p_{1}-p_{1}^{2}-\Tilde{p}_2+p_{3}^{2}v \leq \frac{v-\Tilde{p}_2}{\chi}.
\tag{2'}
\end{align}

\noindent\textbf{Case 1:} $v - \Tilde{p}_2 < 0$.

In this case, we want to show that voting $b$ in the first stage is never
optimal for type $\theta_1$ voter. That is, we want to show condition %
\eqref{inequ_sophisticated_2} never holds for $v < \tilde{p}_2$. To see
this, we can first observe that the RHS is strictly increasing in $\chi$.
Therefore, it suffices to show
\begin{equation*}
2p_{1}-p_{1}^{2}-\Tilde{p}_2+p_{3}^{2}v > v-\Tilde{p}_2.
\end{equation*}
This is true because
\begin{align*}
2p_{1}-p_{1}^{2}-\Tilde{p}_2 + p_{3}^{2}v - \left(v-\Tilde{p}_2\right) &=
2p_{1}-p_{1}^{2} - (1 - p_{3}^{2})v \\
&> 2p_{1}-p_{1}^{2} - (1 + p_{3})p_1 = p_1 p_2 \geq 0
\end{align*}
where the second inequality holds as $v < \frac{p_1}{p_1 + p_2}$.

\bigskip

\noindent\textbf{Case 2:} $v - \Tilde{p}_2 \geq 0$.

Since the RHS of condition \eqref{inequ_sophisticated_2} is greater or equal
to 0, it will weakly increase as $\chi$ decreases. Thus, if condition %
\eqref{inequ_sophisticated_2} holds for some $\chi \in (0,1]$, it will also
hold for all $\chi^{\prime}\leq \chi$. This completes the proof. $%
\blacksquare $

\subsubsection*{Proof of Proposition \protect\ref{prop:voting_sincerely}}

Assuming that all voters vote sincerely in both stages, voter $i$'s $\chi $%
-cursed belief in the second stage upon observing $a_{-i}^{1}=(a,b)$ is
\begin{equation*}
\mu _{i}^{\chi }(\theta _{-i}|a_{-i}^{1}=(a,b))=%
\begin{cases}
p_{1}p_{2}\chi +\frac{p_{1}}{p_{1}+p_{3}}(1-\chi )\qquad \mbox{ if }\theta
_{-i}=(\theta _{1},\theta _{2}) \\
p_{2}p_{3}\chi +\frac{p_{3}}{p_{1}+p_{3}}(1-\chi )\qquad \mbox{ if }\theta
_{-i}=(\theta _{3},\theta _{2}) \\
p_{k}p_{l}\chi \qquad \qquad \qquad \quad \;\;\;\;\;\;\; \mbox{otherwise.}%
\end{cases}%
\end{equation*}

Similar to the proof of Proposition \ref{prop:voting_sophisticatedly}, the
continuation values to a type $\theta _{1}$ voter in the two possible
subgames of the second stage are (let $\Tilde{p}_3 \equiv \textstyle\frac{%
p_{1}}{p_{1}+p_{3}}$):
%\footnote{Note that the continuation value of the $a$ vs $c$ subgame to a type I voter is equal to the voter's $\chi$-cursed belief, conditional on being pivotal, about there being one type I voter among his opponents. %Similarly, the continuation value of the $b$ vs $c$ subgame is equal to the voter's conditional $\chi$-cursed belief about there being one type I or type II voter among his opponents multiplied by $v$.}
\begin{align*}
& a\text{ vs }c:\quad \chi \left( 1-(1-p_{1})^{2}\right) +(1-\chi )\Tilde{p}%
_3 \\
& b\text{ vs }c:\quad \left( 1-p_{3}^{2}\chi \right) v
\end{align*}%
Thus, it is optimal for a type $\theta _{1}$ voter to vote for $a$ in the
first stage if
\begin{align}
& \chi \left( 1-(1-p_{1})^{2}\right) +(1-\chi )\Tilde{p}_3\geq \left(
1-p_{3}^{2}\chi \right) v  \notag \\
\iff & \chi \left( 2p_{1}-p_{1}^{2}-\Tilde{p}_3+p_{3}^{2}v\right) \geq v-%
\Tilde{p}_3.  \label{inequ_sincere}
\end{align}

\noindent\textbf{Case 1:} $v - \Tilde{p}_3 > 0$.

In this case, we want to show that given $p$ and $v$, there exists $\Tilde{%
\chi}$ such that condition \eqref{inequ_sincere} holds if and only if $\chi
\geq \Tilde{\chi}$. Let $\tau \equiv 2p_{1}-p_{1}^{2}-\Tilde{p}_3+p_{3}^{2}v$%
. If $\tau>0$, then condition \eqref{inequ_sincere} holds if and only if $%
\chi \geq \Tilde{\chi} \equiv \frac{v - \Tilde{p}_3}{\tau}$. On the other
hand, if $\tau \leq 0 $, condition \eqref{inequ_sincere} will not hold for
all $\chi \in [0,1]$ and hence we can set $\Tilde{\chi} = 2$.

\bigskip

\noindent\textbf{Case 2:} $v - \Tilde{p}_3 \leq 0$.

In this case, we want to show that given $p$ and $v$, there exists $\Tilde{%
\chi}$ such that condition \eqref{inequ_sincere} holds if and only if $\chi
\leq \Tilde{\chi}$. If $\tau<0$, then condition \eqref{inequ_sincere} holds
if and only if $\chi\leq \frac{v - \Tilde{p}_3}{\tau}$ where the RHS is
greater or equal to 0. On the other hand, if $\tau \geq 0$, then condition %
\eqref{inequ_sincere} will hold for any $\chi \in [0,1]$ and hence we can
again set $\Tilde{\chi} = 2$. This completes the proof. $\blacksquare $

\subsection*{\protect\ref{subsec:dirty} \; The Dirty Faces Game}

\subsubsection*{Proof of Proposition \protect\ref{prop:dirty_static}}

When observing a clean face, a player will know that he has a dirty face
immediately. Therefore, choosing 1 (i.e., choosing $D$ at stage 1) when
observing a clean face is a strictly dominant strategy. In other words, for
any $\chi \in \lbrack 0,1]$, $\hat{\sigma}^{\chi }(O)=1$.

The analysis of the case where the player observes a dirty face is separated
into two cases.

\bigskip

\noindent\textbf{Case 1:} $\chi > \bar{\alpha}$

In this case, we show that $\hat{\sigma}^{\chi }(X)=T+1$ is the only $\chi $%
-CE. If not, suppose $\hat{\sigma}^{\chi }(X)=t$ where $t\leq T$ can be
supported as a $\chi $-CE. We can first notice that $\hat{\sigma}^{\chi
}(X)=1$ cannot be supported as a $\chi $-CE because it is strictly dominated
to choose 1 when observing a dirty face. For $2\leq t\leq T$, given the
other player $-i$ chooses $\hat{\sigma}^{\chi }(X)=t$, we can find player $%
-i $'s \emph{average strategy} is
\begin{equation*}
\bar{\sigma}_{-i}(j)=%
\begin{cases}
1-p\;\;\mbox{ if }\;\;j=1 \\
p\;\;\;\;\;\;\;\;\mbox{ if }\;\;j=t \\
0\;\;\;\;\;\;\;\;\mbox{ if }\;\;j\neq 1,t.%
\end{cases}%
\end{equation*}%
Therefore, the other player $-i$'s $\chi $-cursed strategy is:
\begin{align*}
& \sigma _{-i}^{\chi }(j|x_{i}=O)=%
\begin{cases}
\chi (1-p)+(1-\chi )\;\;\mbox{ if }\;\;j=1 \\
\chi p\qquad \qquad \qquad \;\;\;\;\;\mbox{ if }\;\;j=t \\
0\qquad \qquad \qquad \;\;\;\;\;\;\;\mbox{ if }\;\;j\neq 1,t,%
\end{cases}%
\qquad \mbox{ and } \\
& \sigma _{-i}^{\chi }(j|x_{i}=X)=%
\begin{cases}
\chi (1-p)\qquad \qquad \;\;\;\mbox{ if }\;\;j=1 \\
\chi p+(1-\chi )\quad \;\;\;\;\;\;\;\mbox{ if }\;\;j=t \\
0\qquad \qquad \qquad \;\;\;\;\;\;\;\mbox{ if }\;\;j\neq 1,t.%
\end{cases}%
\end{align*}%
In this case, given (player $i$ perceives that) player $-i$ chooses the $%
\chi $-cursed strategy, player $i$'s expected payoff to choose $2\leq j\leq
t $ when observing a dirty face is:
\begin{equation*}
(1-p)\left[ -\delta ^{j-1}\chi p\right] +p\left\{ \delta ^{j-1}\alpha \left[
\chi p+(1-\chi )\right] \right\} =p\delta ^{j-1}\underbrace{\left[ \alpha
-\chi (1+\alpha )(1-p)\right] }_{<0\iff \chi >\bar{\alpha}}<0.
\end{equation*}%
Hence, given the other player chooses $t$ when observing a dirty face, it is
strictly dominated to choose any $j\leq t$. Therefore, the only $\chi $-CE
is $\hat{\sigma}^{\chi }(X)=T+1$.

\bigskip

\noindent\textbf{Case 2:} $\chi < \bar{\alpha}$

In this case, we want to show that $\hat{\sigma}^{\chi }(X)=2$ is the only $%
\chi $-CE. If not, suppose $\hat{\sigma}(X)=t$ for some $t\geq 3$ can be
supported as a $\chi $-CE. We can again notice that since when observing a
dirty face, it is strictly dominated to choose 1, 1 is never a best
response. Given player $-i$ chooses $\hat{\sigma}^{\chi }(X)=t$, by the same
calculation as in \textbf{Case 1}, the expected payoff to choose $2\leq
j\leq t$ is:
\begin{equation*}
p\delta ^{j-1}\underbrace{\left[ \alpha -\chi (1+\alpha )(1-p)\right] }%
_{>0\iff \chi <\bar{\alpha}}>0,
\end{equation*}%
which is decreasing in $j$. Therefore, the best response to $\hat{\sigma}%
^{\chi }(X)=t$ is to choose 2 when observing a dirty face. As a result, the
only $\chi $-CE in this case is $\hat{\sigma}^{\chi }(X)=2$. This completes
the proof. $\blacksquare $

\subsubsection*{Proof of Proposition \protect\ref{prop:dynamic_dirty}}

When observing a clean face, the player would know that his face is dirty.
Thus, choosing $D$ at stage 1 is a strictly dominant strategy, and $\tilde{%
\sigma}^{\chi }(O)=1$ for all $\chi \in \lbrack 0,1]$. On the other hand,
the analysis for the case where the player observes a dirty face consists of
several steps.

\bigskip

\noindent \textbf{Step 1:} Assume that both players choosing $D$ at some
stage $\Bar{t}$. We claim that at stage $t\leq \Bar{t}$, the cursed belief $%
\mu ^{\chi }(X|t,X)=1-(1-p)\chi ^{t-1}$. We can prove this by induction on $%
t $. At stage $t=1$, the belief about having a dirty face is simply the
prior belief $p$. Hence this establishes the base case. Now suppose the
statement holds for any stage $1\leq t\leq t^{\prime }$ (and $t^{\prime }<%
\Bar{t}$). At stage $t^{\prime }+1$, by Lemma \ref{lemma:rearrange},
\begin{align*}
\mu ^{\chi }(X|t^{\prime }+1,X)& =\chi \mu ^{\chi }(X|t^{\prime },X)+(1-\chi
) \\
& =\chi \left[ 1-(1-p)\chi ^{t^{\prime }-1}\right] +(1-\chi ) \\
& =1-(1-p)\chi ^{t^{\prime }}
\end{align*}%
where the second equality holds by the induction hypothesis. This proves the
claim.

\bigskip

\noindent \textbf{Step 2:} Given the cursed belief computed in the previous
step, the expected payoff to choose $D$ at stage $t$ is:
\begin{align*}
\mu ^{\chi }(X|t,X)\alpha -\left[ 1-\mu ^{\chi }(X|t,X)\right] =& \left[
1-(1-p)\chi ^{t-1}\right] \alpha -\left[ (1-p)\chi ^{t-1}\right] \\
=& \;\alpha -(1-p)(1+\alpha )\chi ^{t-1},
\end{align*}%
which is increasing in $t$. Notice that at the first stage, the expected
payoff is $\alpha -(1-p)(1+\alpha )<0$ by Assumption (\ref{payoff_assumption}%
), so choosing $U$ at stage 1 is strictly dominated. Furthermore, the player
would choose $U$ at every stage when observing a dirty face if and only if
\begin{align*}
\mu ^{\chi }(X|T,X)\alpha -\left[ 1-\mu ^{\chi }(X|T,X)\right] \leq 0& \iff
\alpha -(1-p)(1+\alpha )\chi ^{T-1} \leq 0 \\
& \iff \chi \geq \bar{\alpha}^{\frac{1}{T+1}}.
\end{align*}%
As a result, both players choosing $\tilde{\sigma}^{\chi }(X)=T+1$ is a $%
\chi $-CSE if and only if $\chi \geq \bar{\alpha}^{\frac{1}{T+1}}$.

\bigskip

\noindent \textbf{Step 3:} In this step, we show both players choosing $%
\tilde{\sigma}^{\chi }(X)=2$ is a $\chi $-CSE if and only if $\chi \leq \bar{%
\alpha}$. We can notice that given the other player chooses $D$ at stage 2,
the player would know stage 2 would be the last stage regardless of his face
type. Therefore, it is optimal to choose $D$ at stage 2 as long as the
expected payoff of $D$ at stage 2 is positive. Consequently, both players
choosing $\tilde{\sigma}^{\chi }(X)=2$ is a $\chi $-CSE if and only if
\begin{align*}
\mu ^{\chi }(X|2,X)\alpha -\left[ 1-\mu ^{\chi }(X|2,X)\right] \geq 0& \iff
\alpha -(1-p)(1+\alpha )\chi \\
& \iff \chi \leq \bar{\alpha}.
\end{align*}

\bigskip

\noindent \textbf{Step 4:} Given the other player chooses $\tilde{\sigma}%
^{\chi }(X)>t$, as the game reaches stage $t$, the belief about the other
player choosing $U$ at stage $t$ is:
\begin{align*}
\underbrace{\mu ^{\chi }(X|t,X)}_{\mbox{prob. of dirty}}& \left[ \chi \mu
^{\chi }(X|t,X)+(1-\chi )\right] \\
& +\underbrace{\left[ 1-\mu ^{\chi }(X|t,X)\right] }_{\mbox{prob. of clean}}%
\left[ \chi \mu ^{\chi }(X|t,X)\right] =\mu ^{\chi }(X|t,X).
\end{align*}%
Furthermore, we denote the expected payoff of choosing $D$ at stage $t$ as
\begin{equation*}
\mathbb{E}\left[ u^{\chi }(D|t,X)\right] \equiv \mu ^{\chi }(X|t,X)\alpha
-\left( 1-\mu ^{\chi }(X|t,X)\right) .
\end{equation*}%
In the following, we claim that for any stage $2\leq t\leq T-2$, given the
other player will stop at some stage later than stage $t+2$ or never stop,
if it is optimal to choose $U$ at stage $t+1$, then it is also optimal for
you to choose $U$ at stage $t$. That is,
\begin{align*}
& \mathbb{E}\left[ u^{\chi }(D|t+1,X)\right] <\delta \mu ^{\chi }(X|t+1,X)%
\mathbb{E}\left[ u^{\chi }(D|t+2,X)\right] \\
\implies & \mathbb{E}\left[ u^{\chi }(D|t,X)\right] <\delta \mu ^{\chi
}(X|t,X)\mathbb{E}\left[ u^{\chi }(D|t+1,X)\right] .
\end{align*}%
To prove this claim, first observe that
\begin{align*}
\mathbb{E}\left[ u^{\chi }(D|t+1,X)\right] & <\delta \mu ^{\chi }(X|t+1,X)%
\mathbb{E}\left[ u^{\chi }(D|t+2,X)\right] \\
\iff (1+\alpha )\mu ^{\chi }(X|t+1,X)-1& <\delta \mu ^{\chi }(X|t+1,X)\left[
(1+\alpha )\mu ^{\chi }(X|t+2,X)-1\right] .
\end{align*}%
After rearrangement, the inequality is equivalent to
\begin{equation*}
\delta \chi \left[ \mu ^{\chi }(X|t+1,X)\right] ^{2}+\left[ \delta (1-\chi )-%
\frac{\delta }{1+\alpha }-1\right] \mu ^{\chi }(X|t+1,X)+\frac{1}{1+\alpha }%
>0.
\end{equation*}%
Consider a function $F:[0,1]\rightarrow \mathbb{R}$ where
\begin{equation*}
F(y)=\delta \chi y^{2}+\left[ \delta (1-\chi )-\frac{\delta }{1+\alpha }-1%
\right] y+\frac{1}{1+\alpha }.
\end{equation*}%
Since $\mu ^{\chi }(X|j,X)=1-(1-p)\chi ^{j-1}$ is increasing in $j$, it
suffices to complete the proof of the claim by showing there exists a unique
$y^{\ast }\in (0,1)$ such that $F$ is single-crossing on $[0,1]$ where $%
F(y^{\ast })=0$, $F(y)<0$ for all $y>y^{\ast }$, and $F(y)>0$ for all $%
y<y^{\ast }$. Because $F$ is continuous and

\begin{itemize}
\item $F(0)=\frac{1}{1+\alpha}>0$,

\item $F(1)=\delta \chi +\left[ \delta (1-\chi )-\frac{\delta }{1+\alpha }-1%
\right] +\frac{1}{1+\alpha }=-\frac{\alpha (1-\delta )}{1+\alpha }<0$.
\end{itemize}

By intermediate value theorem, there exists a $y^{\ast }\in (0,1)$ such that
$F(y^{\ast })=0$. Moreover, $y^{\ast }$ is the unique root of $F$ on $[0,1]$
because $F$ is a strictly convex parabola and $F(1)<0$. This establishes the
claim.

\bigskip

\noindent\textbf{Step 5:} For any $3\leq t \leq T$, in this step, we find
the conditions to support both players choosing $\tilde{\sigma}^\chi(X)=t$
as a $\chi$-CSE. We can first notice that both players choosing $\tilde{%
\sigma}^\chi(X)=t$ is a $\chi$-CSE if and only if

\begin{itemize}
\item[1.] $\mathbb{E}\left[u^\chi(D|t,X) \right] \geq 0$

\item[2.] $\mathbb{E}\left[u^\chi(D|t-1,X) \right] \leq
\delta\mu^\chi(X|t-1,X) \mathbb{E}\left[u^\chi(D|t,X) \right]$.
\end{itemize}

Condition 1 is necessary because if it fails, then it is better for the
player to choose $U$ at stage $t$ and get at least $0$. Condition 2 is also
necessary because if the condition doesn't hold, it would be profitable for
the player to choose $D$ before stage $t$. Furthermore, these two conditions
are jointly sufficient to support $\tilde{\sigma}^{\chi }(X)=t$ as a $\chi $%
-CSE by the same argument as \textbf{step 3}.

From condition 1, we can obtain that
\begin{align*}
\mathbb{E}\left[ u^{\chi }(D|t,X)\right] \geq 0\iff & (1+\alpha )\mu ^{\chi
}(X|t,X)-1\geq 0 \\
\iff & 1-(1-p)\chi ^{t-1}\geq \frac{1}{1+\alpha }\iff \chi \leq \bar{\alpha}%
^{\frac{1}{t-1}}.
\end{align*}%
In addition, by the calculation of \textbf{step 4}, we know
\begin{equation*}
\mathbb{E}\left[ u^{\chi }(D|t-1,X)\right] \leq \delta \mu ^{\chi }(X|t-1,X)%
\mathbb{E}\left[ u^{\chi }(D|t,X)\right] \iff F\left( \mu ^{\chi
}(X|t-1,X)\right) \geq 0,
\end{equation*}%
which is equivalent to
\begin{align*}
\mu ^{\chi }(X|t-1,X)\leq & \;\frac{\left[ 1+\frac{\delta }{1+\alpha }%
-\delta (1-\chi )\right] -\sqrt{\left[ 1+\frac{\delta }{1+\alpha }-\delta
(1-\chi )\right] ^{2}-4\delta \chi \left( \frac{1}{1+\alpha }\right) }}{%
2\delta \chi } \\
=& \;\frac{[(1+\alpha )(1+\delta \chi )-\alpha \delta ]-\sqrt{[(1+\alpha
)(1+\delta \chi )-\alpha \delta ]^{2}-4\delta \chi (1+\alpha )}}{2\delta
\chi (1+\alpha )}\equiv \kappa (\chi ).
\end{align*}%
Therefore, condition 2 holds if and only if
\begin{equation*}
1-(1-p)\chi ^{t-2}\leq \kappa (\chi )\iff \chi \geq \left( \frac{1-\kappa
(\chi )}{1-p}\right) ^{\frac{1}{t-2}}.
\end{equation*}%
In summary, both players choosing $\tilde{\sigma}^{\chi }(X)=t$ is a $\chi $%
-CSE if and only if
\begin{equation*}
\left( \frac{1-\kappa (\chi )}{1-p}\right) ^{\frac{1}{t-2}}\leq \chi \leq
\bar{\alpha}^{\frac{1}{t-1}}.
\end{equation*}%
This completes the proof. $\blacksquare $

\end{singlespace}

\end{document}